\RequirePackage{ifpdf}
\documentclass[nohyper,letterpaper,11pt,toc]{JHEP3}

\pdfoutput=1

\usepackage{graphicx}
\usepackage{amsthm}
\usepackage{amsmath}
\usepackage{graphicx,subfigure}
\usepackage{dcolumn}
\usepackage{bm}
\usepackage{slashed}
\usepackage{cancel}
\usepackage{tabularx}

\preprint{}

\newcommand{\be}{\begin{eqnarray}}
\newcommand{\ee}{\end{eqnarray}}
\newcommand{\bea}{\begin{eqnarray}}
\newcommand{\eea}{\end{eqnarray}}

\newcommand{\GeV}{{~\rm GeV}}
\newcommand{\gev}{{~\rm GeV}}

\newcommand{\mX}{m_{_\chi} }

\newcommand{\thetaXS}{\theta_{_\chi}}
\newcommand{\thetaXA}{\theta_{_\chi}}

\newcommand{\LamR}{\Lambda_{_R}}

\newcommand{\ZZ}{{\rm Z}}
\newcommand{\Zst}{\rm Z^{(*)}}

\newcommand{\Wpm}{\mathrm{W}^\pm}
\newcommand{\Wmp}{\mathrm{W}^\mp}
\newcommand{\Wst}{\rm W^{(*)}}
\newcommand{\Wp}{\rm W^+}
\newcommand{\Wm}{\rm W^-}

\newcommand{\SUWeak}{{\rm SU_{_W}(2)}}
\newcommand{\UoneY}{{\rm U_{_Y}(1)}}

\newcommand{\medf}{\psi}
\newcommand{\meds}{\varphi}
\newcommand{\mmedf}{M_{\medf}}
\newcommand{\mmeds}{M_{\meds}}

\title{Looking for new charged states at the LHC: Signatures of Magnetic and Rayleigh Dark Matter}
\author{Jia Liu$^{(a)}$, Brian Shuve$^{(b,c)}$, Neal Weiner$^{(a)}$, and Itay Yavin$^{(b,c)}$\\ \it{(a) Center for Cosmology and Particle Physics, Department of Physics, New York University, New York, NY 10003.} \\ \it{(b) Department of Physics \& Astronomy, McMaster University 1280 Main St. W. Hamilton, Ontario, Canada, L8S 4L8.}\\\it{(c) Perimeter Institute for Theoretical Physics 31 Caroline St. N, Waterloo, Ontario, Canada N2L 2Y5.}}

\abstract{
Magnetic and Rayleigh dark matter are models describing weak interactions of dark matter with electromagnetism through non-renormalizable operators of dimensions 5 and 7, respectively. Such operators motivate the existence of heavier states that couple to dark matter and are also charged under the electroweak  interactions. The recent hints of a gamma-ray line in the Fermi data suggest that these states may be light enough to be produced at the LHC. We categorize such states according to their charges and decay modes, and we examine the corresponding LHC phenomenology. We emphasize unconstrained models that can be discovered in targeted searches at the upgraded LHC run, while also enumerating models excluded by current data. Generally, models with $\SUWeak$-singlet states or models where the charged states decay predominantly to tau leptons and/or gauge bosons are still viable. We propose searches to constrain such models and, in particular, find superior performance over existing proposals for some multi-tau final states. Finally, we note several scenarios, especially those dominated by tau final states, that cannot be probed even with 300/fb at LHC14, motivating the further refinement of tau-lepton searches to improve sensitivity to such final states. 
}

\begin{document}

%
\begin{figure}[tbp]
\centering
\includegraphics[width=0.58\textwidth]{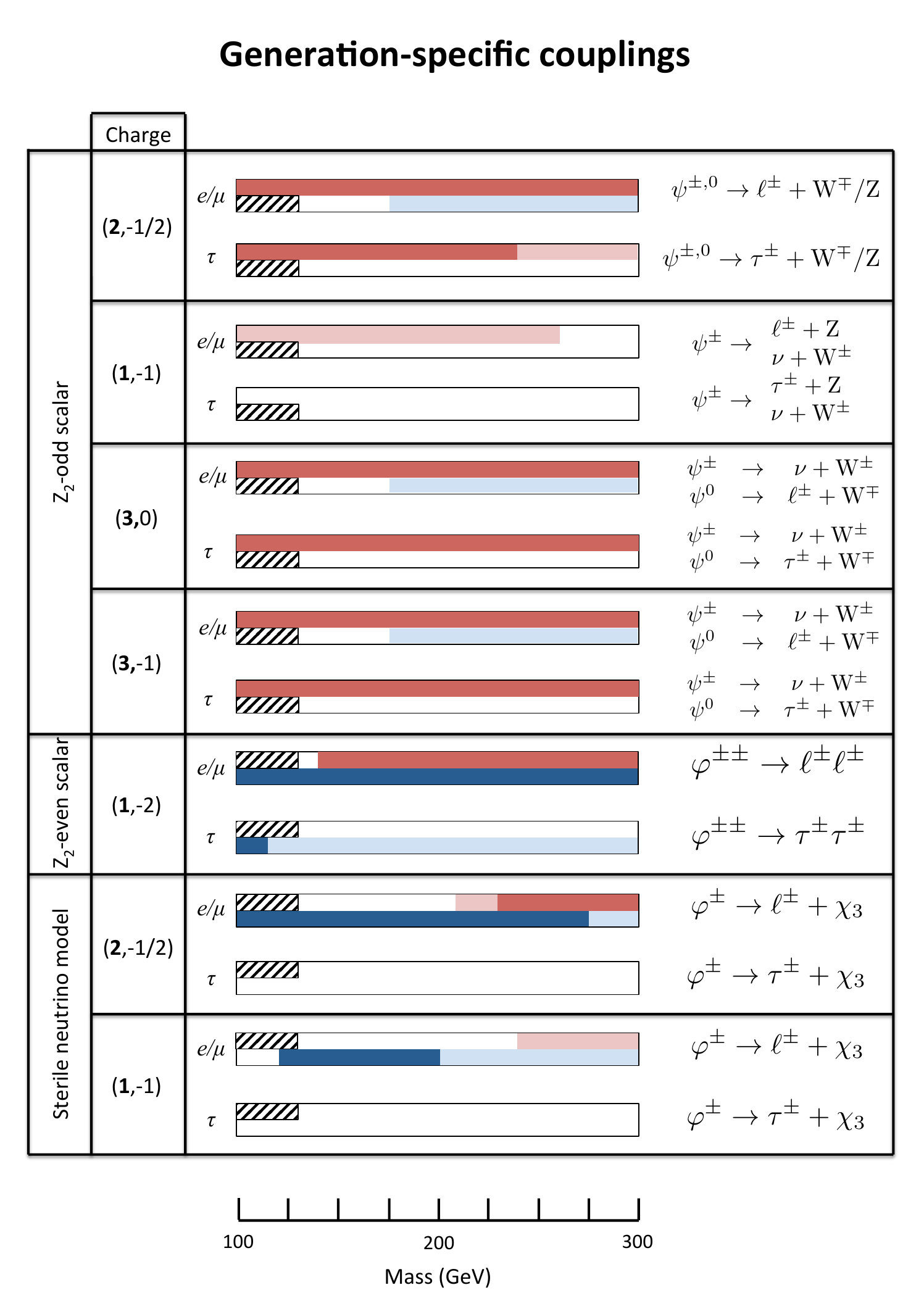}\\
\includegraphics[width=0.58\textwidth]{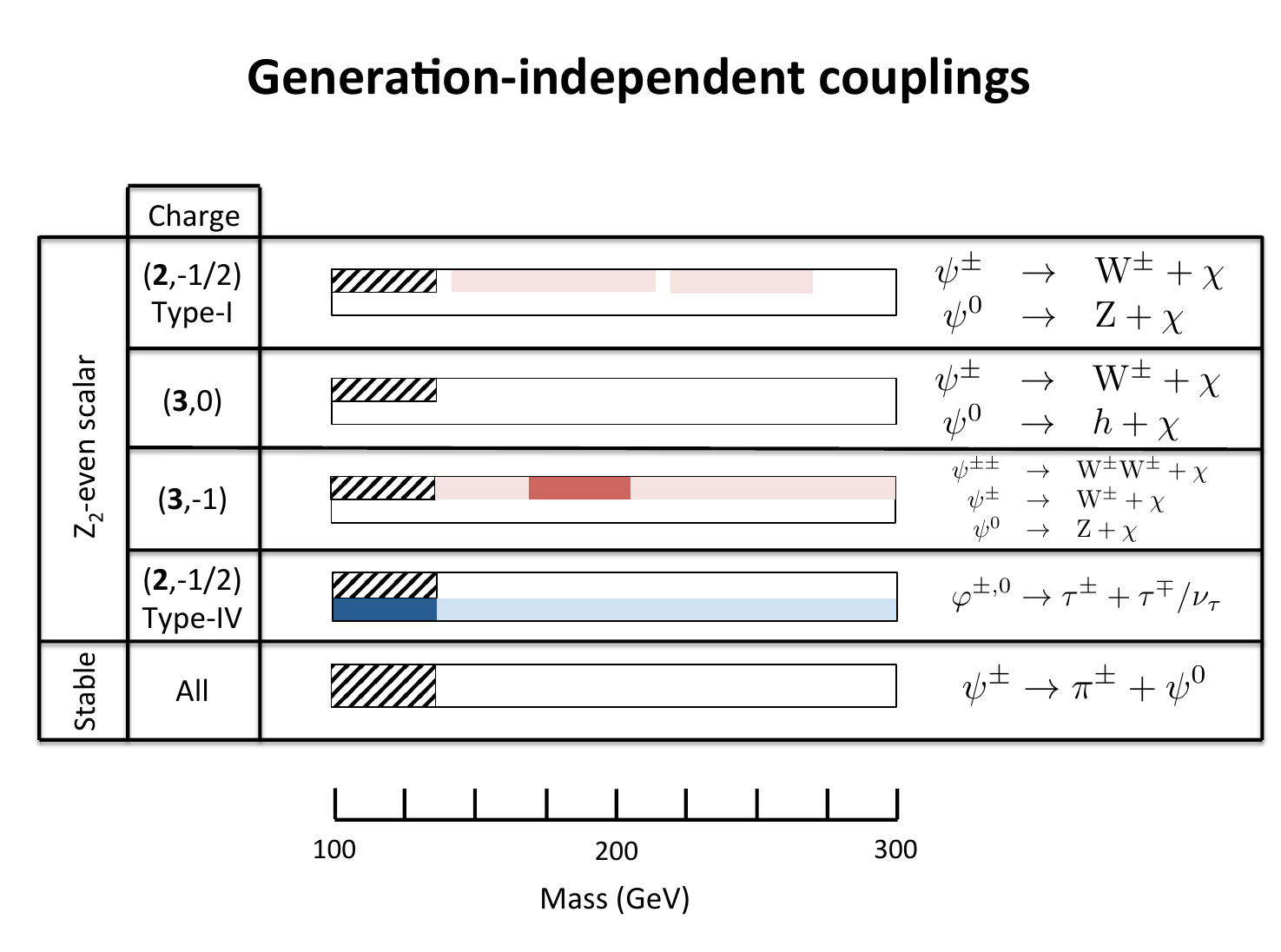}%
\caption{Estimates of model exclusions and discovery reach for fermions (red/upper bar) and scalars (blue/lower bar). Solid color indicates an excluded model, while light shading means that it can be discovered at $5\sigma$ with $<300\,\,\mathrm{fb}^{-1}$ at LHC14. The top (bottom) plot shows models with flavor-dependent (-independent) couplings. The striped regions indicate unstable dark matter. Masses of the Z$_2$-even fields below the dark matter mass are disfavored due to continuum photon constraints and modifications to the dark matter relic abundance, but are shown for completeness. The details of the models used and the searches considered are described in the text.}
\label{fig:summaryplot}
\end{figure}

\clearpage

\section{Introduction and summary}

The weakness of interactions is often understood in field theory as a sign that the corresponding operators are irrelevant. Consequently, the ``darkness'' of dark matter (DM) may be naturally interpreted as a consequence of DM having only irrelevant interactions with light, and more generally with the electroweak gauge bosons. The lowest-dimension irrelevant operator is the dipole operator
\be
\label{eqn:MiDMInteraction}
\mathcal{L}_{\rm dipole}= \left(\frac{\mu_\chi}{2}\right)\bar \chi^* \sigma_{\mu\nu} B^{\mu\nu} \chi + \mathrm{h.c.}
\ee
This operator requires the existence of two separate states $\chi$ and $\chi^*$, or it otherwise vanishes. Here, $\mu_\chi$ is the dipole strength.  In its absence (as is the case for a  Majorana particle), the next most-relevant operator is the Rayleigh operator, which can couple to both hypercharge and $\SUWeak$ gauge bosons. We parameterize these interactions as
\be
\label{eqn:RDMLagrangian}
\mathcal{L}_{\rm Rayleigh} = &\frac{1}{4\LamR^3}&~\Big\{ \bar{\chi}\chi \left( \cos\thetaXS B_{\mu\nu}B^{\mu\nu} + \sin\thetaXS{\rm Tr} W_{\mu\nu}W^{\mu\nu} \right) \\\nonumber  &+&\left. i~\bar{\chi}\gamma_5\chi \left( \cos\thetaXA B_{\mu\nu}\tilde{B}^{\mu\nu} + \sin\thetaXA{\rm Tr} W_{\mu\nu}\tilde{W}^{\mu\nu} \right)\right\}.
\ee
The Rayleigh scale $\LamR$ is a new energy scale similar to the Fermi scale in weak-interactions, which characterizes the strength of the Rayleigh interactions. The angle $\thetaXS$  parameterizes the relative strength between the coupling to hypercharge and the coupling to the weak $\SUWeak$ gauge bosons. 
Both dipole and Rayleigh operators give rise to the scattering, annihilation, and production of DM. The impact of the magnetic operator in a variety of observations has been the subject of several past works ~\cite{Bagnasco:1993st,Pospelov:2000bq,Sigurdson:2004zp,Gardner:2008yn, Masso:2009mu,Cho:2010br,An:2010kc,McDermott:2010pa,Chang:2010en,Banks:2010eh,DelNobile:2012tx}. The Rayleigh operator was studied in more recent dedicated analyses~\cite{Goodman:2010qn,Goodman:2010yf,Goodman:2010ku,Weiner:2012cb}, although older studies already appreciated its relevance for low energy experiments~\cite{Pospelov:2000bq}.

In the past year, a new impetus to these considerations was given by the observation of an excess of gamma-ray events from the center of the galaxy in the Fermi satellite data; see refs.~\cite{Ackermann:2012qk, Bringmann:2012vr,Weniger:2012tx,Tempel:2012ey,Su:2012ft} for early investigations, refs.~\cite{Hektor:2012jc,Hektor:2012kc,Hektor:2012ev,Whiteson:2012hr,Finkbeiner:2012ez,Rao:2012fh} for more detailed follow-ups, and refs.~\cite{Buchmuller:2012rc,Cohen:2012me,Cholis:2012fb,Blanchet:2012vq,Asano:2012zv} for constraints  on continuum emissions. Given the non-renormalizable nature of the above interactions  and their increasing strength at high energies, the LHC is a promising place to look for their signatures. This was already considered to some extent in \cite{Weiner:2012cb}, with a general operator analysis, but it proves difficult to say anything definitive in the most general case. More recently, in ref.~\cite{Weiner:2012gm}, a subset of the authors considered a simple renormalizable theory which gives rise to the interactions (\ref{eqn:MiDMInteraction}) and (\ref{eqn:RDMLagrangian}). Aside from the DM state, this theory includes a charged scalar ($\meds$) and a charged fermion ($\medf$) with a Yukawa interaction of the general form
\be
\label{eqn:Yukawa-type}
\mathcal{L}\supset \lambda\, \bar{\medf}\chi \meds.
\ee
In ref.~\cite{Weiner:2012gm} it was shown that for $\lambda \sim \sqrt{4\pi}$ and messenger masses in the range of several hundred GeV, the resulting dipole moment is $\mu_\chi \sim 10^{-3} $ in units of the nuclear magneton. This is of the right size to obtain the correct relic abundance as was shown in ref.~\cite{Weiner:2012gm} in Fig. 3. Moreover, the resulting Rayleigh scale $\LamR \sim 500\GeV $ is appropriate to explain the excess of gamma-ray events from the center of the galaxy seen in the Fermi satellite data. The details of these operators and their precise numerical values are not particularly important for the purpose of this paper and we refer the reader to refs.~\cite{Weiner:2012cb,Weiner:2012gm} for more details. What is most relevant for the current work is that, if the theory is perturbative with $\lambda \lesssim \sqrt{4\pi}$, then the charged states should have masses below approximately $300$ GeV. Since they are charged under the electroweak group, pair production of these states occurs at the LHC. The purpose of this paper is then to consider the phenomenology of such electroweak production, and to highlight the  LHC searches in which these new states may be discovered. 

We emphasize that, while the current work was motivated by the specific model of \cite{Weiner:2012gm}, it is relevant more generally to the production of $\mathcal{O}(100\,\,\mathrm{GeV})$ weakly charged states. As such, the bounds presented and searches proposed in this paper are applicable more generally to many other models. Indeed, several other models were proposed in connection with the Fermi line~\cite{Dudas:2012pb,Cline:2012nw,Choi:2012ap, Rajaraman:2012db,Buckley:2012ws,Das:2012ys,Heo:2012dk,Park:2012xq,Tulin:2012uq,Cline:2012bz,Bai:2012qy,Bringmann:2012mx,Bergstrom:2012mx,Fan:2012gr,D'Eramo:2012rr,Rajaraman:2012fu,Fan:2013qn}, many of which feature new weakly interacting states at the electroweak scale whether explicitly or implicitly. Other collider studies related to the Fermi line can be found in refs.~\cite{Lee:2012ph,Kopp:2013mi}, but our work studies a wider range of charge assignments and final states, assessing their status and observability.

In this paper, we consider  models where the charged fermion, $\medf$, and the charged boson, $\meds$, carry only electroweak charges and cannot be strongly produced. If $\medf$ and $\meds$ carry color charge, then loop-induced coupling of dark matter to gluons results in a bound of $\mmeds\sim \mmedf\gtrsim500$ GeV from LHC monojet searches and the XENON100 direct detection experiment \cite{Fox:2011pm}, in contradiction with the  requirement from perturbativity that $\mmedf\sim \mmeds\lesssim300$ GeV.

Our findings are presented in Fig.~\ref{fig:summaryplot}, which sums up the current experimental constraints on each model  considered and the discovery prospects for LHC14. We find that generic models with new states decaying into gauge bosons and light-flavor leptons ($e/\mu$) are almost entirely ruled out up to 300 GeV by current multilepton analyses. Models with multiple $\tau$ leptons in the final state, and/or with $\SUWeak$ singlet charge, are less constrained but are mostly within reach of LHC14. Throughout the paper, we use a benchmark of $300\,\,\mathrm{fb}^{-1}$ to determine whether a model can be discovered, although in many instances, considerably less luminosity is required for complete discovery in the 100-300 GeV window (masses below 100 GeV are mostly ruled out by LEP). There are, however, a few models with very challenging signatures (such as $\tau^+\tau^-+\cancel{E}_{\rm T}$ final states, disappearing charged track signatures \cite{Abreu:1999qr,Feng:1999fu,Gunion:1999jr,Ibe:2006de,Asai:2008sk,FileviezPerez:2008bj,Buckley:2009kv}, or entirely hadronic decays) that may not be accessible at LHC14, even with large integrated luminosity.

Table \ref{tbl:analyses} outlines the classes of searches most relevant for the exclusion and discovery of the different models we consider. The most constraining searches to-date at the LHC are multilepton$+\cancel{E}_{\rm T}$ for final states with $e$ and $\mu$, and same-sign dilepton searches with an additional hadronic tau for multi-tau final states. These analyses will continue to be important at 14 TeV for discovering or constraining the models we study, although we suggest in some instances how these searches can be optimized for particular final states by changing kinematic cuts, requiring additional hadronic tau tags, and using resonance reconstruction. In particular, we note that same-sign dilepton $+$ tau searches are ideal for discovering tau-rich final states, and for certain benchmark models (such as a Type IV Two Higgs Doublet model), the searches we propose have improved signal-to-background ratio and significance from other proposed analyses \cite{Kanemura:2011kx} (details can be found in Sections \ref{sec:constraintodd} and \ref{sec:proposaloddfermion}).

\begin{table}[t]
\begin{center}
\begin{tabular}{|l|c|}
\hline\hline
\multicolumn{1}{|c|}{{\bf Collider search}} & {\bf Model}   ~ \\ \hline \hline
 %
 %
$\ell^+\ell^-+\cancel{E}_{\rm T}$  &  Sterile neutrino model ($\meds$ coupled to $e$, $\mu$)~   \\ \hline
 %
%
 $\ge3\ell+\cancel{E}_{\rm T}$~ &  Odd scalar model ($e$ and $\mu$ in final state)~ \\
 & Odd fermion model (Type I 2HDM, triplet)~ \\\hline
Same-sign dilepton $+$ hadronic $\tau$~ &  Odd scalar model (doublet, $\tau$ in final state) \\
& Odd fermion model (Type IV 2HDM)~ \\\hline
Same-sign dilepton $+$ dijet resonance~ &  Odd fermion model (triplet) \\ \hline
Disappearing charged tracks~ &  Stable model~  \\ \hline
No distinctive signature~ & Sterile neutrino model ($\meds$ coupled to $\tau$) \\ 
& Odd scalar model (singlet, $\tau$ in final state) \\ \hline
\end{tabular}
\caption{Overview of collider searches relevant for constraining the models described in Sections \ref{sec:stable} through \ref{sec:sterile}, whether with current or future data. Details of the searches are given for each model in the relevant section.}\label{tbl:analyses}
\end{center}
\end{table}

The paper is organized as follows: readers who are less interested in the details can consult Fig.~\ref{fig:summaryplot} and the conclusions in Section~\ref{sec:conclusions} for a summary of the results of this paper.  Due to the large number of models, we focus throughout the paper on the most constraining signal(s) for each scenario, and we outline the classes of searches relevant for the exclusion and discovery of each in Table \ref{tbl:analyses}.  We begin the paper by outlining our classification of models and the setup of our analysis in Section \ref{sec:setup}. In each of Sections~\ref{sec:stable} through \ref{sec:sterile}, we describe a particular model, the current LHC constraints, and the proposals and prospects for discovery at LHC14.

\section{Setup}\label{sec:setup}

Our motivation is the study of charged states coupled to DM that generate DM-photon couplings as in Eq.~(\ref{eqn:Yukawa-type}). The interaction in Eq.~(\ref{eqn:Yukawa-type}) does not, however, completely fix the phenomenology of the model, since other interactions may allow the states $\meds$ and $\medf$ to decay. There are several phenomenological classes of models, and we discuss the four different possibilities in this section. The implications of each of these scenarios for LHC phenomenology are addressed in the subsequent sections.

The most straightforward and minimal resolution to the problem of new stable charged states is already present in the model itself. The interaction Eq.~(\ref{eqn:Yukawa-type}) respects a $U(1)^2$ symmetry under which the DM and messengers carry charges; this symmetry forbids any other coupling of $\meds$, $\medf$ by themselves to any current made of SM fields, and implies the existence of at least one other stable particle in addition to the dominant dark matter $\chi$. For an $\SUWeak$ multiplet, loop corrections make the electromagnetically charged components of the multiplet slightly heavier. This is similar to the situation occurring in models of anomaly-mediated supersymmetry  (SUSY) breaking~\cite{Randall:1998uk,Giudice:1998xp}, in which the wino state is the lightest supersymmetric particle (LSP). The charged components have a long lifetime and  decay to the neutral components through the emission of an off-shell $\Wpm$-boson. This possibility is particularly intriguing since it  results in more than one type of stable neutral particle.  Most of the DM is typically constituted by the $\chi$ state since it is lightest, but some relic abundance of the neutral components of $\meds$ and $\medf$ may be present. We discuss this model and the cosmological implications in Section \ref{sec:stable}, as well as a related model with a new, light component to DM in Section \ref{sec:sterile}.

Another possibility, with markedly different phenomenology, is to endow one of the messengers ($\medf$ or $\meds$) with some decay mode into SM particles. However, at least one messenger state together with $\chi$ must be odd under a Z$_2$ symmetry; otherwise the DM candidate, $\chi$,  is itself rendered unstable due to the interaction Eq.~(\ref{eqn:Yukawa-type}). We therefore draw the distinction between  models according to which of the charged particles, scalar or fermion, is Z$_2$ odd (i.e. has no direct decay into only SM particles). These models are examined in Sections \ref{sec:oddscalar} and \ref{sec:oddfermion}.

Another discrete choice that must be made is the electroweak charge carried by the messenger fields. The  representations we consider in this paper are shown in Table ~\ref{tbl:representations}. Each charge assignment results in different renormalizable couplings between $\medf/\meds$ and SM fields\footnote{Larger $\SUWeak$ multiplets and hypercharges are possible, but they tend to produce similar final states and the production cross sections are greatly enhanced over those in Table~\ref{tbl:representations}, making constraints much more severe and discovery more straightforward.}.  We consider  various possible gauge charges for each model.

{ \renewcommand{\arraystretch}{1.4}
\begin{table}[t]
\begin{center}
\begin{tabular}{|l|c|c|}
\hline
$\SUWeak \times \UoneY$ charge of $\medf$, $\meds$~ & $\meds$ is Z$_2$ odd & $\medf$ is Z$_2$ odd  ~ \\ \hline
 %
 %
 $(\mathbf{1}, -1)$~  &  $\medf  e^{\rm c},\,\ell H^*\medf^{\rm c}$~  & $\meds ~\epsilon_{ij} \ell_i \ell_j$ \\ \hline
 %
%
$\left(\mathbf{2}, -\frac{1}{2}\right)$~ &  $\medf H^* e^{\rm c},\, \ell\psi^{\rm c}$~ & $V(\meds,h)$~ \\ \hline
$\left(\mathbf{3}, 0\right)$~ &  $(\epsilon H)\,\psi^a\sigma^a \ell$~ & $H^*\meds H$~ \\ \hline
$\left(\mathbf{3}, -1\right)$~ &  $\ell (\psi^{\mathrm{c}})^{a}\sigma^a H^*$~ & $(\epsilon H)\meds^a\sigma^a H, \,(\epsilon \ell) \meds ^*\ell$~ \\ \hline
%
\end{tabular}
\caption{Possible electroweak charge assignments for $\medf$/$\meds$ are listed in the first column.  Higher multiplets and larger hypercharges are also allowed, but are mostly excluded due to their high production cross sections. In the second column, we list the allowed renormalizable interactions with SM fields when the scalar $\meds$ is Z$_2$ odd ($\sigma^a$ are the $\SUWeak$ generators). We ignore operators which are equivalent up to a field redefinition (such as $\meds\chi e^{\rm c}$ for charge $({\bf 1},-1)$). In the third column we list the corresponding operators when the fermion $\medf$  is Z$_2$ odd. We use $V(\meds,h)$ to indicate a generic two-scalar potential coupling $\meds$ to the SM Higgs field.}\label{tbl:representations}
\end{center}
\end{table}

\begin{figure}[tb]
\begin{center}
\includegraphics[width=0.45\textwidth]{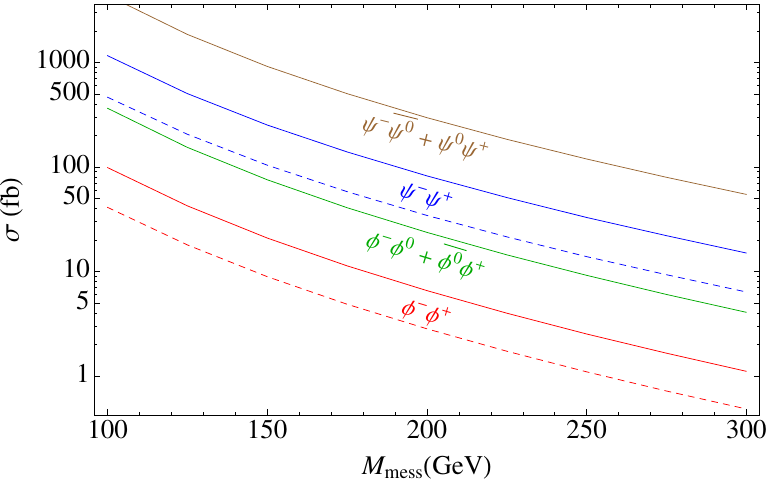}
\includegraphics[width=0.45\textwidth]{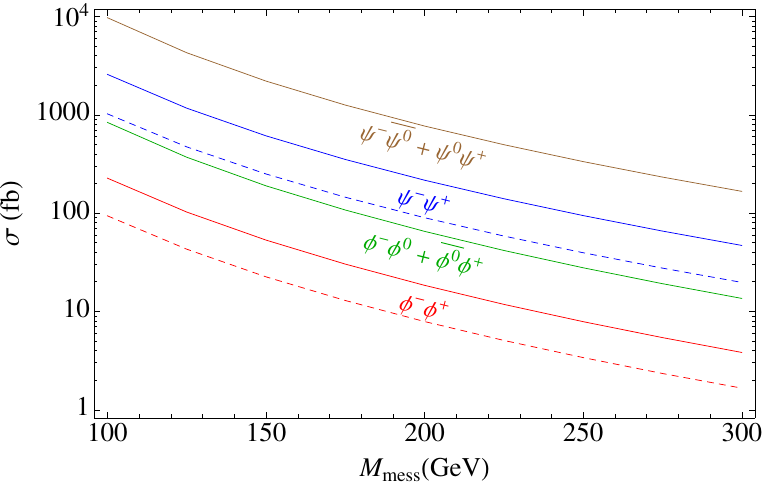}
\end{center}
\caption{LHC production cross sections plotted against the mass of the new messenger electroweak state for (left) 8 TeV and (right) 14 TeV. Solid and dashed lines show cross sections for electroweak charges $(\mathbf{2}, -1/2)$ and $(\mathbf{1}, -1)$, respectively. }
\label{fig:EW_production}
\end{figure}

Searches at the LHC for new electroweak states are hampered by the large QCD and top backgrounds. As a result, the  models most constrained by existing LHC searches are those with multiple ($\ge 2$) leptons in the final state, often accompanied by missing energy, $\cancel{E}_{\rm T}$. New electroweak particles decaying to tau leptons are generally less constrained than their light-flavor counterparts, as are models with only gauge bosons and $\cancel{E}_{\rm T}$ in the final state. The production cross section of the new electroweak states also plays a role, as can be seen in Fig.~\ref{fig:EW_production}: $\SUWeak$ doublets or higher representations have large production cross sections and are more readily constrained, while $\SUWeak$ singlets  easily evade experimental bounds.

We examine the collider bounds for each model in Sections \ref{sec:stable} to \ref{sec:sterile}. The classes of searches relevant for each model were outlined in Table \ref{tbl:analyses}. We find that, generally, the most relevant searches are the dilepton$+\cancel{E}_{\rm T}$, trilepton$+\cancel{E}_{\rm T}$, and same-sign dilepton analyses. 

For the collider studies  performed in our paper, we used \texttt{MadGraph 5} \cite{Alwall:2011uj} to simulate parton-level processes for signal and background, \texttt{Pythia 6} \cite{Sjostrand:2006za} for showering and hadronization, and \texttt{PGS 4} \cite{pgs4} for detector simulation (LHC card, anti-$k_{\rm T}$, $R=0.5$ algorithm for jets). For diboson processes and other analyses sensitive to soft radiation, we used MLM matching involving samples with up to two extra partons in the matrix element. Spectra and model files for the two Higgs doublet scenarios were calculated using \texttt{2HDMC} \cite{Eriksson:2009ws}. We checked the signal efficiency of \texttt{PGS} $\tau$-tagging using $W'\rightarrow\tau\nu$ and $Z'\rightarrow\tau\tau$ events for $m_{Z',W'}=100,\,200,\,300\GeV$; the efficiencies were in the range $40-50\%$, which is comparable to the benchmark values used by ATLAS \cite{ATLAS:2012ht}. We also measured the mistag rate on a dijet sample and found that it is $\approx4-6\%$ depending on $p_{\rm T}$, again comparable to the ATLAS values.

We use leading-order cross sections for signal and background for consistency, as next-to-leading-order corrections do not exist for all of the channels we study. The dominant backgrounds are electroweak, as are the signal processes, and $S/B$ is expected to be comparable for both leading-order and next-to-leading-order analyses.

\newpage
\clearpage

\section{Stable model}\label{sec:stable}

\subsection{The Model}\label{model:stable}

The first model we consider is one where the lightest neutral component of $\meds$ and/or $\medf$ is stable (in addition to the dark matter $\chi$). For concreteness, we consider $\medf^0$ as the lightest component, although our results extend easily to the scalar case. The neutral fermion $\medf^0$ is a thermal relic with abundance determined by its annihilation to $\chi\bar\chi$ through the Yukawa coupling in Eq.~(\ref{eqn:Yukawa-type})\footnote{The relic densities of $\meds$ and $\medf$ are also affected to a small degree by the messengers' interactions with electroweak gauge bosons, akin to the Higgsino in the MSSM.}. The interaction strength, $\lambda$, is required to be large ($\sim\sqrt{4\pi}$ for $\mmedf\sim \mmeds\sim100$ GeV) in order to account for the observed Fermi gamma-ray line, and this ensures that $\medf^0$ does not overclose the universe. Nevertheless, we wish to determine whether the $\medf^0$ relic is still large enough to violate experimental constraints on DM, particularly direct detection bounds that strongly constrain the  DM fraction coupling to the Z. We start by determining the $\medf^0$ relic abundance. Assuming $s$-wave-dominated annihilation, the thermally-averaged $\medf^0$  annihilation cross section is
\be
\langle\sigma\,v\rangle \approx \frac{|\lambda|^4}{64\pi}\,\frac{2\mmedf^2-m_\chi^2}{(m_\chi^2-\mmedf^2-\mmeds^2)^2}\,\sqrt{1-\frac{m_\chi^2}{\mmedf^2}}.
\ee
We calculate the relic abundance $\Omega_{\medf}\,h^2$ following the derivation of \cite{Kolb:1990vq}, and subsequently compute the relic number density of $\medf^0$ relative to that of a weakly interacting massive particle (WIMP) of the same mass which saturates the DM component. We find that the number density of $\medf^0$ is suppressed relative to DM of the same mass by a factor of $5\times10^{-7}$ to $10^{-3}$, depending on the precise values of $\lambda$, $\mmedf$, and $\mmeds$.

For $\mmedf$, $\mmeds\sim130-300$ GeV, which is our range of interest, a thermal relic with a $(\mathbf{2}, -1/2)$ electroweak charge is constrained by XENON100 to have a density that is $\lesssim10^{-8}$ times that of dark matter \cite{Aprile:2012nq}. As a result, essentially all of the models with stable $\medf^0$ are ruled out \emph{if} they couple to the Z, even though they make up only a tiny fraction of DM. Therefore, the most viable models with stable $\medf^0$ are those where the messenger fields have zero hypercharge, and as a consequence, the neutral states do not couple to the Z. The authors of \cite{Kopp:2013mi} found a similar result, even though $\medf^0$ annihilation proceeded through a different channel in that case.\footnote{If the Dirac fermion is split into a pseudodirac state, or the complex scalar is split into non-degenerate real and imaginary pieces, the Z-coupling can be suppressed \cite{Han:1997wn,Hall:1997ah,TuckerSmith:2001hy}. However, this requires a higher dimension operator, implying additional fields. For economy, we neglect this case.}

\begin{figure}[tbp]
\begin{center}
\includegraphics[width=0.45\textwidth]{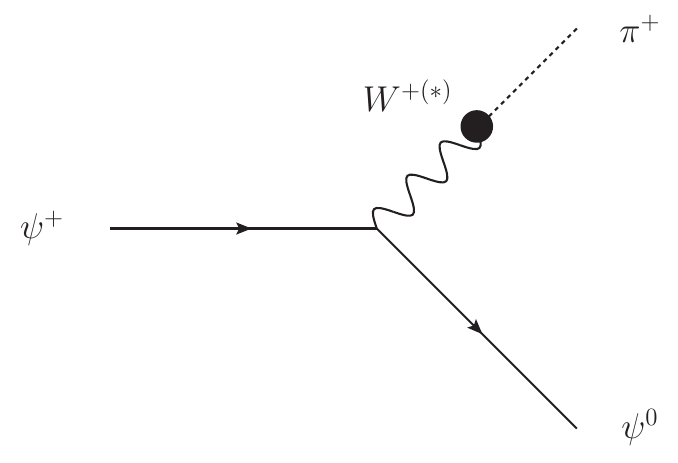}
\includegraphics[width=0.45\textwidth]{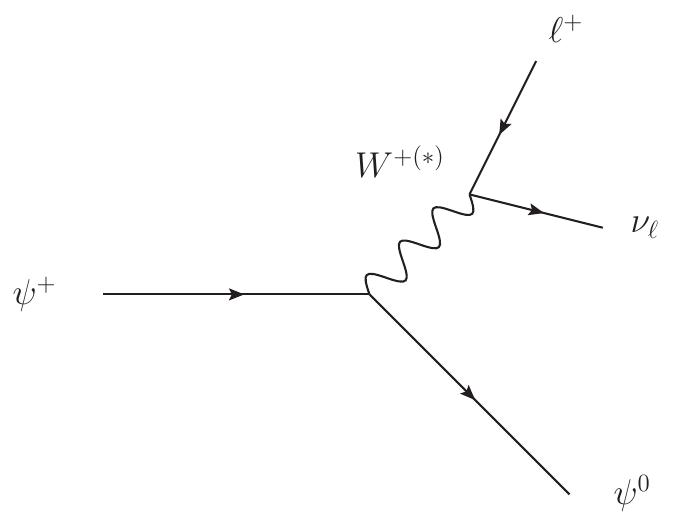}
\end{center}
\caption{Decays of the charged state $\medf^+$ into the stable neutral state $\psi^0$. The left (right) decay dominates when the mass difference $M_{\medf^+}-M_{\medf^0}>m_\pi$ ($< m_\pi$).}
\label{fig:stable_decays}
\end{figure}

When $\medf^0$ is stable, the heavier charged components in the $\SUWeak$ multiplet decay into it. The mass splitting between charged and neutral particles in the multiplet is generated radiatively after electroweak symmetry breaking, and for $\mmedf\sim100$ GeV, the splitting is $\mathcal O(100\,\,\mathrm{MeV})$. For a splitting induced entirely by electroweak gauge boson loops, $\Delta m\equiv m_{\psi^+}-m_{\psi^-}>m_\pi$, and production of $\medf^\pm$ produces $\medf^0$, along with a soft pion. Tree-level modifications to the splitting could increase the lifetime and can give $\Delta m < m_\pi$, leading to  a decay to a soft lepton and neutrino. We show the decays in Fig.~\ref{fig:stable_decays}.

\subsection{LHC Constraints} \label{collider:stable}
In models where the lightest neutral component of the new charged matter is stable, collider bounds are extremely weak. The phenomenology of such models has already been discussed in the literature \cite{Abreu:1999qr,Feng:1999fu,Gunion:1999jr,Ibe:2006de,Asai:2008sk,FileviezPerez:2008bj,Buckley:2009kv}, and there has been a recent paper applying these searches to a particular model of the Fermi line \cite{Kopp:2013mi}. As a result, we  sketch only a few relevant details here.

The lightest neutral component of $\meds$/$\medf$ leaves missing energy in the detector and can, in principle, be the subject of monojet searches. However, the production cross section is electroweak in strength and is typically several orders of magnitude (at least 3-5) below the current bounds \cite{Kopp:2013mi}. Pair production of the charged components, $\meds^\pm/\medf^\pm$, can give a more striking signature. When the mass splitting between charged and neutral states is $\sim m_\pi$, the charged particles travel some finite distance through the detector ($\sim10$ cm for an $\SUWeak$ triplet) before decaying. The signature consists of charged particle tracks that disappear midway through the detector, and monojets and monophotons can be used as a trigger \cite{Abreu:1999qr,Feng:1999fu,Gunion:1999jr,Ibe:2006de,Asai:2008sk,FileviezPerez:2008bj,Buckley:2009kv}. In principle, there is no SM background that gives this signature. There is, however, a significant contribution from SM combinatoric background where tracks are mis-reconstructed to give the illusion of a disappearing charged particle.

ATLAS has performed a search for anomaly-mediated SUSY breaking via the direct production of winos, which are $\SUWeak$ triplets displaying such a disappearing charged track signature. They rule out such triplets with masses $<100$ GeV when the mass splitting between the charged and neutral state arises solely from electroweak gauge boson loops, as is expected in a minimal model \cite{ATLAS:2012jp}. Therefore, triplet $\medf$ with no tree-level corrections to the mass splitting are ruled out up to the lower limit of our window of interest, while other gauge charges for $\medf$ have no constraints due to shorter track lengths.

\subsection{Proposals and Prospects for Future Searches}\label{search:stable}
As discussed in Section \ref{collider:stable}, the dominant way of searching for the new states $\psi$ and $\phi$ is via production of the charged components of the $\SUWeak$ multiplet, and performing either a monojet or a disappearing-charged-track search. We mention briefly the prospects for such searches at LHC14.

While some improvements in the monojet and monophoton bounds are to be expected with higher center-of-mass energy and luminosity, the current analyses are limited by systematic effects. Suggested improvements to such analyses (such as using the razor variables \cite{Fox:2012ee}) are  predicted to modify the bounds by a factor of $\sim2$, which is substantial but not sufficient for monojet and monophoton searches to constrain this model due to the small electroweak production rate of $\meds/\medf$.

Searches for disappearing charged tracks are more promising, as existing ATLAS searches at LHC7 already constrain some models with masses $\lesssim100$ GeV \cite{ATLAS:2012jp}.  An improvement in the bounds by a factor of two in the $\psi^\pm-\psi^0$ mass splitting could result in $\mmedf\lesssim200$ being ruled out for an $\SUWeak$ triplet. As such searches are still in their infancy, an increase in cross section and luminosity in the 8 TeV run, along with a better understanding of systematic errors,  could be sufficient to yield such an improvement in the bound; however, we refrain from speculating as to what the expected improvement will be. On the other hand, $\SUWeak$ doublets have lifetimes that are an order of magnitude shorter due to the larger mass splitting between lightest charged and neutral states; there is no bound achievable in the foreseeable future, and a great improvement in disappearing track searches would be necessary to probe such models.

\newpage
\clearpage

\section{Odd scalar model}\label{sec:oddscalar}

\subsection{The Model}\label{model:oddscalar}

When the charged scalar messenger $\meds$ is odd under the Z$_2$ symmetry, the fermion is Z$_2$-even and can  mix with the leptons of the SM,  rendering it unstable. The relevant operators are shown in the second column of Table~\ref{tbl:representations} for the different charge assignments. The mixing is either a direct bilinear mixing or  is achieved after the Higgs boson develops a vacuum expectation value (VEV)\footnote{After electroweak symmetry breaking, the two are equivalent up to field and coupling redefinitions.}. We discuss in some detail the ({\bf 2}, -1/2) charge assignment, and comment on the other possibilities below.

The possible interactions are
\be
\mathcal{L} \supset \mmedf \medf \medf^{\rm c} + \lambda^e \ell h e^{\rm c} +  y \medf h e^{\rm c} + m \ell \medf^{\rm c},
\ee
where $h$ is the SM Higgs, $\ell$ ($e^{\rm c}$) is any one of the left (right)-handed leptons, and flavor indices are suppressed. After electroweak symmetry breaking, this results in a mixing between the SM charged leptons and the heavy $\SUWeak$ doublet fields $\medf = \left(\medf^0,\medf^-\right)$ and $\medf^c = \left(\medf^{\rm c0},\medf^{+}\right)$,
\be
\mathcal{L}\supset \mmedf \medf^0\medf^{\rm c0} +\mmedf \medf^- \medf^+ + y v ~\medf^- e^{\rm c} + \lambda^e v~ e e^{\rm c} + m \nu \medf^{\rm c0} + m e \medf^+.
\ee
The neutral component is easily unmixed by defining
\be
\psi^{0'} = \frac{1}{\sqrt{\mmedf^2+m^2}}\left(\mmedf \psi^0+ m\nu\right).
\ee
This introduces a coupling between the WIMP ($\chi$) and the neutral scalar messenger ($\meds^0$) to the light neutrinos $\nu$ through the Yukawa vertex Eq.~(\ref{eqn:Yukawa-type}). It allows the neutral scalar component to decay via $\meds^0 \rightarrow \chi~+ \nu$. The charged fermion component similarly mixes with the SM electrons and allows for the decay of the charged scalar component $\meds^\pm \rightarrow \chi~+ e^\pm $. The mass mixing of  the fermions with the SM leptons allows the fermions to decay through $\Wpm$, $\ZZ$, or $h$ emission into SM leptons. These decays are shown in Fig.~\ref{fig:Heavy_fermion_mixing_decays}. We note that the mixing must be fairly small; otherwise, it introduces a new annihilation channel of DM into leptons that can easily dominate over the annihilation into photons~\cite{Weiner:2012gm}. Similarly, couplings to electrons and muons have to satisfy $y\lesssim10^{-3}$ to avoid excessively large contributions to  anomalous magnetic dipole moments.

\begin{figure}[tb]
\begin{center}
\includegraphics[width=0.9\textwidth]{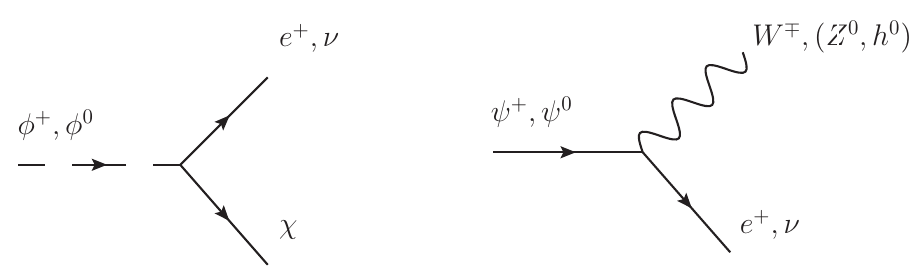}
\end{center}
\caption{Decays of the new charged states when the heavy fermions mix with SM leptons. On the left, the scalars decay into the WIMP candidate, $\chi$, and a SM lepton. On the right the fermions decay directly into a gauge-boson and a SM lepton.}
\label{fig:Heavy_fermion_mixing_decays}
\end{figure}

If the charged fermion is a singlet under $\SUWeak$,  the mixing can be done in an analogous fashion, although mixing now occurs with the right-handed electrons. The main difference is that there are no heavy neutral states in this case. Finally, in our discussion so far we have neglected the issue of flavor. Flavor-changing neutral current constraints place very strong bounds on models where $\medf$ mixes with different flavors of leptons. We take a phenomenological approach and assume that $\medf$ couples exclusively to one flavor of lepton (or, alternatively, that the couplings are aligned with the SM Yukawa matrices \cite{Ali:1999we,Buras:2000dm,D'Ambrosio:2002ex}). If a signal is seen in any channel, it would  be important to determine how the result is compatible with flavor observables.

\subsection{LHC Constraints} \label{collider:oddscalar}
When $\meds$ is odd and $\medf$ mixes with SM leptons, the components of $\medf$ decay to $\nu/\ell$ together with a boson $\Wpm/\ZZ/h$, with branching fractions determined by the $\SUWeak$ and $\UoneY$ charges of $\medf$. Signal events can therefore be rich in leptons and missing energy, giving good prospects for discovery at the LHC. Indeed, most scenarios with $\medf$ decaying to leptons plus gauge bosons are already ruled out.
The only models still allowed in the $\mmedf\lesssim300$ GeV range are those with small production cross sections (such as $\SUWeak$ singlets) and those decaying predominantly to $\tau$ leptons. Although constraints on $\tau$ final states are weak, triplets and higher multiplets decaying to taus are  ruled out by same-sign dilepton $+$ hadronic tau searches due to the large production cross section, as we discuss below. There are currently no constraints on the scalar decay $\meds\rightarrow\chi+\ell/\nu$  because of the small production cross section and the squeezed spectrum for $\meds$, leading to low acceptances for collider cuts.

Since the phenomenology of the mixed $\medf$-lepton models depends  on the gauge charges and the flavor structure, we detail each type of charge and flavor coupling in turn. We refer the reader to Appendix \ref{app:evenfermion} for the fermion branching ratios of each charge.\\

\noindent{\bf (2,\,-1/2):} The doublet fermion consists of charged and neutral components, $\medf^{\pm}$, $\medf^0$. The neutral components decay  exclusively to $\Wpm\ell^{\mp}$, while the charged component $\medf^{\pm}$ decays predominantly to $\ZZ\ell^{\pm}$, with branching fraction ranging from $0.6-1$ depending on $\mmedf$. LHC production of $\medf^+\medf^-$, $\medf^{\pm}\medf^0$, and $\medf^0\medf^0$ leads to trilepton final states when one of the gauge bosons decays leptonically. ATLAS trilepton $+\cancel{E}_{\rm T}$ searches at 8 TeV with $13\,\,\mathrm{fb}^{-1}$ \cite{ATLAS-CONF-2012-154} give the strongest bound on the model when $\ell=e$, $\mu$: associated production of $\medf^{\pm}\medf^0$ is ruled out for $M_{\medf^+}=M_{\medf^0} \lesssim 350$ GeV.

If $\medf$ decays predominantly to $\tau$V, the final state for $\medf^0\overline{\medf}^0$ production is $\Wp\Wm\tau^+\tau^-$ and $\Wpm\ZZ\tau^+\tau^-$, leading to same-sign dilepton + tau signatures. The strongest constraints come from the CMS search for same-sign dilepton  plus a hadronic  tau  at 8 TeV with $9.2\,\,\mathrm{fb}^{-1}$ \cite{CMS-PAS-SUS-12-022}, which constrains $M_{\medf^\pm}=M_{\medf^0} \lesssim 240$ GeV.\\

\noindent{\bf (1,\,-1):} There exists one new charged fermion, $\medf^\pm$, which decays dominantly via $\medf^\pm \rightarrow \Wpm\nu$, and with secondary decay $\medf^\pm\rightarrow \ZZ\ell^\pm$. The states $\medf^\pm$ are pair-produced, and the most common LHC signature is $\mathrm{W}^+\mathrm{W}^-+\cancel{E}_{\rm T}$, which is challenging to disentangle from the large $\mathrm{W}^+\mathrm{W}^-$ background. The $\Wpm\ZZ+\ell+\cancel{E}_{\rm T}$ final state is more promising, but current LHC multilepton analyses do not yet constrain this scenario as the fermion pair production cross section is small for a singlet, and most 4-lepton analyses either veto all Z bosons in an event or allow for 2 Zs, leading to much larger backgrounds.

If $\medf$ decays dominantly to $\tau$, the most promising final state is WZ$+\tau+\cancel{E}_{\rm T}$, where the gauge bosons decay leptonically and the $\tau$ hadronically. Due to the small cross section, current bounds from same-sign lepton + $\tau$ searches do not apply, and the prospects do not improve significantly at 14 TeV.\\

\noindent{\bf (3,\,0):} Analogously to the doublet model, the neutral components of the triplet decay almost exclusively to $\Wpm\ell^{\mp}$, while in contrast with the doublet, charged components decay dominantly to $\Wpm\nu$. As there is no Z coupling to $\medf^0\overline{\medf}^0$, the strongest constraint comes from pair production of $\medf^{\pm}\medf^0$, giving trilepton$+\cancel{E}_{\rm T}$ final states when the two W bosons decay leptonically. The ATLAS trilepton search \cite{ATLAS-CONF-2012-154} rules out such a scenario when $\ell=e$, $\mu$ and $M_{\medf^\pm}=M_{\medf^0} \lesssim 350$ GeV.

If $\medf$ decays dominantly to $\tau$, the CMS same-sign lepton plus hadronic tau search \cite{CMS-PAS-SUS-12-022} constrains the final states $\medf^{\pm}\medf^0\rightarrow \Wpm \Wpm\tau^\mp\bar \nu$. The masses $M_{\medf^\pm}=M_{\medf^0} \lesssim 330$ GeV are ruled out.\\

\noindent{\bf (3,\,-1):} The triplet fermion consists of doubly charged, charged and neutral components, $\medf^{--}$, $\medf^{-}$ and $\medf^0$. Similar to fermions with charge (3,0), the fermions  decay dominantly to W + lepton or neutrino. The pair production cross section is the highest among the possible electroweak charges, thus it receives the strongest constraint. For $\ell=e$, $\mu$,  $M_{\medf^\pm} \lesssim 500$ GeV is ruled out, while for $\ell=\tau$, $M_{\medf^\pm} \lesssim 400$ GeV is ruled out.

\subsection{Proposals and Prospects for Future Searches} \label{search:odd scalar}

The models above that are unconstrained by current searches are those in which the charged fermion mixes with $\tau$ leptons, and those where the new states are $\SUWeak$ singlets with correspondingly suppressed production cross sections. The Z$_2$-odd scalars themselves are also currently unconstrained with present data. They decay as $\meds^\pm\rightarrow\ell^\pm+\chi$ and therefore resemble sleptons decaying into leptons + a neutralino with mass 130 GeV. Since such scenarios are well-studied, we simply extrapolate from ref.~\cite{Andreev:2004qq} and quote that $(\mathbf{2}, -1/2)$ scalars can be discovered in the range 175-300 GeV at LHC14, while $(\mathbf{1}, -1)$ scalars and doublets decaying to taus cannot be discovered with $5\sigma$ significance.

 We now turn to our main study: singlet $\medf$ decaying to all flavors of leptons, and doublet $\medf$ mixed with taus. We show that  these scenarios can be discovered with $5\sigma$ significance at LHC14, with the exception of an $\SUWeak$ singlet that mixes with $\tau$ leptons.\\

\noindent {\bf (2,-1/2):} If the charged fermion mixes with the $\tau$ lepton then the dominant final states for production of $\medf^\pm$ and/or $\medf^0$ are $\tau^+\tau^-+$VV, where V$=$W, Z. Two final states give the distinctive signature of same-sign dileptons plus a hadronic tau: $\medf^0\bar\medf^0\rightarrow\Wp\Wm\tau^+\tau^-$ and $\psi^\pm\psi^0\rightarrow\Wpm \ZZ\tau^+\tau^-$, where the same-sign W and $\tau$ decay leptonically. The final state also includes a dijet resonance from the hadronically decaying gauge boson. The dominant SM backgrounds are $t\bar t+$V and WZ$+$jets. We include only prompt lepton backgrounds, while according to the recent ATLAS search for same-sign leptons \cite{ATLAS:2012mn}, the prompt background constitutes only one-third of the total SM background for $e^\pm e^\pm$ and half of the total SM background for $e^\pm\mu^\pm$. To account for the effects of non-prompt backgrounds (such as heavy flavor decays or external photon conversion), we run three analyses where we multiply the background normalization by factors of 1, 2, and 3, respectively. While non-prompt backgrounds are not necessarily well-modeled by an overall renormalization of the background, they are typically softer than leptons from signal or prompt background, and a rescaling of the background renormalization is therefore a conservative way of characterizing their effects.

We propose the following search, which is similar to the CMS same-sign dilepton + hadronic tau search \cite{CMS-PAS-SUS-12-022}, but with an added dijet resonance tag:
%
\begin{enumerate}
\item Exactly two same-sign leptons with $p_{\rm T}>20$ GeV and $|\eta|<2.5$ (electrons should have $p_{\rm T}>25$ GeV)
\item Exactly one hadronic tau with $p_{\rm T}>20$ GeV and $|\eta|<2.5$
\item $\cancel{E}_{\rm T}>40$ GeV
\item Z veto: all same-flavor lepton pairs\footnote{We perform a Z-veto on all same-flavor pairs (and not just opposite-sign same-flavor pairs) to suppress backgrounds from charge misidentification.} should satisfy $|m_{\ell\ell}-m_{\ZZ}| > 10$ GeV
\item At least two jets with $p_{\rm T}>20$ GeV and with one pair having an invariant mass in the Z/W mass window ($65$ GeV $<m_{\rm jj}<105$ GeV)
\end{enumerate}
 We consider the cross section for a benchmark point, $m_\psi=240$ GeV, which is just above current collider bounds. The signal cross section after all cuts is $1.0$ fb, while the background cross section from Monte Carlo is $0.18$ fb at LHC14. We show in the left pane of Fig.~\ref{fig:tautauVV-SSlepton} the dijet resonances associated with the hadronically decaying gauge boson. In the right pane of Fig.~\ref{fig:tautauVV-SSlepton}, we show the integrated luminosity needed at LHC14 for a $5\sigma$ discovery of $\psi$ as a function of $m_\psi$, with the three curves corresponding to renormalization of the background by factors of 1 (bottom), 2, and 3 (top). We see that $\psi$ can be discovered over the entire range of interest ($m_\psi\lesssim300$ GeV) with $20-70\,\,\mathrm{fb}^{-1}$ of data, depending on the effects of non-prompt backgrounds.\\

\begin{figure}[t]
\centering
\includegraphics[width=0.45 \textwidth]{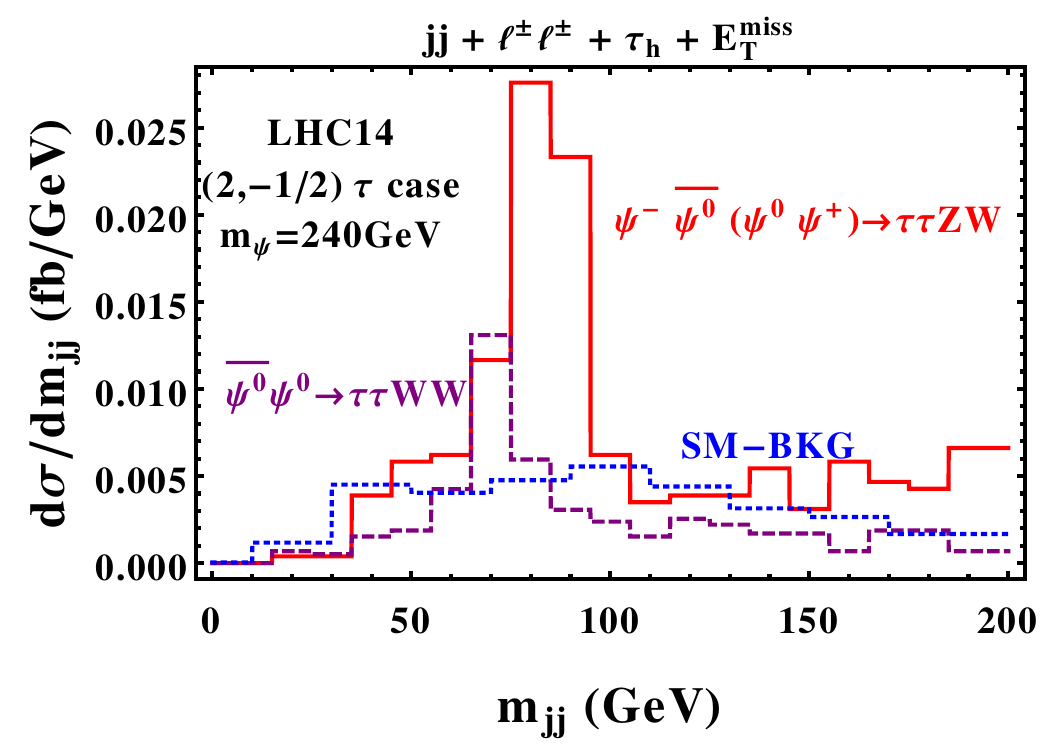}%
\includegraphics[width=0.45 \textwidth]{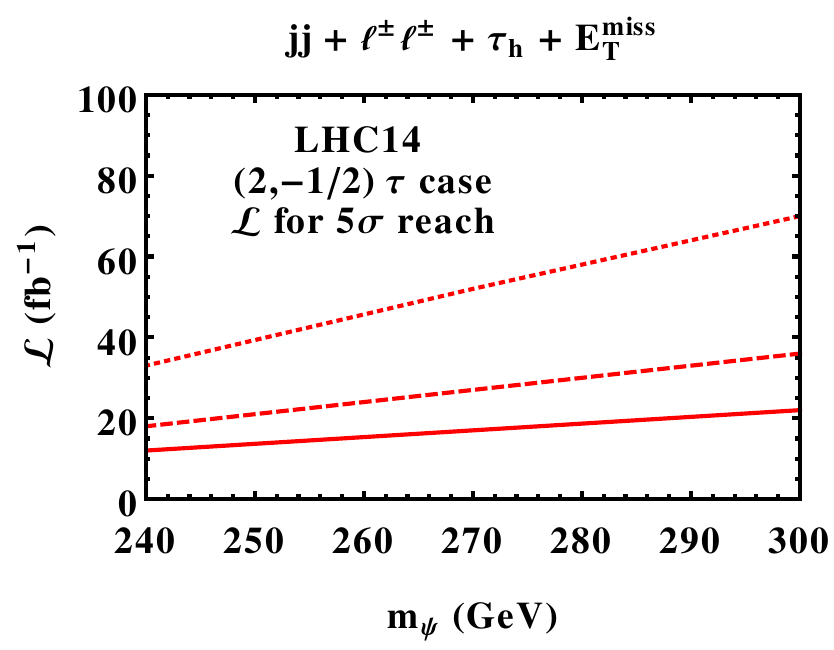}%
\caption{In the left pane, we show the dijet invariant mass ($m_{\rm jj} $) distribution in the doublet Z$_2$-odd scalar model where $\medf$ mixes with taus. The production modes at LHC14 are $pp\rightarrow\medf^\pm\medf^0/\medf^0\overline{\medf}^0\rightarrow\mathrm{jj}+\ell^\pm\ell^\pm+\tau_{\rm h}+\cancel{E}_{\rm T}$. All cuts were applied except those involving $m_{\rm jj} $. The plot on the right is the luminosity for the $5\sigma$ discovery of $\medf$ at LHC14 as a function of fermion mass. The three curves correspond to background normalizations corresponding to 1 (bottom), 2, and 3 (top) times the value from Monte Carlo to account for non-prompt lepton backgrounds. The plot begins at 240 GeV, which is the current bound.}\label{fig:tautauVV-SSlepton}
\end{figure}

\noindent{\bf (1,-1):} Because of the low production cross section and  branching ratio to charged leptons relative to higher multiplets, $\SUWeak$ singlet  fermions $\medf$ are not currently constrained. We propose a four-lepton search that allows discovery of $\mmedf\lesssim260$ GeV with $300\,\,\mathrm{fb}^{-1}$ when $\medf$ mixes with light-flavor leptons. When $\medf$ mixes with taus, however, we find that even with aggressive cuts, the signal-to-background ratio is still too small for the search to be viable.

The process we study is the production of $\medf^+\medf^-$, where one fermion decays to W$\nu$ (branching ratio $\approx0.7$), while the other decays to Z$\ell$ (branching ratio $\approx0.3$). The gauge bosons are required to decay leptonically, and the final state is four leptons plus $\cancel{E}_{\rm T}$. A distinctive feature of this final state is the existence of a Z$\ell$ resonance, and we exploit this by cutting on the three-lepton mass, $M_{\ell'\ell'\ell}$, where $\ell'\ell'$ are the leptons from the Z decay. We scan over $\mmedf$ and apply the following cuts:
\begin{enumerate}
\item Exactly four leptons with $p_{\rm T}>20$ GeV and $|\eta|<2.5$ (electrons should have $p_{\rm T}>25$ GeV)
\item Exactly one pair of same-flavor leptons with $|m_{\ell\ell}-m_\ZZ | < 10$ GeV
\item $\cancel{E}_{\rm T}>50$ GeV
\item At least one triplet of leptons with $|M_{\ell'\ell'\ell}-\mmedf|/\mmedf < 0.2$, where $m_{\ell'\ell'}$ reconstructs the Z mass
\end{enumerate}
The main SM backgrounds for such a process include diboson (ZZ) and triboson (ZWW) processes, where the gauge bosons decay leptonically. Some of the lepton pairs could come from off-shell photon final state radiation instead of a Z, and we also include this internal photon conversion to dileptons. We have checked our background Monte Carlo against  the 8 TeV CMS searches for four leptons with missing energy \cite{CMS-PAS-SUS-12-026}.

In Fig.~\ref{fig:1m1signal}, we show the trilepton invariant mass $M_{\ell' \ell' \ell }$ distribution (both combinations) after cuts 1-3 at LHC14, as well as the luminosity for a $5\sigma$ discovery. For a benchmark point of $m_\psi=180$ GeV,  the signal cross section after all cuts is $0.2$ fb, while the SM background is $0.05$ fb. Such a search allows for the discovery of $\medf^\pm$ in the mass range up to 260 GeV with $\mathcal{O}(300\,\,\mathrm{fb}^{-1})$.

\begin{figure}[h]
\centering
\includegraphics[width=0.45 \textwidth]{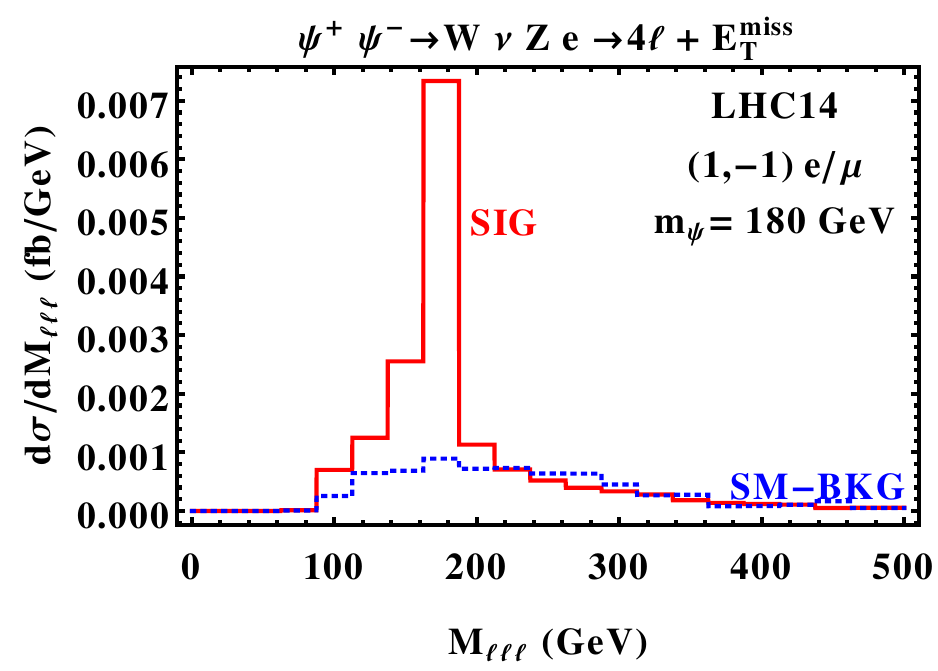}%
\includegraphics[width=0.45 \textwidth]{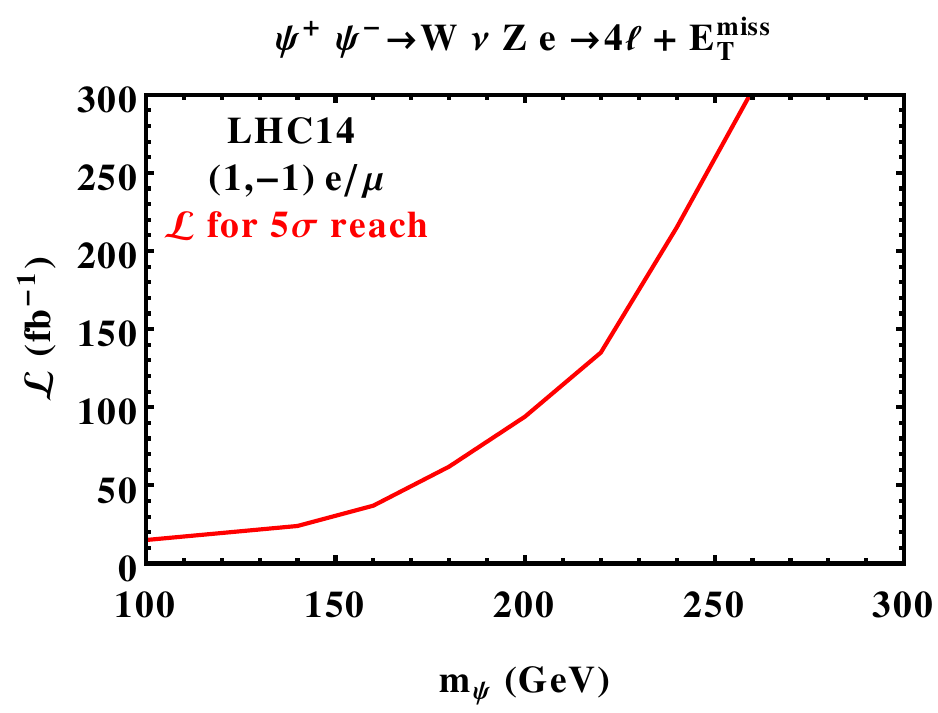}%
\caption{In the left pane, we show the trilepton invariant mass $M_{\ell' \ell' \ell }$ distribution at LHC14 as described in the text for the singlet Z$_2$-odd scalar model, with production modes $pp\rightarrow\medf^+\medf^-\rightarrow 4\ell+\cancel{E}_{\rm T}$.  On the right is the luminosity needed for $5\sigma$ discovery at LHC14 as a function of the fermion mass. }
\label{fig:1m1signal}
\end{figure}

We attempted a similar analysis for models where the fermion decays to $\tau\,+\,$V instead of $e/\mu\,+\,$V. The most promising final state from $\psi^+\psi^-$ pair production is therefore WZ$\tau\nu_\tau$, with WZ decaying leptonically and $\tau$ hadronically. Requiring that there be three light leptons and one hadronic tau in the event with criteria similar to the above search, the backgrounds are substantially higher because of fake taus from diboson+jet events. After similar cuts to the above four-lepton analysis, we find $S/B\sim0.02$. It is unlikely that uncertainties on estimates of hadronic $\tau$-tagging and fake lepton rates will allow multilepton $+$ fake $\tau$ backgrounds to have a systematic uncertainty below 2\%, making it very challenging to discover new electroweak states in this channel.

\vspace{5mm}
\hspace{-3mm}{\bf Summary}
\vspace{5mm}

We summarize the current constraints and prospects for detection of  the Z$_2$-even fermions with various SM charges as a function of $\mmedf$ in Table~\ref{detecttable}. We organize the table according to the SM charges  and final state flavors for each model, as well as the status of the signal.

\begin{table}
    \begin{tabularx}{\textwidth}{|l|l|l|l|X|}
        \hline
         \multicolumn{1}{|c|}{$\medf$ Charge} &  \multicolumn{1}{c|}{Flavor} &  \multicolumn{1}{c|}{Constraints} &  \multicolumn{1}{c|}{Reach ($5\sigma@$14TeV)} & \multicolumn{1}{c|}{Dominant Signal(s)} \\ \hline
        \multicolumn{5}{|c|}{{\bf Already excluded}} \\ \hline
        $\left(\mathbf{2},-\frac{1}{2}\right)$~ & $e,\mu$~ & $\mmedf \gtrsim 350 \gev $~ &  N/A & $\psi^\pm\psi^0\rightarrow\ell^+\ell^-\Wp\ZZ\rightarrow 4\ell+\cancel{E}_{\rm T}$ \\ \hline
              $(\mathbf{3},0)$~ & $e,\mu$~ & $\mmedf \gtrsim 350\gev$~ & N/A  & $\psi^\pm\psi^0\rightarrow\Wp\Wm\ell\nu\rightarrow 3\ell+\cancel{E}_{\rm T}$ \\ \hline
        $(\mathbf{3},0)$~ & $\tau$~  & $\mmedf \gtrsim 330\gev$~ & N/A  & $\psi^\pm\psi^0\rightarrow \Wpm \Wpm\tau^\mp\nu\rightarrow \ell^\pm\ell^\pm+\tau_{\rm h}+\cancel{E}_{\rm T}$
\\ \hline
        $(\mathbf{3},-1)$~ & $e,\mu$~ & $\mmedf \gtrsim 500\gev$~ & N/A & $\geq 3\ell$ plus $\cancel{E}_{\rm T}$ (multiple channels) \\ \hline
        $(\mathbf{3},-1)$~ & $\tau$~  & $\mmedf \gtrsim 400\gev$~ & N/A &  $\ell ^{\pm }\ell ^{\pm }$ plus $\tau _h$ (multiple channels) \\ \hline
                \multicolumn{5}{|c|}{{\bf Proposed search}} \\ \hline
                      $(\mathbf{1},-1)$~ & $e,\mu$~ & none~ & $\begin{array}{c}\mmedf<260\,\,\mathrm{GeV}\\ (\mathcal{L}\approx300\,\,\mathrm{fb}^{-1})\end{array} $& $\psi ^-\psi ^+\rightarrow \Wpm\nu  \ZZ\ell\to  4\ell+\cancel{E}_{\rm T}$ (no Z) \\ \hline
                          $\left(\mathbf{2},-\frac{1}{2}\right)$~ & $\tau$~  & $\mmedf \gtrsim 240 \gev $~ & $\begin{array}{c}\mmedf < 300\,\,\mathrm{GeV}\\ (\mathcal{L}\approx40\,\,\mathrm{fb}^{-1})\end{array} $& 
$\begin{array}{c}
\hspace{-1cm}\psi ^\pm \psi ^0  \rightarrow  \ZZ \tau  \tau  \Wpm \rightarrow \\ \hspace{1cm} \rm{jj} + \ell ^{\pm}\ell ^{\pm} + \tau_{\rm h} + \cancel{E}_{\rm T}\\
\hspace{-1cm}\psi ^0 \overline\psi ^0  \rightarrow  \Wpm \tau  \tau  \Wmp \rightarrow\\ \hspace{1cm} \rm{jj} + \ell ^{\pm}\ell ^{\pm} + \tau_{\rm h} + \cancel{E}_{\rm T}
\end{array} $ \\ \hline
                        \multicolumn{5}{|c|}{{\bf Very low sensitivity}} \\ \hline
  
        $(\mathbf{1},-1)$~ & $\tau$~  & none~ & none & $\psi^-\psi^+\rightarrow \Wpm\nu \ZZ\tau^\mp\rightarrow 3\ell+\tau_{\rm h}+\cancel{E}_{\rm T}$\\ \hline

    \end{tabularx} \caption{Summary table for Z$_2$-even fermions with the indicated  electroweak charges and flavor couplings. When relevant, the mass reach indicates the highest $\mmedf$ that can be discovered in our window of interest (up to 300 GeV) and the luminosity required at LHC14 for a $5\sigma$ discovery.}\label{detecttable}
\end{table}

\newpage
\clearpage

\section{Odd fermion model}\label{sec:oddfermion}

\subsection{The Model}\label{model:oddfermion}

When $\medf$ is odd under the Z$_2$ symmetry, operators exist that allow $\meds$ to decay into SM fields. Most such models couple $\meds$  to the SM through the scalar potential including the Higgs boson, $V(h,\meds)$. Depending on the terms in the potential and the quantum numbers for $\meds$, components of $\meds$ can mix with the Higgs doublet components $h$, and the neutral component can develop a VEV. A VEV for $\meds$ is constrained since it induces a Dirac mass term mixing $\chi$ with $\medf$ through the Yukawa coupling $\lambda\langle \meds\rangle\chi\medf$. If $\meds$ violates custodial $\mathrm{SU}(2)$ symmetry, more stringent constraints on $\langle\meds\rangle$   arise from bounds on the $\rho$ parameter~\cite{Blank:1997qa,Chen:2006pb}. Below, we consider models where $\meds$ mixes with the SM Higgs as well as models where it couples directly to fermions, and we give constraints on $\langle\meds\rangle$ in models where it acquires a VEV.  

\vspace{5mm}
\hspace{-3mm}{\bf $\meds$ mixes with the Higgs boson}
\vspace{5mm}

The scalar $\meds$ can develop a VEV in one of two ways: through a negative mass-squared term or through a linear term in the potential. The former is challenging to realize in a model: taking as a potential
\be\label{eq:phipotential}
V(\meds) = \frac{\lambda_\meds}{4}\left(\left|\meds\right|^2-\langle\meds\rangle^2\right)^2,
\ee 
the VEV and physical mass $\mmeds$ are related by
\be
\mmeds^2=\frac{\lambda_\meds\,\langle \meds\rangle^2}{2}.
\ee 
As we show below,  the VEV of $\meds$ is constrained to be much smaller than the SM Higgs VEV, and thus $\mmeds^2\ll m_h^2$ even when $\lambda_\meds\sim\mathcal O(1)$. Since fields coupling to the $\ZZ$ with mass $<m_{\ZZ}/2$ are generally excluded, this is not a consistent way to induce a $\meds$ VEV. The other possibility is to have a source for $\meds$, and we now consider several models with this feature.\\

\noindent {\bf (2,-1/2):} The scalars $\meds$ and $h$ comprise a Two Higgs Doublet Model (2HDM). The two most important phenomenological parameters for our purposes are the mixing angles of the VEVs ($\tan\beta$) and the CP-even component (($\sin\alpha$)  of the scalar $h$ and $\meds$. We work in the decoupling limit where both mixing angles are  small, and we show below that this is a consistent approximation. The mixing angles are controlled by terms in the Lagrangian of the form $\meds^*h$ added to the potential (\ref{eq:phipotential}). If the effective mixing term is  $\mu_{\mathrm{mix}}^2\meds^* h$,  then the mixing angles are
\be
\tan\beta \equiv \frac{\langle\meds\rangle}{v} \approx \frac{\mu_{\mathrm{mix}}^2}{\mmeds^2}
\ee
and
\be
\sin\alpha \approx \frac{\mu_{\mathrm{mix}}^2}{m_h^2+\mmeds^2}.
\ee
In the mass basis, there are three heavy scalar states: using conventional 2HDM terminology, these are the heavy CP-even and CP-odd scalars, $H$ and $A$, and the charged scalar $H^\pm$. 

When $\langle\meds\rangle\neq0$, the lightest DM mass eigenstate acquires a doublet fraction $f\sim \lambda\langle\meds\rangle/\sqrt{2}(\mmedf-m_\chi)$ through the Yukawa coupling Eq.~(\ref{eqn:Yukawa-type}). The doublet fraction of DM is constrained to  be $f\lesssim0.2-0.45$; otherwise, $\chi\bar\chi\rightarrow\mathrm{W}^+\mathrm{W}^-$ would violate constraints from the Fermi continuum gamma ray spectrum\footnote{The most conservative bound comes from background saturation, while a shape analysis improves constraints on the annihilation cross section by an order of magnitude.}~\cite{Buchmuller:2012rc,Cohen:2012me,Cholis:2012fb,Blanchet:2012vq,Asano:2012zv}. With $m_\chi=130$ GeV, $\mmedf$ at the weak scale, and the Yukawa coupling $\lambda$ being sufficiently large to generate appropriate Magnetic and Rayleigh operators ($\lambda \sim \sqrt{4\pi}$), this implies that 
\begin{equation}\label{eq:vevconstraint}
\tan\beta  = \frac{f\,(\mmedf-m_\chi)}{\lambda\,v} \lesssim 0.07\left(\frac{f}{0.4}\right)\left(\frac{\sqrt{4\pi}}{\lambda}\right)\left(\frac{\mmedf-m_\chi}{100\,\,\mathrm{GeV}}\right),
\end{equation}
where $v=246$ GeV is the SM Higgs VEV. Therefore, if $\meds$ develops a VEV, $\langle\meds\rangle$ is  constrained to be much smaller than the SM Higgs VEV. The constraints on $\langle\meds\rangle$ from (\ref{eq:vevconstraint}) are satisfied as long as $\mu_{\rm mix}^2\ll \mmeds^2$. As we mentioned above, this scenario is really the decoupling limit of the 2HDM where the CP-even mass eigenstates strongly correlate with the interaction eigenstates $h$ (SM Higgs boson) and $\meds$ (heavy doublet).

\begin{figure}[tb]
\begin{center}
\includegraphics[width=0.3\textwidth]{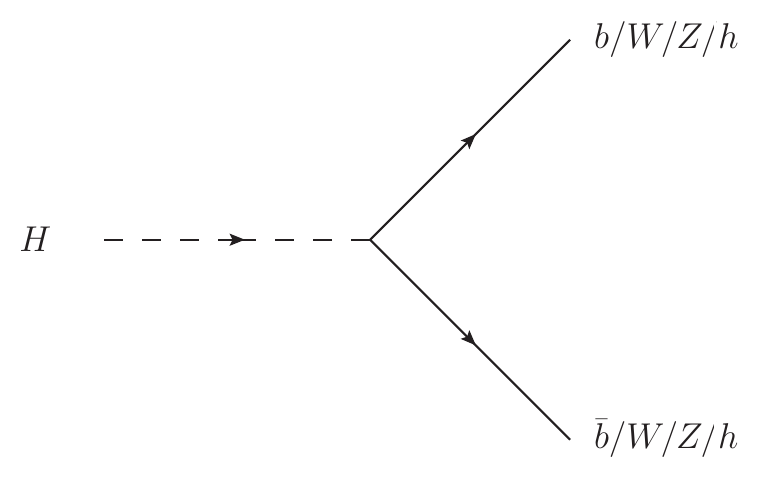}
\includegraphics[width=0.3\textwidth]{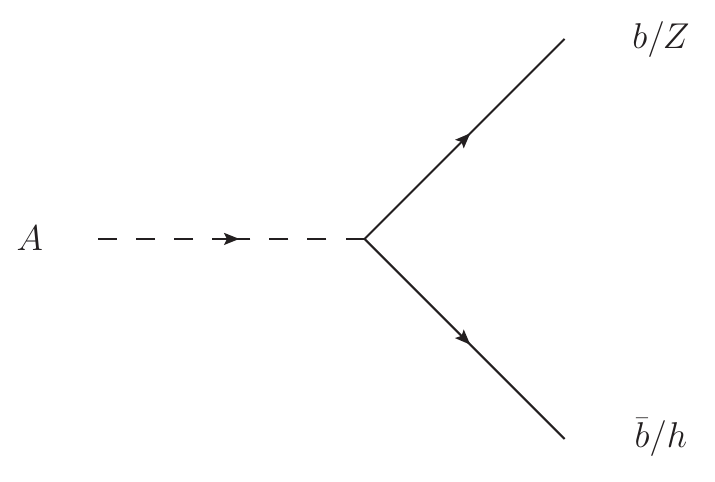}
\includegraphics[width=0.3\textwidth]{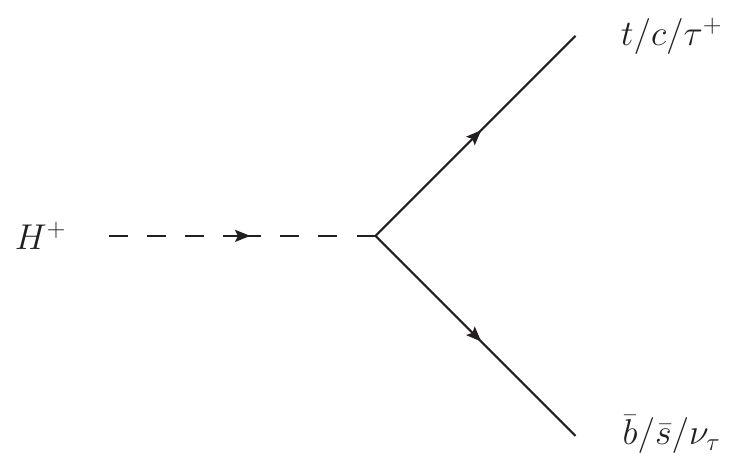}\\
\vspace{1cm}
\includegraphics[scale=0.5]{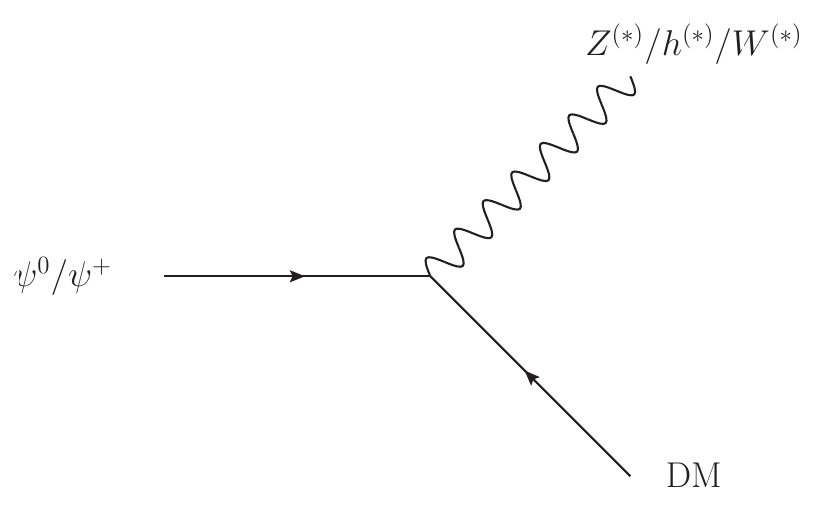}
\end{center}
\caption{Decays of the heavy scalars (top) and fermions (bottom) in the Type I Two Higgs Doublet model. Scalar branching ratios depend sensitively on which final states are kinematically allowed.}
\label{fig:typei}
\end{figure}

In viable 2HDMs, the scalar sector is charged under an additional Z$_2$ symmetry \cite{Glashow:1976nt}, ensuring that only one particular Higgs couples to each of down quarks/up quarks/leptons and avoiding dangerous flavor-changing neutral currents. These are described by the different ``Types'' of 2HDM, and can also be described by models with Yukawa alignment due to a minimal-flavor-violating hypothesis \cite{Ali:1999we,Buras:2000dm,D'Ambrosio:2002ex}. In what follows we focus on the most unconstrained 2HDMs: Type I, in which $\meds$ has no tree-level Yukawa couplings to SM fermions, which derive their mass entirely through coupling to $h$; and Type IV, in which $\meds$ couples exclusively to leptons and $h$ couples exclusively to quarks.

\begin{figure}[tb]
\begin{center}
\includegraphics[width=0.45\textwidth]{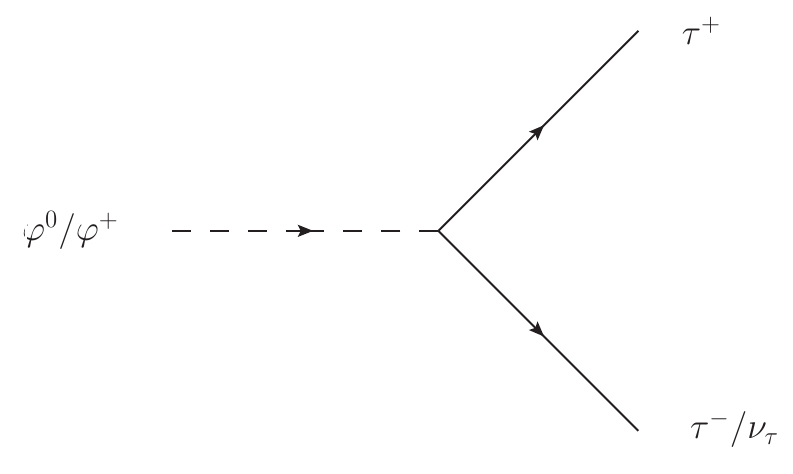}
\includegraphics[width=0.45\textwidth]{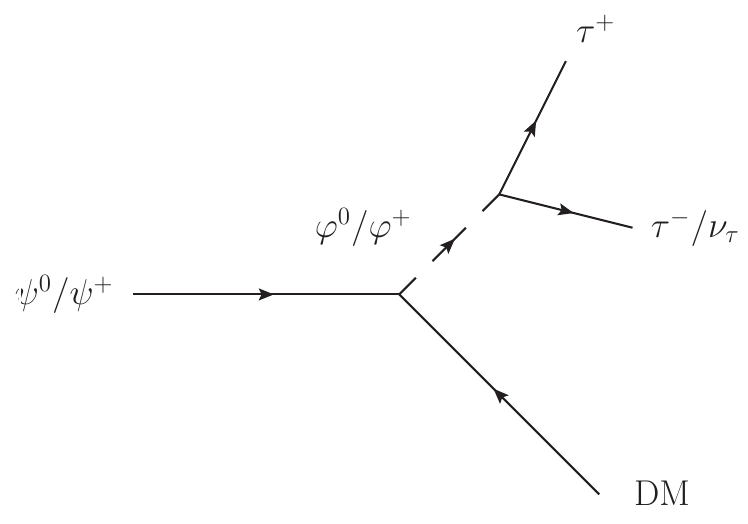}
\end{center}
\caption{Decays of the heavy scalars (left) and fermions (right) in the Type IV (leptophilic) two Higgs doublet model. Scalar branching ratios depend sensitively on which final states are kinematically allowed.}
\label{fig:typeiv}
\end{figure}

In the {\bf Type I model}, with Yukawa couplings
\be
\mathcal L_{\rm Type\, I} \supset \lambda_u\,Qhu^{\rm c} + \lambda_d\,Qh^*d^{\rm c}+\lambda_\ell\,Lh^*\ell^{\rm c},
\ee
the light CP-even Higgs is almost exactly SM-like due to the SM-Yukawa structure and small mixing with the CP-even component of $\meds$. The heavy CP-even scalar $H$ decays through mixing with the SM Higgs to $b\bar b$, $\mathrm{W}^+\mathrm{W}^-$, $\ZZ\ZZ$, and $hh$ when kinematically allowed, while the CP-odd scalar $A$ decays predominantly to $h\ZZ$ and $b\bar b$. The charged heavy Higgs, $H^\pm$, decays  to $t\bar b$ when allowed, and $\tau\nu$ otherwise. The Z$_2$-odd fermion $\medf$ can decay to a vector boson and the WIMP, $\chi$, due to the mixing induced by $\langle\meds\rangle$. It can also decay to $h\chi$ due to the mixing between $\meds$ and the SM Higgs. At low masses, the decays of $\medf$ to DM through off-shell gauge bosons dominates, while for $\mmedf\gtrsim220$ GeV, $\medf$ decays predominantly to on-shell  bosons. The decays are shown in Fig.~\ref{fig:typei}.

For {\bf Type IV models}, with Yukawa couplings
\be
\mathcal L_{\rm Type\, IV} \supset \lambda_u\,Qhu^{\rm c} + \lambda_d\,Qh^*d^{\rm c}+\lambda_\ell\,L\meds^*\ell^{\rm c},
\ee
 the scalar field $\meds$ couples exclusively to leptons, which obtain their mass through $\langle\meds\rangle$\footnote{We use the convention of \texttt{2HDMC} \cite{Eriksson:2009ws}; some sources invert the ordering of Types III and IV or use the alternative terminology of Types X and Y.}.  Consequently, the $\tau$ Yukawa coupling is $\mathcal{O}(1)$. The light Higgs is SM-like, although the coupling to $\tau$ is modified. Somewhat counterintuitively, the $h\rightarrow\tau^+\tau^-$ rate can actually be enhanced by an $\mathcal O(1)$ factor because the large Yukawa coupling of $\tau$ to $\meds$ compensates for the small mixing between $h$ and $\meds$. The decay modes of the heavy Higgses $H$, $A$, and $H^\pm$ are  almost exclusively to $\tau$ final states. The Z$_2$-odd fermion $\medf$ decays via $\medf^\pm\rightarrow \tau^\pm\nu_\tau \chi$ and $\medf^0\rightarrow \tau^+\tau^-\chi$. The decays are shown in Fig.~\ref{fig:typeiv}.

We briefly comment on 2HDM of {\bf Types II and III}, where $\meds$ couples predominantly to down-type quarks. These models predict enhanced couplings of heavy Higgses to $b$ quarks at large $\tan\beta$. Generally, heavy Higgses are excluded by $b\rightarrow s\gamma$ constraints in the range of interest of our models ($m_{H^\pm}\lesssim300$ GeV) for all values of $\tan\beta$ \cite{Mahmoudi:2009zx}. Therefore, we do not consider these models further.\\

\noindent {\bf (3,0):}  The term $h^* \meds^a\sigma^a h$ appears in the potential (\ref{eq:phipotential}) ($\sigma^a$ are the $\SUWeak$ generators), mixing the neutral component of $\meds$ and $h$. Unlike  a 2HDM, however, there is no direct coupling of $\meds$ to the Z, leading to a tree-level correction to the $\rho$ parameter when $\langle\meds\rangle\neq0$. This means that the size of the mixing must be small ($\langle\meds\rangle\lesssim8\GeV$). The lack of a tree-level coupling of $\meds$ to SM fermions means that it decays through Higgs mixing or, at high masses, to longitudinal $\mathrm{W}^+\mathrm{W}^-$ and $\ZZ\ZZ$, and to $hh$. The fermions $\medf^\pm/\medf^0$ decay predominantly through the scalar mixing to $\Wpm/h+\chi$.\\

\noindent {\bf (3,-1):}  The term $(\epsilon h) \meds^a\sigma^a h$ appears in the potential (\ref{eq:phipotential}). The symmetries also allow a Yuakwa coupling $(\epsilon\ell) \meds^{*a}\sigma^a\ell$ ($\ell$ is a SM lepton doublet), which would induce an unacceptably large Majorana neutrino mass. This possibility is therefore ruled out unless some symmetry  forbids the $\meds\ell\ell$ interaction. As in the $(\mathbf{3},0)$ scenario, the mixing between $\meds^0$ and $h$ must be small ($\langle\meds\rangle\lesssim8\GeV$) to avoid large corrections to the $\rho$ parameter. The $\meds$ and $\medf$ decays in this model are similar to the Type I 2HDM discussed above, where the $\meds$ decays predominantly to SM gauge bosons and Higgs bosons when kinematically allowed, and to $\tau$ and $b$ final states through the Higgs mixing otherwise. Additionally, there exist doubly charged states $\meds^{\pm\pm}$ ($\medf^{\pm\pm}$), which decay to $\Wpm\Wpm$ ($\Wpm\Wpm+\chi$) final states. This is an interesting and uncommon signature which improves the detection prospects as we discuss below in Section~\ref{sec:proposaloddfermion}.

\vspace{5mm}
\hspace{-3mm}{\bf $\meds$ couples to fermions}
\vspace{5mm}

It is  possible that the charge of $\meds$ does not allow it to mix with the SM Higgs. An example is the charge $(\mathbf{1}, -2)$; the only  renormalizable interaction leading to $\meds$ decay is
\begin{equation}
\mathcal L \supset \meds\, e^{\rm c}e^{\rm c},
\end{equation}
where $e^{\rm c}$ are right-handed SM leptons. The scalar $\meds$ is  a particle of charge $-2$ decaying into same-sign dileptons, giving a striking LHC signature. Other interactions are allowed with more exotic charges, particularly when non-renormalizable decay modes are allowed. Such interactions are beyond the scope of this paper, but we note that when $\meds$ is sufficiently long-lived and decays on detector length scales, the model becomes similar to the stable models we considered above in Section \ref{sec:stable}.

\subsection{LHC Constraints}\label{sec:constraintodd}
When $\medf$ is odd under the Z$_2$ symmetry stabilizing dark matter, $\meds$ can decay promptly into SM states. The phenomenology is then dictated by the couplings of $\meds$, which can interact either directly with SM fermions, or indirectly through mixing with the SM Higgs boson. When $\meds$ mixes with the Higgs, its phenomenology is largely dictated by the allowed Yukawa couplings to SM fermions. If $\meds$ has no tree-level couplings to SM fermions, the models are unconstrained by current LHC data for $\mmedf=100-300$ GeV, with the exception of states with charge $(\mathbf{3},-1)$, which are excluded for $\mmedf=170-210\GeV$. On the other hand, if $\meds$ is responsible for giving mass to the leptons (Type IV 2HDM), it is constrained to have a mass $\mmeds\gtrsim130$ GeV. Multilepton signatures of more general 2HDMs have  been considered  in \cite{Craig:2012pu}.

Finally, we consider a scenario where a doubly-charged scalar $\meds^{--}$ decays directly into two same-sign SM leptons through a Yukawa interaction; this scenario is excluded completely in the window of interest unless $\meds^{--}$ decays dominantly into taus, in which case $M_{\meds^{--}}>110$ GeV for a 100\% branching ratio into taus. For $M_{\meds^{--}}\approx100$ GeV, a sizeable branching fraction  into taus ($\sim97\%$) is still allowed.

\vspace{5mm}
\hspace{-3mm}{\bf $\meds$ mixes with the Higgs boson}
\vspace{5mm}

\noindent {\bf (2, -1/2) Type I 2HDM:} After mixing with the SM Higgs, the heavy scalar eigenstates $H$, $A$, and $H^\pm$, as well as the fermions $\medf$, are directly produced at the LHC via the electroweak gauge interactions and decay through mixings with the SM Higgs. LEP has  ruled out $m_{H^\pm}\lesssim80$ GeV in searches for $\tau\nu$ and $c\bar s$ \cite{Holzner:2001tv}, and this remains the strongest bound to-date. Searches for direct production of $H^+H^-\rightarrow \tau^+\tau^-\nu_\tau\bar\nu_\tau$ are difficult due to the challenging final state (see Section \ref{sec:lhcsterile} below for a similar model). The lack of a direct coupling between the top and $H^\pm$ renders searches in $t\rightarrow H^\pm b$ ineffectual, while for $m_{H^\pm}>m_t$, the dominant decay is to $t\bar b$; the enormous top backgrounds limit the viability of this search.

The heavy neutral Higgs bosons are produced in association with one another at the LHC: $pp\rightarrow \ZZ^{(*)}\rightarrow AH$. In the kinematic regime probed at LEP, both $A$ and $H$ decay to $b\bar b$, and searches for fully hadronic decays of Higgs bosons exclude the masses up to $m_A\approx m_H\approx70$ GeV in the $4b$ final state \cite{LEPHiggsWorking:2001ab}. This search channel suffers from large hadronic backgrounds at the Tevatron and LHC and there are no stronger bounds. At higher masses ($m_H\gtrsim160$ GeV), the $H\rightarrow$WW mode dominates, although $t\bar t$ background swamps the signal. The most promising search region is $m_H\gtrsim250$ GeV, where the $H\rightarrow hh$ and $A\rightarrow h$Z modes dominate, but the cross sections are too small to have been probed at LHC8.

The fermion $\medf$ also appears in the spectrum and can be pair-produced,  typically decaying through the mixing induced between $\medf$ and $\chi$: $\medf^\pm \rightarrow \Wpm+\chi$ and $\medf^0\rightarrow \ZZ+\chi$. The branching ratios of $\medf$ are given in Appendix \ref{app:oddfermion}. When $\mmedf\gtrsim220$ GeV, the production of the fermions results in diboson + $\cancel{E}_{\rm T}$ signal. The leading ATLAS and CMS searches are not yet sensitive to $\medf$ production in this regime.

For $\mmedf\lesssim220\GeV$, the most visible signatures of these models are associated production  ($\medf^\pm\medf^0$) and pair production $\medf^0\overline{\medf}^0$, leading to $\ge3\ell+\cancel{E}_{\rm T}$ and no Z. Current LHC multilepton + $\cancel{E}_{\rm T}$ analyses do not yet have the sensitivity to probe these models, due to the low $\cancel{E}_{\rm T}$ and lepton $p_{\rm T}$ when $\medf$ has a mass approaching its lower limit of 130 GeV, and due to the low production cross section for higher masses. In Section \ref{sec:proposaloddfermion}, however, we show the reach of such a search at LHC14.\\

\noindent {\bf (2, -1/2) Type IV 2HDM:} The particle content is the same as for Type I above, but the new charged particles decay almost exclusively to $\tau$ leptons. Production of $HA$, $AH^\pm$, and $H^+H^-$ leads to $4\tau$, $3\tau+\cancel{E}_{\rm T}$, and $2\tau+\cancel{E}_{\rm T}$ final states. The Z$_2$-odd fermions $\medf$ decay into the same final states, along with extra missing energy carried away by the WIMP, $\chi$.

The most promising channels are $4\tau\rightarrow \ell^\pm\ell^\pm+2\tau_{\rm h}+\cancel{E}_{\rm T}$ and $3\tau\rightarrow \ell^\pm\ell^\pm+\tau_{\rm h}+\cancel{E}_{\rm T}$, where $\tau_{\rm h}$ is a hadronically tagged tau.
The strongest constraints currently come from the $3\tau$ mode due to the larger production cross section. The CMS  collaboration has searched for same-sign leptons and one hadronic tau at 8 TeV and $9.2\,\,\mathrm{fb}^{-1}$ \cite{CMS-PAS-SUS-12-022} , and this excludes heavy scalar masses below $130 \gev$, which is a superior bound than one would find with other proposed techniques for Type IV 2HDM. In particular, we find that using the same-sign lepton + tau signature for the $3\tau$ mode, the signal-to-background ratio is a factor of 4 larger and the significance is approximately $15\%$ higher for the $3\tau$ mode as compared with the quoted values for the benchmark LHC14 analysis in \cite{Kanemura:2011kx}, while the performance for the $4\tau$ mode is comparable for both channels. The authors of \cite{Kanemura:2011kx} find the highest significance in events with 3-4 hadronic tau tags but neglect pure QCD and all-hadronic $V+$jets backgrounds; because we rely on smaller numbers of hadronic tau tags (one or two), our analysis is less sensitive to such all-hadronic backgrounds and should consequently have lower systematic uncertainties as well. In Section \ref{sec:proposaloddfermion} below, we propose related searches in the same-sign dilepton + tau final states, and discuss the reach of such models at LHC14. Finally, we find that there are no useful constraints on the fermion $\medf$ in this scenario due to the squeezed spectrum and consequent limited phase space available in its decays.\\

\noindent {\bf (3, 0):} As in the 2HDMs discussed above, the neutral and charged components of $\meds$ mix after electroweak symmetry breaking. With no direct coupling of $\meds$ to SM fermions, the decays of the scalar proceed predominantly through Higgs mixing to WW, ZZ, and hh. The scalar production cross section is very small, however, and such scalars are  not currently constrained by LHC data. The production cross section of the fermion $\medf$ is larger, with decays $\medf^\pm\rightarrow\Wpm+\chi$ and $\medf^0\rightarrow h+\chi$. However, none of the production modes is  constrained by existing searches due to large SM top and diboson backgrounds\footnote{In SUSY models, the associated production of triplet fermions $\tilde\chi_2^0\tilde\chi_1^\pm$ is strongly constrained through the decays $\tilde\chi_2^0\rightarrow\ZZ\tilde\chi_1^0$ and $\tilde\chi_1^\pm\rightarrow\Wpm\tilde\chi_1^0$. In such models, the neutralino decay to $\ZZ$  occurs through Higgsino mixing, whereas in the models we consider, the only allowed decay is to $h$, and the SUSY constraints  do not apply.}.\\

\noindent {\bf (3, -1):} The spectrum and decays are similar to the $(\mathbf{3},0)$ model, with the exceptions that $\meds^0$ couples to  $\ZZ\ZZ$ and there exist doubly charged states $\meds^{--}$, $\medf^{--}$. The production rate of the scalars is too small for observation, but the fermions decay to several highly-constrained final states, including $\medf^\pm\medf^0\rightarrow \Wpm\ZZ+\chi\bar\chi$ and $\medf^{\pm\pm}\medf^\mp\rightarrow \Wpm\Wpm\Wmp+\chi\bar\chi$. The strongest constraints are on the $\Wpm\ZZ+\cancel{E}_{\rm T}$ final state with leptonic decays, and $170\GeV<\mmedf<210\GeV$ is excluded \cite{ATLAS-CONF-2012-154}.

\vspace{5mm}
\hspace{-3mm}{\bf $\meds$ does not mix with the Higgs boson}
\vspace{5mm}

When the scalar $\meds$ does not mix with the Higgs boson it decays through its couplings to leptons. The model we consider involves the coupling $\meds^{--} e^{\rm c} e^{\rm c}$. The scalar $\meds^{--}$ therefore appears as a same-sign dilepton resonance and is identical to a right-handed, doubly-charged Higgs boson $H_{\rm R}^{--}$. ATLAS has presented bounds on this model, and assuming 100\% branching ratio into $\mu^{-}\mu^{-}$ or $e^{-} e^{-}$, $\meds^{--}$ is ruled out through the entire region of interest for our models, $\mmeds \lesssim300$ GeV. Decays of $\medf^{--}\rightarrow \ell^-\ell^-\chi$ are similarly excluded. However, if $\meds^{--}$ decays predominantly to $\tau$ leptons then the bounds are significantly weaker ($\mmeds>110$ GeV).

\subsection{Proposals and Prospects for Future Searches} \label{sec:proposaloddfermion}

\vspace{5mm}
\hspace{-3mm}{\bf $\meds$ mixes with the Higgs boson}
\vspace{5mm}

\noindent {\bf (2, -1/2) Type I 2HDM:}
In a Type I 2HDM, the doublet scalar decays through mixing with the Higgs, while the Z$_2$-odd $\medf$ decays to dark matter and a (possibly off-shell) W/Z/$h$. We focus on the phenomenology of $\medf$ decays, as they are a unique prediction of our model and are not present in a generic 2HDM. The charged fermion, $\medf^\pm$, decays exclusively to $\Wst\chi$, while $\medf^0$ decays to both $\Zst\chi$ and $h\chi$, with the Z mode dominating up to $\mmedf = m_h+m_\chi \approx 260\GeV$, at which point both modes become  equally important (in the range $260\GeV\lesssim \mmedf\lesssim300\GeV$, the branching fraction to $\ZZ\chi$ is about 60\%). We discuss the phenomenology of the scalars at the end of this section.

We focus on two kinematic regions: when $\mmedf\lesssim220\GeV$ and the gauge boson decays are off-shell, and $\mmedf\gtrsim220\GeV$, when the decays are on-shell. The signature with the best prospects for discovery in the lower-mass region is the production of $\psi^0\overline\psi^0\rightarrow 4\ell+\cancel{E}_{\rm T}$. This is similar to neutralino production in SUSY models, where the heavier neutralino decays into an off-shell $\ZZ$ boson and the LSP. The dominant SM background is diboson production (ZZ), and we also include triboson (WWZ), both with fully leptonic decays. The backgrounds are similar to those used in the charge $(\mathbf{1}, -1)$ Z$_2$-even fermion search in Section \ref{search:odd scalar} above.  For $\mmedf \lesssim 220$ GeV, the  backgrounds are efficiently suppressed with a Z-veto. We apply the following cuts (which are  similar to current multilepton analyses):
\begin{enumerate}
\item Exactly four leptons with $p_{\rm T}>10$ GeV and $|\eta|<2.5$ (leading lepton has $p_{\rm T}>20$ GeV)
\item No pair of same-flavor leptons with $|m_{\ell\ell}-m_\ZZ| < 15$ GeV
\item To remove leptons from photon conversions, we veto events where three leptons satisfy $|M_{\ell\ell\ell}-m_\ZZ|<15$ GeV
\item $\cancel{E}_{\rm T}>50$ GeV
\end{enumerate}
The $p_{\rm T}$ cuts are relaxed relative to our earlier $4\ell$ analysis because backgrounds are very small, and discovery potential is consequently optimized by maximizing the signal efficiency. We present the $\cancel{E}_{\rm T}$ distribution after  cuts 1-3 and the luminosity needed for $5\sigma$ discovery at LHC14  in Fig.~\ref{fig:4lplusMET}. For a benchmark point of $m_\psi=200$ GeV, the signal cross section after cuts is $0.11$ fb, while the SM background is $0.017$ fb.   Discovery can begin for $\mmedf\approx180$ GeV with an integrated luminosity of $\approx80\,\,\mathrm{fb}^{-1}$. Discovery of $\medf$ below this mass is more challenging because most of the final state energy is taken by the dark matter mass and there is little available phase space to pass the kinematic cuts\footnote{At $\mmedf\approx150$ GeV, there is a feature in the curve of luminosity required for discovery in Fig.~\ref{fig:4lplusMET}. We have confirmed that this is  the result of  the convolution of the falling production cross section, a rising efficiency of passing some kinematic cuts (such as lepton $p_{\rm T}$ and $\cancel{E}_{\rm T}$), and a falling efficiency of the Z-veto at higher masses.}.

\begin{figure}[t]
\centering
\includegraphics[width=0.45 \textwidth]{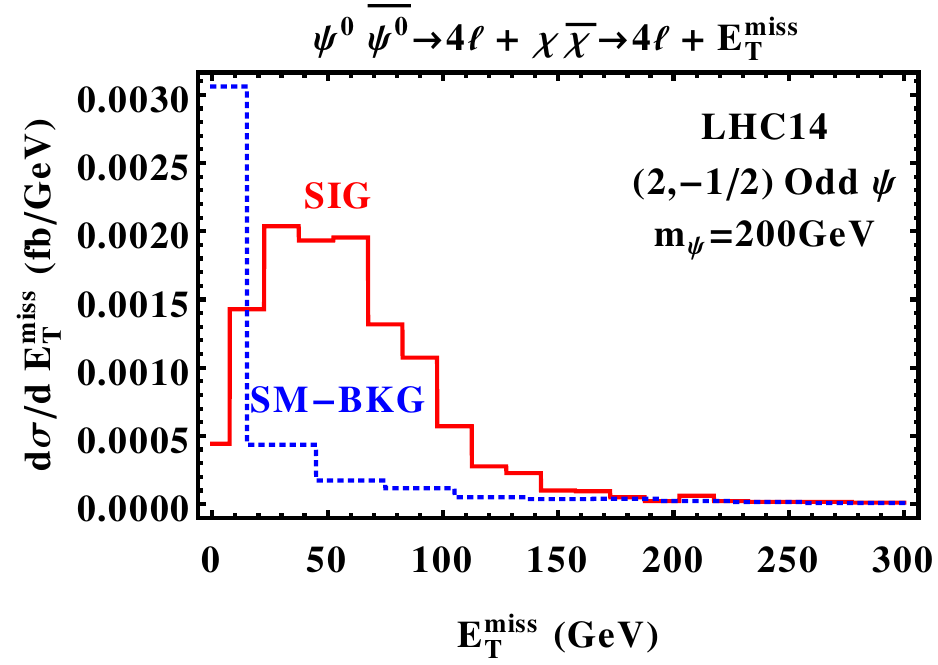}%
\includegraphics[width=0.45 \textwidth]{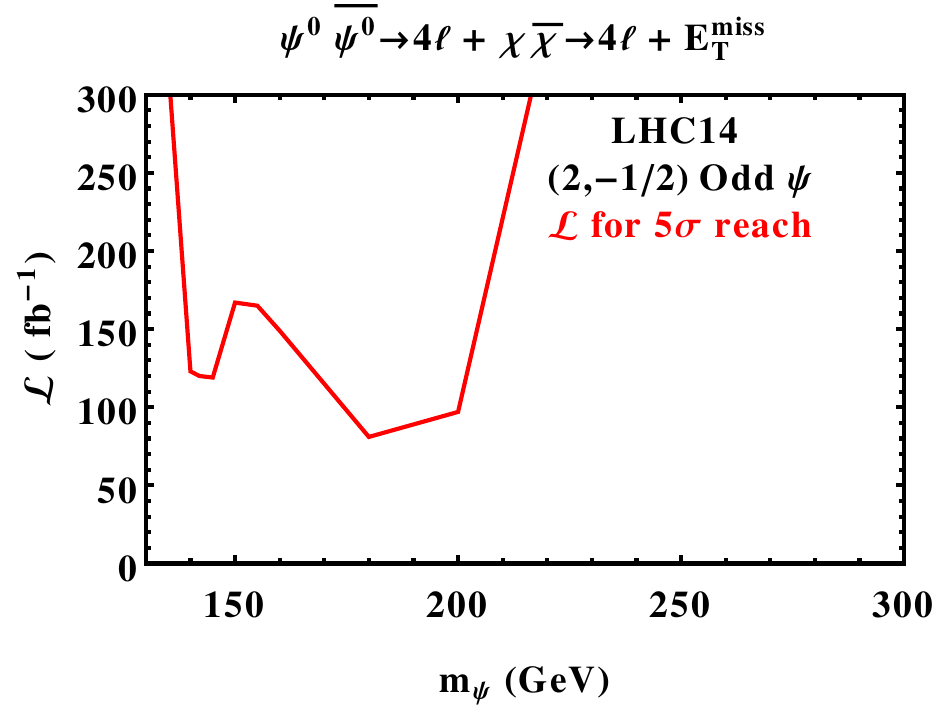}%
\caption{We consider the Type I 2HDM in the $\ZZ\ZZ+\cancel{E}_{\rm T}$ final state. In the left pane, we show the $\cancel{E}_{\rm T}$ distribution at LHC14 after all cuts (except the $\cancel{E}_{\rm T}$ cut). The production mode is  $pp\rightarrow\psi ^\pm \psi ^0\rightarrow4\ell+\cancel{E}_{\rm T}$. On the right is the luminosity needed for a $5\sigma$ discovery at 14 TeV is given as a function of $\mmedf$. This search mode targets $\mmedf\lesssim220\GeV$.}\label{fig:4lplusMET}
\end{figure}

For $\mmedf\gtrsim220$ GeV, the $\ZZ$ in $\medf^0$ decay goes on-shell, and $\ZZ\ZZ$ backgrounds dominate over the signal. A more effective search strategy considers the associated production mode $\medf^\pm\medf^0\rightarrow\Wpm\ZZ+\chi\bar\chi$. We now require that two leptons reconstruct the $\ZZ$, and the missing energy from $\chi\bar\chi$ in the final state allows for signal discrimination through a cut on the transverse mass ($m_{\rm T} \equiv \sqrt{E_{\rm T}^2 - p_{\rm T}^2} $) of the missing energy and the lepton not associated with the $\ZZ$. We propose the following cuts:
\begin{enumerate}
\item Exactly three leptons with $p_{\rm T}>10\GeV$ and $|\eta|<2.5$ (leading lepton has $p_{\rm T}>20\GeV$)
\item No hadronic tau with $p_{\rm T}>20\GeV$
\item One pair of opposite-sign, same-flavor leptons with $|m_{\ell\ell}-m_\ZZ|<15\GeV$
\item $\cancel{E}_{\rm T}>50\GeV$
\item $m_{\rm T}>120\GeV$
\end{enumerate}
The dominant background is $\Wpm\ZZ$. We present the luminosity needed for $5\sigma$ discovery at LHC14 in Fig.~\ref{2-half-OddF-3lMET}. For a benchmark point of $\mmedf = 250\GeV$, the signal cross section after cuts is 0.8 fb and the background is 3.4 fb. Discovery is possible for $230\GeV\lesssim \mmedf\lesssim265\GeV$; for lower masses, there is insufficient energy to pass the stringent $m_{\rm T}$ cut, while at higher masses, signal/background discrimination is, in principle, possible but the rate is too small.

\begin{figure}[t]
\centering
\includegraphics[width=0.6 \textwidth]{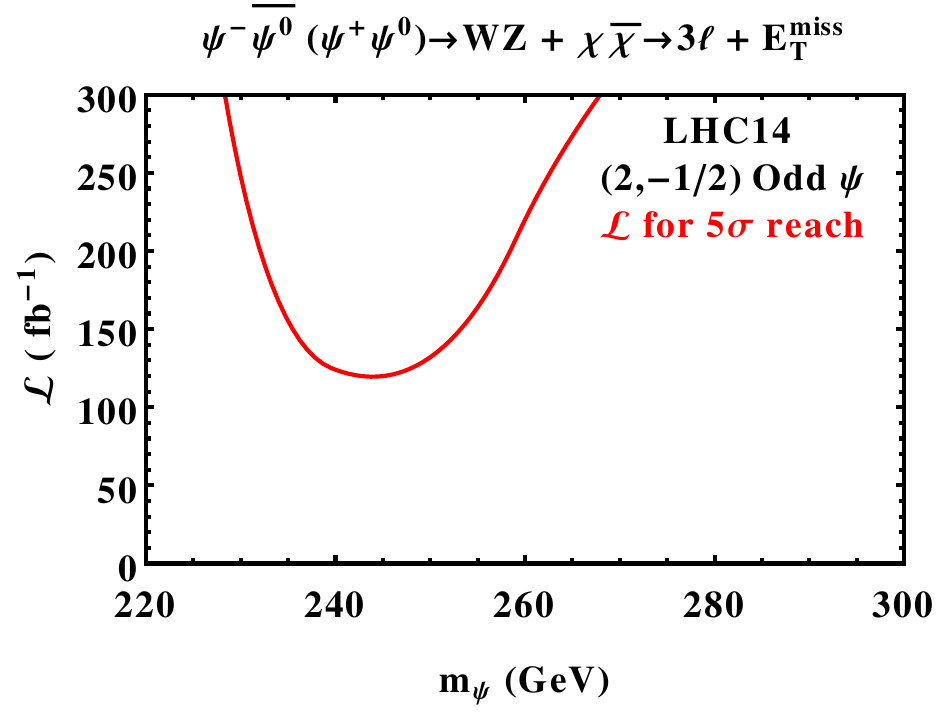}%
\caption{ We show the luminosity needed for a $5\sigma$ discovery in $\medf^\pm\medf^0\rightarrow\Wpm\ZZ+\cancel{E}_{\rm T}$ at 14 TeV in the Type 1 2HDM. This search mode targets $\mmedf\gtrsim220\GeV$.}\label{2-half-OddF-3lMET}
\end{figure}

The scalars $H$, $A$, and $H^\pm$ are also abundantly produced, although their discovery prospects are dim, as we now discuss. These scalar decay to $b$ quarks, gauge bosons, and the Higgs boson $h$, depending on their respective masses. In general, these signatures are difficult to disentangle from QCD backgrounds, and we touch only briefly on the possibilities for detection. Below two times the mass of the $\Wpm$ bosons, the most distinctive signatures are $HA\rightarrow 4b$ with a leading-order (LO) cross section of 235 fb for $m_H=m_A=100$ GeV. This is still about three orders of magnitude below the $b\bar b$ background, and observing the double-resonance structure of the signal events is still challenging to distinguish from the massive background.

At higher masses $m_H=m_A\sim200$ GeV, other decay modes are available to the heavy Higgs, $H\rightarrow$WW, ZZ while $A$ still decays to  $b\bar b$. The final state WW$b\bar b$ looks exactly like $t\bar t$, but with a cross section four orders of magnitude smaller, and this also is effectively invisible. Similarly, the signal ZZ$b\bar b$ can give distinctive $4\ell+2b$ signatures, but suffers significantly from branching fractions (leptonic and $H\rightarrow$ZZ, which is $\sim20\%$). Indeed, we find that this is about an order of magnitude smaller than the background for 4 leptons from Z bosons and $\ge1b$ jet. Finally, at masses above 250 GeV, we get the final state $HA\rightarrow hhh$Z. This can give spectacular $2\ell+6b$ signatures, but the cross section of such processes is minute ($\sim6$ ab), and such a search is not feasible except at perhaps the very highest integrated luminosities of LHC14.

The prospects are not any better for $H^\pm$ final states. For $m_H<m_t$, $H^\pm$ decays to $\tau\nu$, and signal events are swamped by W backgrounds. Above $m_t$, $H^\pm$ decays to $t\bar b$, giving rise to $2t+2b$ and $t\bar b$VV final states. Once again, signal discrimination from the enormous top backgrounds is most likely impossible without relying on extremely subtle features of the signal kinematics. At this point, it does not appear that the backgrounds are sufficiently well-understood for such channels to be viable.\\

\noindent {\bf (2, -1/2) Type IV 2HDM:} In the Type IV model, the charged Higgs and neutral Higgs decay predominantly to taus, which provide tau-rich final states. Because of the large contamination of VV$+$jets backgrounds to final states with leptons and taus, we concentrate on the relatively clean signature where two same-sign taus decay leptonically and at least one other tau decays hadronically. This leads to same-sign dilepton $+$ hadronic tau signatures similar to those studied in Section \ref{search:odd scalar}. The most-constrained process is the associated production of heavy charged/neutral Higgs bosons, with the production of $H^\pm+A/H\rightarrow3\tau+\nu$. We find that the most powerful discriminant is the invariant mass between each of the leptons and the tau, $m_{\ell\tau}$, which is much larger than in the SM background due to the higher mass scales present in the signal production process. The SM electroweak backgrounds are ZZ and WZ. As in Section \ref{search:odd scalar}, there are also contributions from  non-prompt lepton backgrounds, such as heavy flavor decays and photon conversions. Since these are poorly modeled by MC, we again run three analyses where we multiply the background normalization by factors of 1, 2, and 3, respectively, to provide an estimate as a function of increasing non-prompt lepton backgrounds.  We use the following cuts:
\begin{enumerate}
\item Exactly two same-sign leptons with $p_{\rm T}>20$ GeV and $|\eta|<2.5$ (electrons have $p_{\rm T}>25$ GeV)
\item Exactly one hadronic tau  satisfying $p_{\rm T}>20$ GeV and $|\eta|<2.5$
\item $\cancel{E}_{\rm T}>100$ GeV
\item The invariant mass of one lepton and the tau must satisfy $m_{\ell\tau}>100$ GeV for all possible combination
\end{enumerate}
For a benchmark point of $m_H=m_A=m_{H^\pm}=200$ GeV, we show in Fig.~\ref{fig:type4dist} the invariant lepton-tau mass $m_{\ell \tau }$ distribution after cuts 1-3,  and the luminosity needed for $5\sigma$ discovery as a function of heavy Higgs branching ratio to taus. After all cuts,  the signal cross section for $m_H=200$ GeV is $0.5$ fb, while the SM background is $0.16$ fb at LHC14.  We see that this search is powerful and probes 2HDM beyond Type IV: it allows for discovery of  heavy scalars with branching ratios to taus as low as 0.2 in the LHC late running.

\begin{figure}[t]
\centering
\includegraphics[width=0.45 \textwidth]{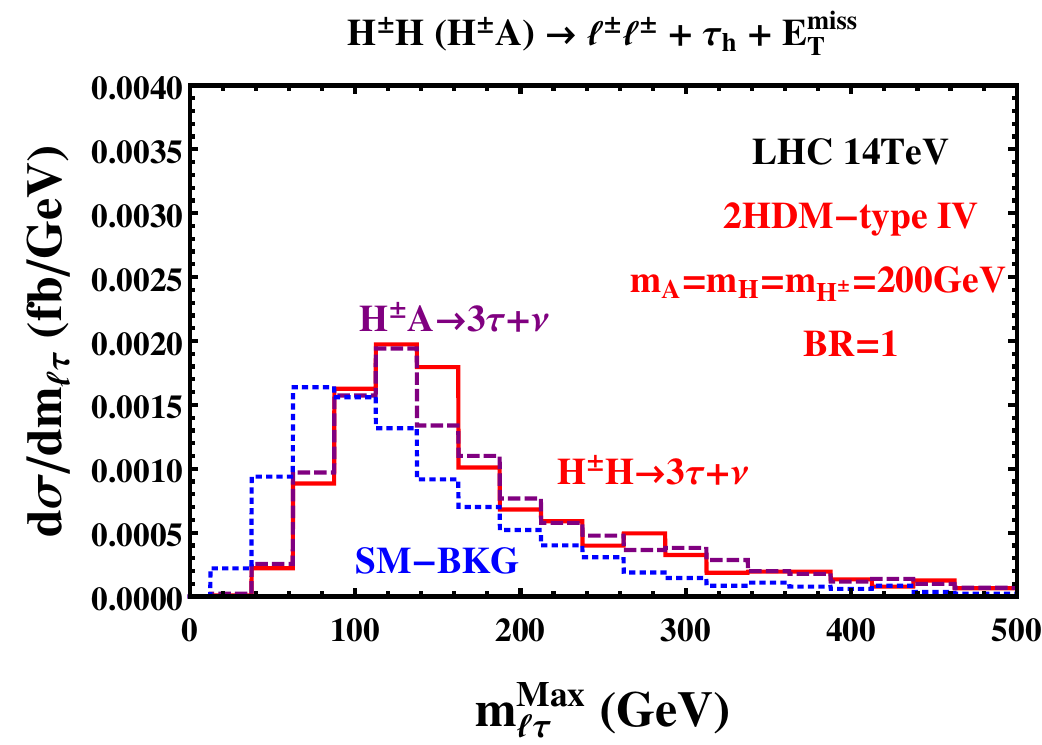}%
\includegraphics[width=0.45 \textwidth]{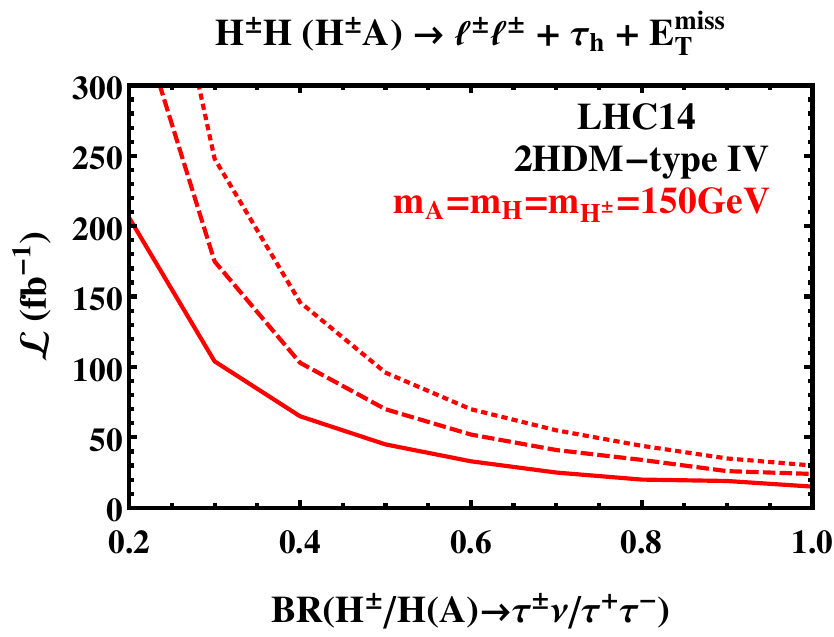}%
\caption{In the left pane, we show the $\mathrm{max}(m_{\ell \tau })$ distribution at LHC14 after all cuts (except the $m_{\ell \tau }$ cut) in the Type IV 2HDM. The production modes are  $pp\rightarrow H^{\pm }H/H^{\pm }A\rightarrow \ell^\pm\ell^\pm+\tau_{\rm h}+\cancel{E}_{\rm T}$. The masses of the heavy Higgses are set to 200 GeV for left panel (150 GeV for right panel), and they decay $100 \%$ to  taus. On the right, the luminosity  needed for a $5\sigma$ discovery at 14TeV is given for different decay branching ratios to taus, along with three different prefactors renormalizing the Monte Carlo background: 1 (bottom), 2, and 3 (top).} \label{fig:type4dist}
\end{figure}

There are also final states  $HA\rightarrow4\tau$, and the above search can be applied to this scenario with the following modifications:
\begin{enumerate}
\item Exactly two hadronic taus\footnote{To be conservative, we multiply the signal efficiency by 50\% when using $>1$ tau tag to account for possible reductions in $\tau$-tagging efficiencies, mis-modeling of tau tagging in \texttt{PGS}\cite{pgs4}, and/or enhancements in systematic uncertainties associated with having multiple hadronic $\tau$ leptons in the final state.}
\item $\cancel{E}_{\rm T}>100$ GeV
\item All combinations of lepton + tau must satisfy $m_{\ell\tau}>50$ GeV
\end{enumerate}
A more relaxed $m_{\ell\tau}$ cut is chosen because almost all backgrounds are eliminated by the second tau tag, and using a $m_{\ell\tau}>100$ GeV cut has very poor statistics in our Monte Carlo study. Background would be even further reduced, but there would be significant systematic uncertainties associated with its estimation, and since the analysis is predominantly limited by the small signal statistics anyway, we consider  $m_{\ell\tau}>50$ GeV.

\begin{figure}[t]
\centering
\includegraphics[width=0.45 \textwidth]{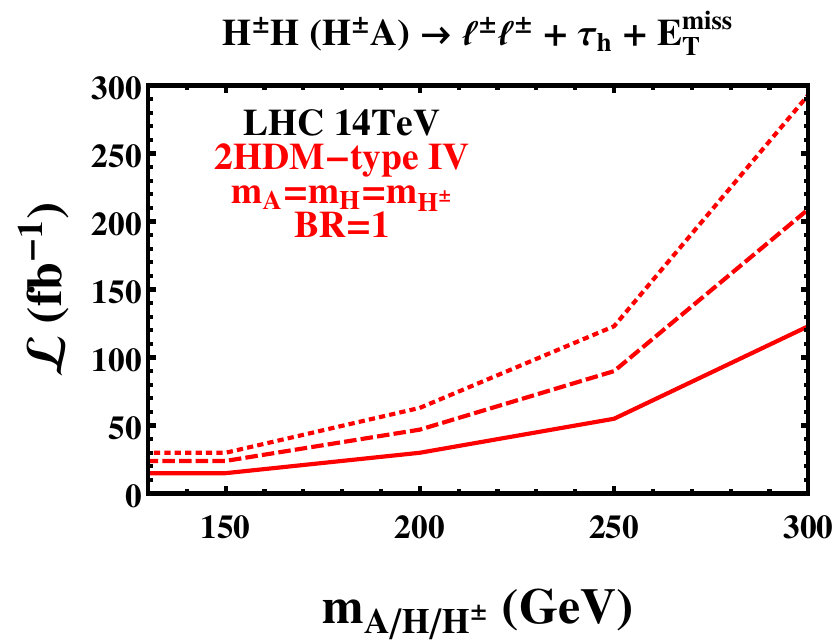}%
\includegraphics[width=0.45 \textwidth]{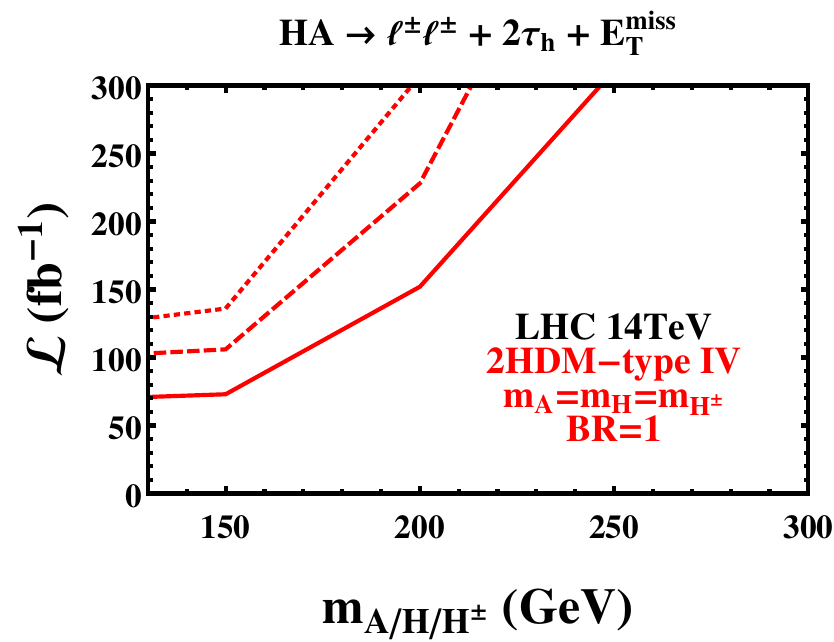}%
\caption{We show the luminosity  needed for a $5\sigma$ discovery in $3\tau$ (left pane) and $4\tau$ (right pane) searches at 14 TeV as a function of the heavy Higgs mass in the Type IV 2HDM with final states $\ell^\pm\ell^\pm+(1-2)\tau_{\rm h}+\cancel{E}_{\rm T}$. In this plot, we assume that $H^{\pm },H,A$ decay to taus with $100\%$ BR. The three curves have the same meaning as those depicted in the right pane of Fig.~\ref{fig:type4dist}.}\label{fig:type4sign}
\end{figure}

In  Fig.~\ref{fig:type4sign}, we plot the luminosity needed for a $5\sigma$ discovery of heavy scalars as a function of the scalar mass assuming $100 \%$ branching ratio to taus for the $3\tau$ (left) and $4\tau$ (right) final states. Although the $3\tau$ search is more powerful, the $3\tau$ and $4\tau$ analyses can serve as independent cross-checks of one another, and in the $3\tau$ channel with $125-300\,\,\mathrm{fb}^{-1}$ at LHC14, the scalars can be discovered over our entire region of interest.

The prospects for the discovery of the Z$_2$-odd $\medf$ are poorer. The available phase space from $\medf$ decay is limited by the squeezed spectrum from the nearly degenerate DM $\chi$ in the final state, and the neutrinos from $\tau$ decay carry off some of the remaining kinetic energy. As a result, we find that the efficiencies of kinematic cuts are lower for $\medf$ and, in spite of the larger production cross section compared to $H/A$, the above $3\tau$ and $4\tau$ analyses do not allow for $5\sigma$ discovery of $\medf$ in the Type IV 2HDM with $300\,\,\mathrm{fb}^{-1}$.  \\

\noindent {\bf (3,0):} The scalars in the $\left(\mathbf{3},0\right)$ model decay through Higgs mixing with similar branching fractions as the Type I 2HDM, and are therefore challenging to discover at the LHC.

The $\medf$ production modes are $\medf^0\overline{\medf}^0\rightarrow hh+\chi\bar\chi$, $\medf^\pm\medf^0\rightarrow \Wpm h+\chi\bar\chi$, and $\medf^\pm\medf^\mp\rightarrow \Wpm\Wmp+\chi\bar\chi$. All of these final states suffer from either large backgrounds, or are clean signatures but have very tiny rates due to small branching fractions (ex.~$hh\rightarrow\ZZ\ZZ\ZZ\ZZ\rightarrow >4\ell$). The most promising signature we find is $\Wpm h+\cancel{E}_{\rm T}$, with $h\rightarrow\gamma\gamma$; this is still suppressed by the small $h\rightarrow\gamma\gamma$ and leptonic W branching fractions, but is one of the cleanest Higgs decay modes to study. The resulting final state is $\ell+\gamma\gamma+\cancel{E}_{\rm T}$, where the photons reconstruct $m_h$ if $h$ is on-shell in $\medf^0$ decay (i.e.~$\mmedf\gtrsim250\GeV$). This final state is similar to the conventional Higgs-strahlung process $\Wpm h$, although with more invisible particles, and the two processes have comparable cross sections at 14 TeV ($\sim1-2\,\,\mathrm{pb}$, depending on $\mmedf$). 

Based on extrapolations from current $\Wpm h\rightarrow\ell^\pm+\nu_\ell+\gamma\gamma$ searches \cite{ATLAS-CONF-2013-012}, it is possible that the $\medf$ could be discovered  over QCD and Higgs-strahlung backgrounds, particularly with cuts that are more targeted to the kinematics of $\medf^0$ decay, but this would likely require a high integrated luminosity. A precise understanding of the backgrounds, systematic uncertainties, and the consequent optimization of a $\medf$ search, are beyond the scope of this paper, but the prospects should be studied in more detail by the experimental collaborations.  \\

\noindent {\bf (3,-1):} The spectrum and decays in this scenario are similar to the Type I 2HDM. With $\approx100\,\,\mathrm{fb}^{-1}$ of integrated luminosity at LHC14, $\medf$ can be discovered at $5\sigma$ for 100-300 GeV masses. For $\mmedf\lesssim220\GeV$, $\medf$ decays to off-shell $\mathrm{V}+\chi$, and multilepton searches with a Z-veto are effective. Considering in particular the production mode $pp\rightarrow\medf^0\overline{\medf}^0\rightarrow \Zst\Zst+\cancel{E}_{\rm T}$, we apply the $4\ell+\cancel{E}_{\rm T}$ analysis from the Type I 2HDM to $\medf$ with charge $(\mathbf{3},-1)$. The results are shown in Fig.~\ref{fig:3minus1_4l}. Discovery of $\medf$ can begin at $\approx15\,\,\mathrm{fb}^{-1}$ at LHC14, and the entire range up to $\mmedf=220\GeV$ can be discovered at $5\sigma$ with $50\,\,\mathrm{fb}^{-1}$, except in a small window where $\medf$ and $\chi$ are  degenerate.

\begin{figure}[t]
\centering
\includegraphics[width=0.6 \textwidth]{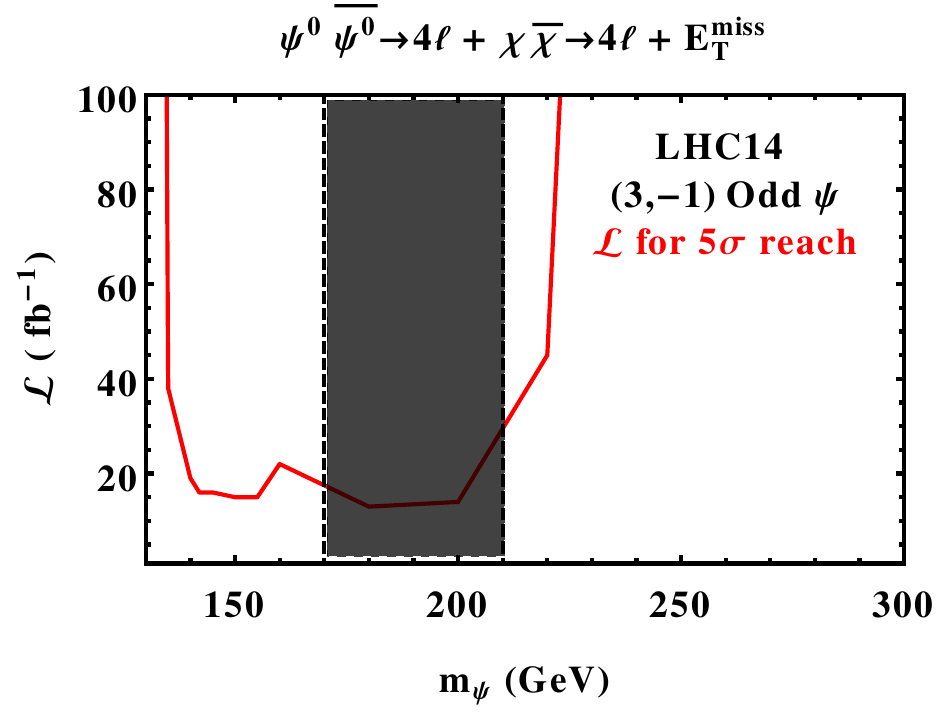}%
\caption{We show the luminosity  needed for a $5\sigma$ discovery in $\medf^0\overline{\medf}^0\rightarrow4\ell+\cancel{E}_{\rm T}$  searches at 14 TeV for the Z$_2$-odd fermion with charge $\left(\mathbf{3},-1\right)$. The shaded region is already excluded. This search mode targets $\mmedf\lesssim220\GeV$.}\label{fig:3minus1_4l}
\end{figure}

For $\mmedf \gtrsim220\GeV$, the gauge bosons from $\medf$ decay are on-shell, and there are two dominant production modes: $\medf^\pm\medf^0\rightarrow \Wpm\ZZ+\chi\bar\chi$, as in the Type I 2HDM, and $\medf^{\pm\pm}\medf^{\mp}\rightarrow \Wpm\Wpm\Wmp+\chi\bar\chi$. For the $\Wpm\ZZ+\cancel{E}_{\rm T}$ final state, the trilepton analysis proposed for the Type I 2HDM provides the best prospect for discovery, while the production of the doubly charged fermion leads to an $\ell^\pm\ell^\pm+2\mathrm{j}+\mathrm{E}_{\rm T}$ final state. We restrict ourselves to same-sign dimuon signatures, since muons have small non-prompt backgrounds, and we consider diboson, $\Wpm\Wpm$, and $\bar t t\Wpm$ backgrounds. The cuts we propose are:
\begin{enumerate}
\item Exactly two same-sign muons each with $p_{\rm T}>20\GeV$ and a combined invariant mass $m_{\mu\mu}>12\GeV$
\item No additional reconstructed leptons, hadronic taus with $p_{\rm T} >20\GeV$, or $b$-jets with $p_{\rm T}>20\GeV$ (assuming a 70\% tagging efficiency and 1\% mistagging rate)
\item Same-sign leptons do not reconstruct a Z ($\left| m_{\mu\mu}-m_\ZZ\right| > 15\GeV$)
\item At least two jets with $p_{\rm T}>20\GeV$ and $\left|m_{jj}-m_{\rm W}\right| <15\GeV$
\end{enumerate}
We show in Fig.~\ref{fig:3minus1_3l_www}  the luminosity needed for $5\sigma$ discovery as a function of $\mmedf$, considering the trilepton search in the left pane and the same-sign muon search in the right pane.   We see that both searches can discover $\medf$ between 220 and 300 GeV. A discovery in the $\approx230\GeV$ range is possible with $50\,\,\mathrm{fb}^{-1}$ with the trilepton search, while a discovery in the $\approx300\GeV$ range is possible with $20\,\,\mathrm{fb}^{-1}$ with the same-sign dimuon search. Both discovery modes complement one another and could be used to confirm any observed excess and probe the underlying model.

\begin{figure}[t]
\centering
\includegraphics[width=0.45 \textwidth]{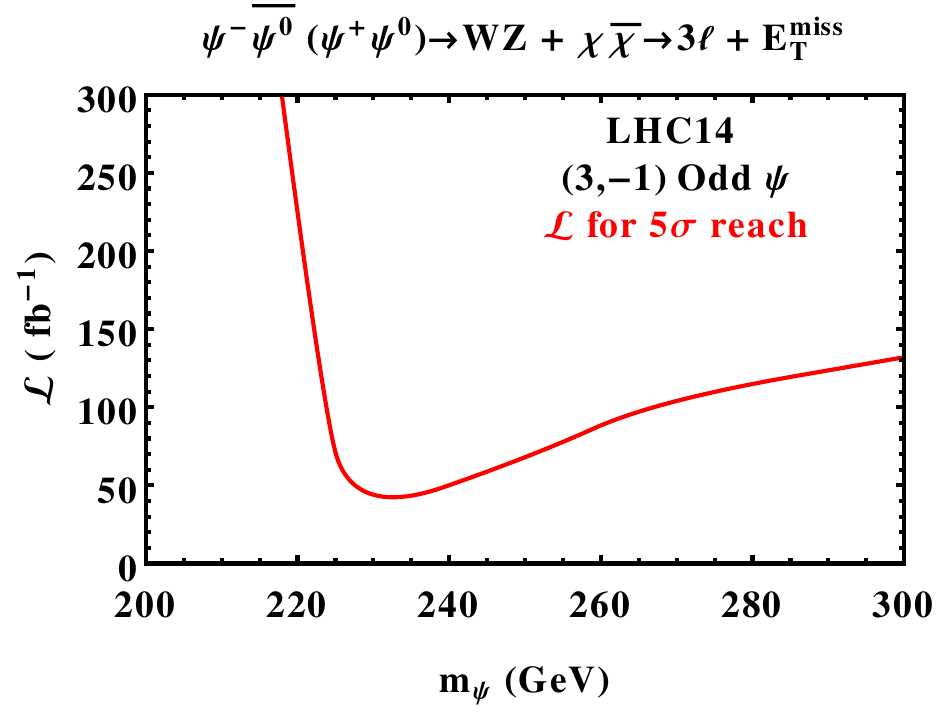}%
\includegraphics[width=0.45 \textwidth]{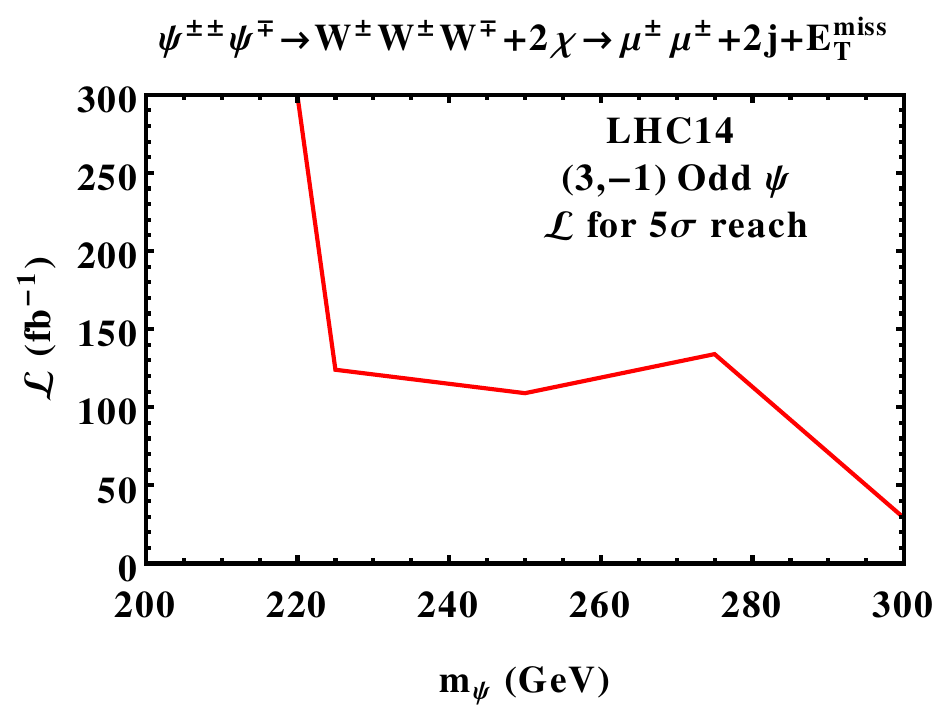}%
\caption{We show the luminosity  needed for a $5\sigma$ discovery in $\medf^\pm\medf^0\rightarrow3\ell+\cancel{E}_{\rm T}$ (left pane) and $\medf^{\pm\pm}\medf^\mp\rightarrow \mu^\pm\mu^\pm+2\mathrm{j}+\cancel{E}_{\rm T}$ (right pane) searches at 14 TeV for the Z$_2$-odd fermion with charge $\left(\mathbf{3},-1\right)$. These search modes target $\mmedf\gtrsim220\GeV$.}\label{fig:3minus1_3l_www}
\end{figure}
%

\vspace{5mm}
\hspace{-3mm}{\bf $\meds$ does not mix with the Higgs boson}
\vspace{5mm}

One model we considered in Section \ref{sec:constraintodd} included a doubly charged scalar that decayed into same-sign leptons through a $\meds^{--} e^{\rm c}e^{\rm c}$ operator. Decays of $\meds$ to electrons and muons are already ruled out unless the branching ratio is $<1\%$. This leaves the decay of $\meds^{--}$ into $\tau^-\tau^-$. Pair production of $\meds^{--}$ gives a $4\tau$ final state, and this scenario can be constrained using the same $4\tau$ search we applied to the Type IV 2HDM above. Because the details are  similar, we do not repeat the $4\tau$ analysis in full, but simply re-scale the signal cross section to match the Drell-Yan production of $\meds^{--}$. We find that the ratio of cross sections, $\sigma(pp\rightarrow\meds^{++}\meds^{--})/\sigma(pp\rightarrow HA)$, is approximately 1.5 at $M_{\meds^{--}}=100$ GeV and 1.9 at $M_{\meds^{--}}=300$ GeV . Applying this to the results of Fig.~\ref{fig:type4sign}, we find that new discoveries can begin with $\sim8\,\,\mathrm{fb}^{-1}$ of luminosity, and this scenario can be completely excluded in the range $M_{\meds^{--}}=100-300$ GeV at LHC14 with $\approx75-125\,\,\mathrm{fb}^{-1}$.

\vspace{5mm}
\hspace{-3mm}{\bf Summary}
\vspace{5mm}

We summarize the current constraints and prospects for detection of  the Z$_2$-even scalars and their partner fermions  in Table~\ref{detecttableodd}. We organize the table according to the SM charges  and final state flavors for each model, as well as the status of the signal.

\begin{table}
    \begin{tabularx}{\textwidth}{|l|l|l|l|X|}
        \hline
         \multicolumn{1}{|c|}{Charge} &  \multicolumn{1}{c|}{Flavor} &  \multicolumn{1}{c|}{Constraints} &  \multicolumn{1}{c|}{Reach ($5\sigma@$14TeV)} & \multicolumn{1}{c|}{Dominant Signal(s)} \\ \hline
                         \multicolumn{5}{|c|}{{\bf Already excluded}} \\ \hline
                         $(\mathbf{1},-2)$~ & $e,\mu$~ & $\mmeds \gtrsim 300\gev$~ & N/A  & $\meds^{--}\meds^{++}\rightarrow4\ell$ (no Z) \\ \hline
                \multicolumn{5}{|c|}{{\bf Proposed search}} \\ \hline
                 $\left(\mathbf{2},-\frac{1}{2}\right)$~ & N/A & none ~ &  $\mmedf < 265$ GeV & $\medf^0\overline{\medf}^0\rightarrow\Zst\Zst+\chi\bar\chi\rightarrow 4\ell+\cancel{E}_{\rm T}$ \\ 
                                Type I & & & $(\mathcal{L}=300\,\,\mathrm{fb}^{-1})$  & $\medf^\pm\medf^0\rightarrow\Wpm\ZZ+\chi\bar\chi\rightarrow 3\ell+\cancel{E}_{\rm T}$  \\ \hline
        $\left(\mathbf{2},-\frac{1}{2}\right)$~ & N/A & $m_H \gtrsim 130 \gev $~ &  $m_H < 300$ GeV & $H^\pm+H/A\rightarrow 3\tau+\nu\rightarrow \ell^\pm\ell^\pm+\tau_{\rm h}+\cancel{E}_{\rm T}$ \\ 
        Type IV & & & $(\mathcal{L} \approx 150\,\,\mathrm{fb}^{-1})$ & \\ 
        \hline
          $\left(\mathbf{3},-1\right)$ & N/A & $\mmedf\notin(170,210)\GeV$ & $\mmedf < 300\GeV$ & $\medf^0\overline{\medf}^0\rightarrow\Zst\Zst+\chi\bar\chi\rightarrow4\ell+\cancel{E}_{\rm T}$ \\ 
       &  &  & ($\mathcal{L} \approx 100\,\,\mathrm{fb}^{-1}$) & $\medf^\pm\medf^0\rightarrow \Wpm\ZZ+\chi\bar\chi\rightarrow 3\ell+\cancel{E}_{\rm T}$ \\ 
              &  &  &  & $\medf^{\pm\pm}\medf^{\mp} \rightarrow 3\mathrm{W}+\chi\bar\chi\rightarrow\mu^\pm\mu^\pm+2\mathrm{j}+\cancel{E}_{\rm T}$ \\ 
        \hline
                                 $(\mathbf{1},-2)$~ & $\tau$~ & $\mmeds \gtrsim 110\gev$~ & $\mmeds < 300$ GeV  & $\meds^{--}\meds^{++}\rightarrow4\tau\rightarrow \ell^\pm\ell^\pm+2\tau_{\rm h}+\cancel{E}_{\rm T}$ \\
                          & & &  $(\mathcal{L} \sim 100\,\,\mathrm{fb}^{-1})$ &\\  \hline
                           \multicolumn{5}{|c|}{{\bf Very low sensitivity}} \\ \hline
                             $\left(\mathbf{3},0\right)$ & N/A & none & none & $\medf^\pm\medf^0\rightarrow \Wpm h+\chi\bar\chi\rightarrow \ell^\pm+\gamma\gamma+\cancel{E}_{\rm T}$ \\ 
        \hline


    \end{tabularx} \caption{Summary table for Z$_2$-odd fermions and partner scalars with the indicated  electroweak charges and flavor couplings. When relevant, the mass reach indicates the highest mass ($\mmedf$ and/or $\mmeds$) that can be discovered in our window of interest (up to 300 GeV), and the luminosity required at LHC14 for a $5\sigma$ discovery over the possible range.}\label{detecttableodd}
\end{table}

\newpage
\clearpage

\section{Sterile neutrino model}\label{sec:sterile}

\subsection{The Model}\label{model:sterile}

The final model we consider departs from the minimal models described above. It contains an additional light state with the quantum numbers of a sterile neutrino. So far we considered the DM state to be made up of at most two Weyl fermions, $\chi_1$ and $\chi_2$, as in Eq.~(\ref{eqn:MiDMInteraction}). It is natural to try to extend the model to three states which behave like three right-handed neutrinos, and so we add a third state, $\chi_3$. We consider a Z$_2$ symmetry under which $\medf$, $\medf^c$, $\chi_1$ and $\chi_2$ are odd and a mass structure which results in a pseudo-Dirac dark matter pair, $\chi_1$ and $\chi_2$, together with a single Majorana fermion $\chi_3$, 
\be
M\chi_1 \chi_2 + \tfrac{1}{2}m_1 \chi_1\chi_1+ \tfrac{1}{2}m_2 \chi_2\chi_2 + \tfrac{1}{2}m_3 \chi_3\chi_3.
\ee
Introducing a heavy Dirac pair of $\SUWeak$ doublets, $\medf$ and $\medf^c$ of mass $\mmedf$ and a new scalar doublet $\meds$ with mass $\mmeds$, the Yukawa coupling is
\be\label{eq:sterileyukawa}
\mathcal{L} \supset \lambda_1 \medf \meds \chi_1 + \lambda_2^* \bar{\medf}^c \meds^\dag \bar{\chi}_2 + \rm{h.c.}
\ee
These terms  combine into a Yukawa coupling between the Dirac pairs $\left(\medf,\bar{\medf}^c\right)$ and $\left(\chi_1,\bar{\chi}_2\right)$, generating the operators (\ref{eqn:MiDMInteraction}) and (\ref{eqn:RDMLagrangian}).

Consistent with the symmetries we can write a Yukawa term between $\meds$, the normal lepton doublets, and the sterile $\chi_3$,
\be\label{eq:steriledecay}
\mathcal{L} \supset \lambda_3 \ell \meds\chi_3 + \rm{h.c.}
\ee
This allows the scalar Higgs to decay on shell to a SM lepton and the sterile state $\chi_3$. It also allows the heavy leptonic doublets, $\medf$ and $\medf^c$, to decay to a WIMP, a SM lepton and the sterile state $\chi_3$ through an off-shell scalar $\meds$. These decays are shown in Fig.~\ref{fig:Decays_with_sterile}.

With the interactions (\ref{eq:sterileyukawa}) and (\ref{eq:steriledecay}), the model has an additional Z$_2$ symmetry under which $\medf$, $\meds$, and $\chi_3$ are odd. This symmetry can be weakly broken, for instance by mixing with the SM neutrinos through the operator $h\ell\chi_3$. In this case $\chi_3$ may decay through an electromagnetic dipole moment transition with the SM neutrinos as is usually the case for sterile neutrinos~\cite{Kusenko:2009up}. If $m_{\chi_3} \gtrsim \GeV$ then it decays before Big-Bang nucleosynthesis, and cosmological constraints do not apply.  On the other hand, if the symmetry is unbroken then $\chi_3$ is a stable thermal relic that is a component of dark matter, and its interactions are bounded by overclosure constraints.  This component, $\chi_3$, annihilates to leptons through effective operators like $\lambda_3^2(\ell\chi_3)(\ell\chi_3)^\dagger/\mmeds^2$. For electrons $\ell=e$, LEP monophoton searches constrain $\lambda_3^2/\mmeds^2$ to be small enough that the $\chi_3$ relic abundance overcloses the universe unless $m_{\chi_3}\gtrsim20-50$ GeV, depending on the Lorentz structure of the interaction \cite{Fox:2011fx}. If $\chi_3$ is a WIMP with a thermal relic abundance, it must therefore either have a weak scale mass or couple predominantly to muons and/or taus.

\begin{figure}[tb]
\begin{center}
\includegraphics[width=0.9\textwidth]{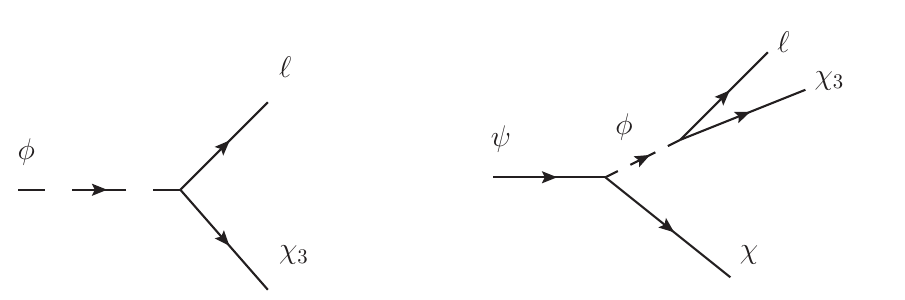}
\end{center}
\caption{Decays of the heavy charged states when a light sterile state, $\chi_3$, is present. On the left the scalar decays into a SM lepton and the sterile state. On the right the fermion decays into the WIMP candidate $\chi$, a SM lepton, and the sterile state through the scalar.  }
\label{fig:Decays_with_sterile}
\end{figure}

Alternatively, $\chi_3$ can be extremely light or massless, in which case it contributes to the relativistic degrees of freedom at the times of Big-Bang nucleosynthesis and recombination. Its contribution depends on the temperature at which it decouples from the thermal bath, which depends in turn on the magnitude of $\lambda_3$ and the lepton species to which it couples. In the most constrained case, $\chi_3$ couples to electrons with $\mathcal O(1)$ coupling, and it contributes a full effective neutrino degree of freedom ($\Delta N_{\rm eff}\sim1$). The predicted contribution to $\Delta N_{\rm eff}$ drops significantly if $\lambda_3\ll1$ or couples dominantly to taus.  We show in Fig.~\ref{fig:Neff} the additional effective neutrino degrees of freedom $\Delta N_{\rm eff}$ resulting from different values of $\lambda_3/\mmeds$ for the most constraining scenario where the lepton in Eq.~(\ref{eq:steriledecay}) is an electron.

\begin{figure}[tb]
\begin{center}
\includegraphics[width=0.5\textwidth]{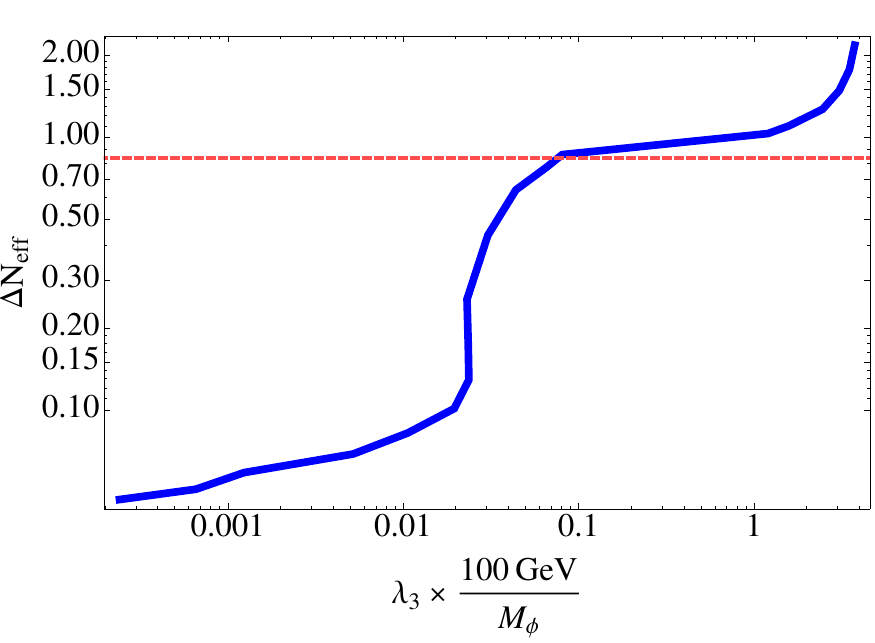}
\end{center}
\caption{Contribution of $\chi_3$ to the effective neutrino degrees of freedom ($\Delta N_{\rm eff}$) as a function of the Yukawa coupling $\lambda_3$ and the messenger mass $\mmeds$. The red dashed line shows the current $1\sigma$ upper limit from Planck ($\Delta N_{\rm eff}\lesssim0.84$) \cite{Ade:2013zuv}.  }
\label{fig:Neff}
\end{figure}

\subsection{LHC Constraints} \label{sec:lhcsterile}
In models with a sterile state $\chi_3$, the scalar messenger $\meds$ decays into a SM lepton and $\chi_3$ as in Fig~\ref{fig:Decays_with_sterile}. The signature is identical to slepton production in SUSY models. The phenomenology depends on the mass of $\chi_3$, which is a free parameter, but for concreteness we choose to let $m_{\chi_3}=0$ for all analyses that follow. If $\chi_3$ has a weak-scale mass, the constraints and search prospects consequently weaken. It is customary for ATLAS and CMS to quote slepton constraints as a function of the neutralino mass, and it is straightforward to adapt such constraints to the sterile neutrino model with non-zero $m_{\chi_3}$.

Pair production of $\meds^\pm$ gives a dilepton$+\cancel{E}_{\rm T}$ signature at colliders. LEP places a bound of $\mmeds \gtrsim90$ GeV, regardless of the final state lepton flavor or $\SUWeak$ charge of $\meds$. At the LHC,  searches in the dielectron+$\cancel{E}_{\rm T}$ and dimuon$+\cancel{E}_{\rm T}$ channels place constraints on the $\meds^\pm$ mass provided it decays into light-flavor leptons (electrons or muons). They exclude $\meds^\pm$ lighter than 275 GeV if it has electroweak charge $(\mathbf{2}, -1/2)$, while $\meds^\pm$ is excluded between $120-200$ GeV if it has charge $(\mathbf{1}, -1)$. There remains an open window for the $\SUWeak$ singlet between 90-120 GeV; while the $\meds$ production rate is relatively large in this window, the events kinematically resemble fully leptonic WW production, a background that is two orders of magnitude larger. Given a mild excess already observed in WW production~\cite{Curtin:2012nn} at 7 and 8 TeV, and that the boosted WW fraction will rise at 14 TeV, it will be a challenge to close this gap.

Na\"ively, similar bounds coming from dilepton$+\cancel{E}_{\rm T}$ searches can be applied to the fermion $\medf^\pm$ since it decays through an off-shell $\meds^\pm$ into a 3-body final state $\chi+ \ell^\pm + \chi_3$. However, when $\mmedf\sim \mX$,  the spectrum is squeezed and the bounds on $\medf^\pm$ are considerably weaker. When $\medf^\pm$ is heavy, however, the missing energy of its decay is quite large due to contributions from both $\chi_3$ and dark matter, and the bounds are correspondingly strong. If $\medf^\pm$ is an $\SUWeak$ doublet decaying to electrons or muons plus missing energy, it is excluded from 230-415 GeV. An $\SUWeak$ singlet is not constrained by current data, but bounds on the model will likely result over a similar interval at LHC14, while bounds on the doublet will be  extended slightly down to $\sim200$ GeV.

Which constraints are most relevant for this scenario depends on the mass of the scalars. When the scalar is in the mass range $100\GeV \lesssim \mmeds \lesssim 250\GeV$  its decays provide the strongest bounds. Those are summarized in Table~\ref{tab:sterile}. On the other hand, when the scalar $\meds^\pm$ is heavy ($\mmeds \gtrsim 300\GeV$) the strongest bounds on the model come from constraints on the fermion $\medf^\pm$, whose production cross section is enhanced relative to $\meds^\pm$.  The bounds associated with the fermion $\medf$ are summarized in Table \ref{tab:sterile2}.

If decays are dominantly to $\tau^\pm\, +\, \cancel{E}_{\rm T}$, however, there are no constraints. The leptonic branching fraction of taus is sufficiently small to make the dilepton$+\cancel{E}_{\rm T}$ bounds irrelevant, while hadronic $\tau$-tagging algorithms have sufficiently large mistag rates that Z$+$jet events dwarf any ditau$+\cancel{E}_{\rm T}$ signal. As a result, the prospects for discovering $\meds^\pm$ at the LHC are very poor when it preferentially decays into tau, and motivates the development of high-purity tau taggers.

\begin{table}
    \begin{tabularx}{\textwidth}{|l|l|l|c|X|}
        \hline
         \multicolumn{1}{|c|}{Charge} &  \multicolumn{1}{c|}{Flavor} &  \multicolumn{1}{c|}{Constraints} &  \multicolumn{1}{c|}{Discover $\mmeds=300$ GeV?} & \multicolumn{1}{c|}{Dominant Signal} \\ \hline
        $\left(\mathbf{2},-\frac{1}{2}\right)$~ & $e,\mu$~ & $\mmeds \gtrsim 275 \gev $~ & yes & $\meds^+\meds^-\rightarrow\ell^+\ell^-+\cancel{E}_{\rm T}$ \\  \hline
              $(\mathbf{1},-1)$~ & $e,\mu$~ & $\mmeds < 120$ GeV and~ & yes &  $\meds^+\meds^-\rightarrow\ell^+\ell^-+\cancel{E}_{\rm T}$ \\
              & & $\mmeds > 200$ GeV & & \\\hline
                      $\left(\mathbf{2},-\frac{1}{2}\right)$~ & $\tau$~ & none~ & no~ & $\meds^+\meds^-\rightarrow\tau^+\tau^-+\cancel{E}_{\rm T}$ \\  \hline
              $(\mathbf{1},-1)$~ & $\tau$~ & none~ & no~ &  $\meds^+\meds^-\rightarrow\tau^+\tau^-+\cancel{E}_{\rm T}$ \\\hline
      
    \end{tabularx} \caption{Summary table for bounds on the scalar messenger $\meds^\pm$ in the sterile neutrino model according to weak charges and flavor couplings, as well as whether  $\meds^\pm$ can be discovered at $5\sigma$ at LHC14 with $\mathcal{L}<300\,\,\mathrm{fb}^{-1}$. When both the scalar $\meds$ and the fermion  $\medf^\pm$ are  in the mass range 100 - 250 GeV the bounds on $\medf^\pm$ are typically weaker because the spectrum is squeezed.}\label{tab:sterile}
\end{table}

\begin{table}
    \begin{tabularx}{\textwidth}{|l|l|l|c|X|}
        \hline
         \multicolumn{1}{|c|}{Charge} &  \multicolumn{1}{c|}{Flavor} &  \multicolumn{1}{c|}{Constraints} &  \multicolumn{1}{c|}{Discovery potential} & \multicolumn{1}{c|}{Dominant Signal} \\ \hline
        $\left(\mathbf{2},-\frac{1}{2}\right)$~ & $e,\mu$~ &  $230 \lesssim m_ \medf \lesssim 415 \gev $~ & \parbox{100pt}{~ \\ Extend range down to 200 GeV and above 415 GeV \\} & $ \medf^+ \medf^-\rightarrow\ell^+\ell^-+\cancel{E}_{\rm T}$ \\  \hline
              $(\mathbf{1},-1)$~ & $e,\mu$~ & none & $250 \lesssim m_ \medf \lesssim 400 \gev $ &  $ \medf^+ \medf^-\rightarrow\ell^+\ell^-+\cancel{E}_{\rm T}$ \\
              \hline
                      $\left(\mathbf{2},-\frac{1}{2}\right)$~ & $\tau$~ & none~ & no~ & $ \medf^+ \medf^-\rightarrow\tau^+\tau^-+\cancel{E}_{\rm T}$ \\  \hline
              $(\mathbf{1},-1)$~ & $\tau$~ & none~ & no~ &  $ \medf^+ \medf^-\rightarrow\tau^+\tau^-+\cancel{E}_{\rm T}$ \\\hline
      
    \end{tabularx} \caption{Summary table for bounds on the fermion messenger $ \medf^\pm$ in the sterile neutrino model according to weak charges and flavor couplings, as well as whether  $ \medf^\pm$ can be discovered at $5\sigma$ at LHC14 with $\mathcal{L}<300\,\,\mathrm{fb}^{-1}$. }\label{tab:sterile2}
\end{table}

\newpage
\clearpage

\section{Discussion and Conclusions}
\label{sec:conclusions}

We have examined the bounds and search prospects from the LHC on a variety of models exhibiting light ($\sim$ 100 GeV - 300 GeV) charged states. These models are directly motivated by the recent observation of a 130 GeV gamma-ray line in the Fermi data, although the resulting collider signatures apply more generally to searches for $\mathcal{O}(100\,\,\mathrm{GeV})$ electroweak states. We have classified the results according to the Z$_2$ charge of each particle and their electroweak gauge charges. A summary of the results was shown in Fig.~\ref{fig:summaryplot}.

Generally, electroweak doublets (or higher electroweak multiplets) decaying to light-flavor leptons with gauge bosons are completely excluded by current searches. Electroweak singlets and final states with multiple tau leptons or gauge bosons have weaker constraints, but most of these scenarios can be probed at LHC14 with up to $300\,\,\mathrm{fb}^{-1}$ luminosity. We have proposed searches for LHC14 that are modifications of existing multilepton analyses and that can discover the new charged fermion, $\medf$, and charged scalar, $\meds$. In particular, same-sign lepton + hadronic tau searches are identified as excellent probes of $3\tau$ and $4\tau$ final states, and can improve the signal-to-background ratio and significance of such models  over other proposed analyses at fixed luminosity. Finally, we have found that there are some models which require more exotic searches (such as the disappearing charged track signature of the stable model), or whose signatures are completely buried in SM backgrounds (due to small production cross section or $\tau\tau+\cancel{E}_{\rm T}$ final states); such models are very challenging to see in hadronic machines, and a lepton collider such as the International Linear Collider (ILC) may be needed to probe them. Overall, many models of electroweak physics generating a strong gamma ray line through dipole and Rayleigh operators are either excluded, or can be probed with moderate luminosity at LHC14.

\begin{acknowledgments}
We would like to thank the anonymous referee for useful comments and suggestions. BS is supported in part by the Canadian Institute of Particle Physics. NW is supported by NSF grant \#0947827. IY is supported in part by funds from the Natural Sciences and Engineering Research Council (NSERC) of Canada. Research at the Perimeter Institute is supported in part by the Government of Canada through Industry Canada, and by the Province of Ontario through the Ministry of Research and Information (MRI).
\end{acknowledgments}

\appendix


\section{The  branching ratios for Z$_2$-even fermions}\label{app:evenfermion}

We provide details of the branching ratio calculations for Z$_2$-even fermions with charge $(\mathbf{1}, -1)$; the other calculations proceed analogously, and we show only the results.

\vspace{5mm}
\hspace{-3mm}{\bf The decay of Z$_2$-even fermions with charge $(\mathbf{1}, -1)$}
\vspace{5mm}

The electroweak Lagrangian for the SM lepton fields ($\ell$, $e^{\rm c}$) and messenger fields ($\medf$, $\medf^{\rm c}$) is:

\begin{equation}
\begin{array}{c}
 \mathcal{L}_Z=\frac{g}{\mathrm{c}_{\rm W}}Z_{\mu }\left[\left(\frac{-1}{2}+\mathrm{s}_{\rm W}^2\right)\overline{\ell}\gamma ^{\mu }\ell+\mathrm{s}_{\rm W}^2\overline{e^{\rm c}}\gamma ^{\mu }e^{\rm c}+\mathrm{s}_{\rm W}^2\overline{\medf}\gamma ^{\mu }\medf\right], \\
 \mathcal{L}_W=\frac{g}{\sqrt{2}}W_{\mu }^-\overline{\ell}\gamma ^{\mu }\nu_\ell+\mathrm{h.c.}, \\
 \mathcal{L}_H=\lambda_e\,\ell H^* e^{\rm c} + \lambda _2\,\ell H^*\medf^{\rm c}-M_{\psi }\,\psi\psi^{\rm c} +\mathrm{h.c.} ,
\end{array}
\end{equation}
where $H$ is the SM $\SUWeak$ Higgs doublet, and we have defined $\mathrm{s}_{\rm W} = \sin\theta_{\rm W}$, etc. We diagonalize the mass matrix and find a left-handed charged fermion mixing angle of  $\tan\theta_{\rm L}=\frac{-\lambda _2 }{\sqrt{2}}\frac{\upsilon}{M_{\psi }}$, while the right-handed mixing angle is $\tan\theta_{\rm R}=\frac{-\lambda _2\lambda _e }{2}\frac{\upsilon^2}{M_{\psi }^2}$; the left-handed mixing is largest. The mass of the charged fermion is $m_{\psi^\pm}=\sqrt{M_{\psi }^2+\left.\left(\lambda _2\upsilon \right){}^2\right/2}$ to leading-order. The decay widths of the fermion components are:
\begin{eqnarray}
 \Gamma \left(\psi ^-\rightarrow Z+e^-\right)&=&\frac{g^2}{128\pi}\sin^2\theta_{\rm L}\cos^2\theta_{\rm L}\frac{m_{\psi ^-}^3}{m_W^2}\left(1-\frac{m_Z^2}{m_{\psi ^-}^2}\right)\left(1+\frac{m_Z^2}{m_{\psi ^-}^2}-\frac{2m_Z^4}{m_{\psi ^-}^4}\right), \\
 \Gamma \left(\psi ^-\rightarrow W^-+\nu \right)&=&\frac{g^2}{64\pi}\sin^2\theta_{\rm L}\frac{m_{\psi ^-}^3}{m_W^2}\left(1-\frac{m_W^2}{m_{\psi ^-}^2}\right)\left(1+\frac{m_W^2}{m_{\psi ^-}^2}-\frac{2m_W^4}{m_{\psi ^-}^4}\right), \\
 \Gamma \left(\psi ^-\rightarrow h+e^-\right)&=&\left(\lambda _2\cos\theta_{\rm L}\cos\theta_{\rm R}\right)^2\frac{m_{\psi ^-}}{64\pi }\left(1-\frac{m_h^2}{m_{\psi ^-}^2}\right){}^2.
\end{eqnarray}
We plot the branching ratios in the left plot of Fig.~\ref{fig:model12}.
\begin{figure}[h]
\centering
\includegraphics[width=0.45 \textwidth]{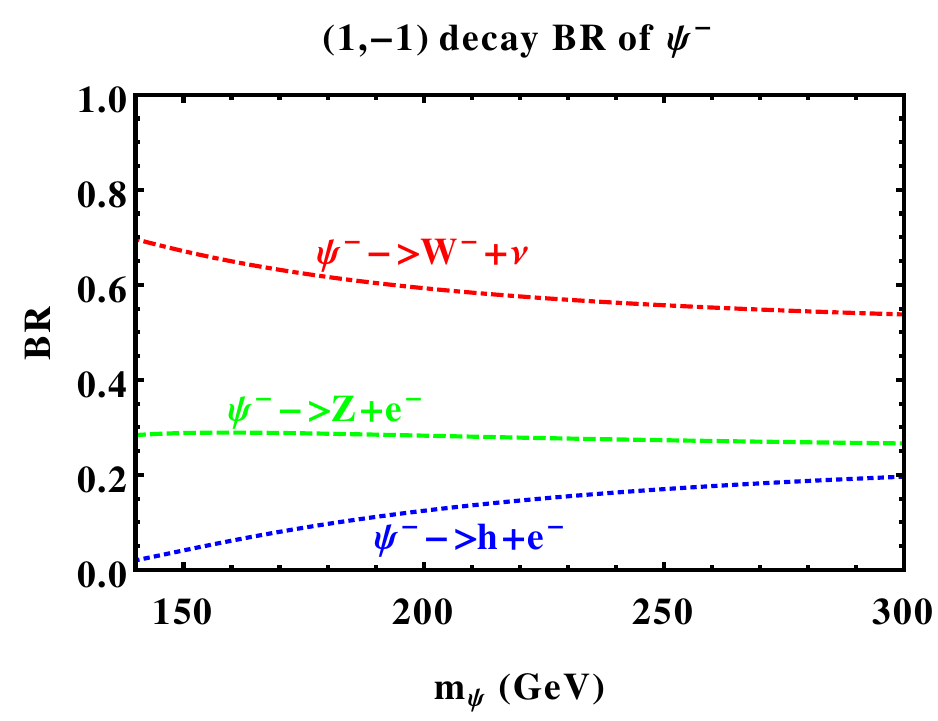}%
\includegraphics[width=0.45 \textwidth]{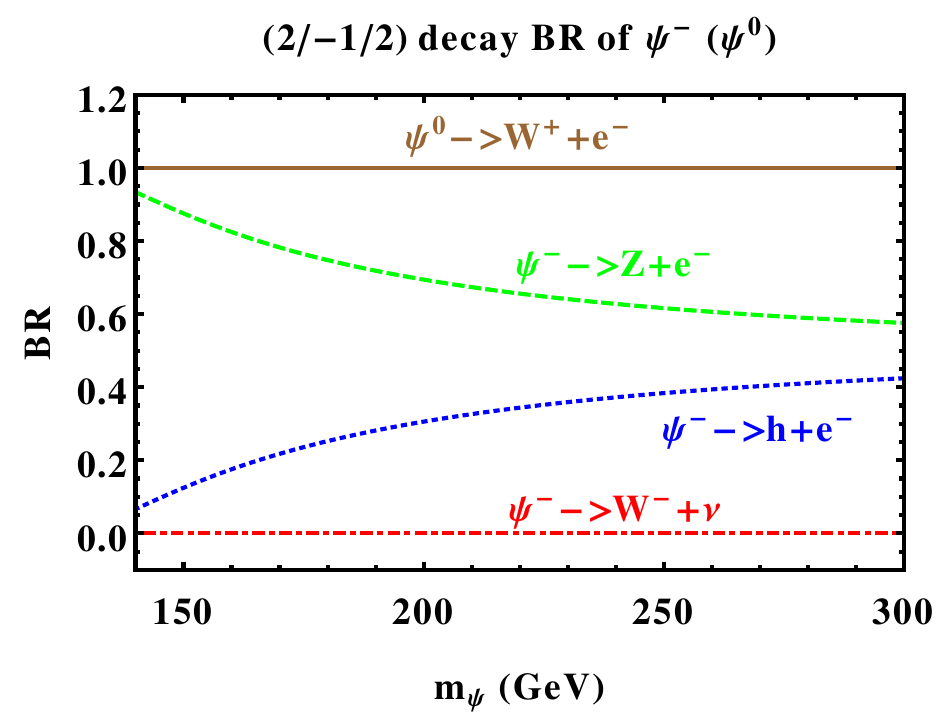}%
\caption{The decay branching ratios of charged and neutral fermions with electroweak charges $(\mathbf{1}, -1)$ and $(\mathbf{2}, -1/2)$.}\label{fig:model12}
\end{figure}
%

\vspace{5mm}
\hspace{-3mm}{\bf The decay of Z$_2$-even fermions with charge $(\mathbf{2}, -1/2)$}
\vspace{5mm}

The branching ratios are shown in the right plot of Fig.~\ref{fig:model12}.

\vspace{5mm}
\hspace{-3mm}{\bf The decay of Z$_2$-even fermions with charge $(\mathbf{3}, 0)$}
\vspace{5mm}

The triplet vector fermion $\psi$ has three components: $\left(\psi ^+,\psi ^0,\psi ^-\right)$. We show the branching ratios in the left plot of Fig.~\ref{fig:model-tri}.

\begin{figure}[h]
\centering
\includegraphics[width=0.45 \textwidth]{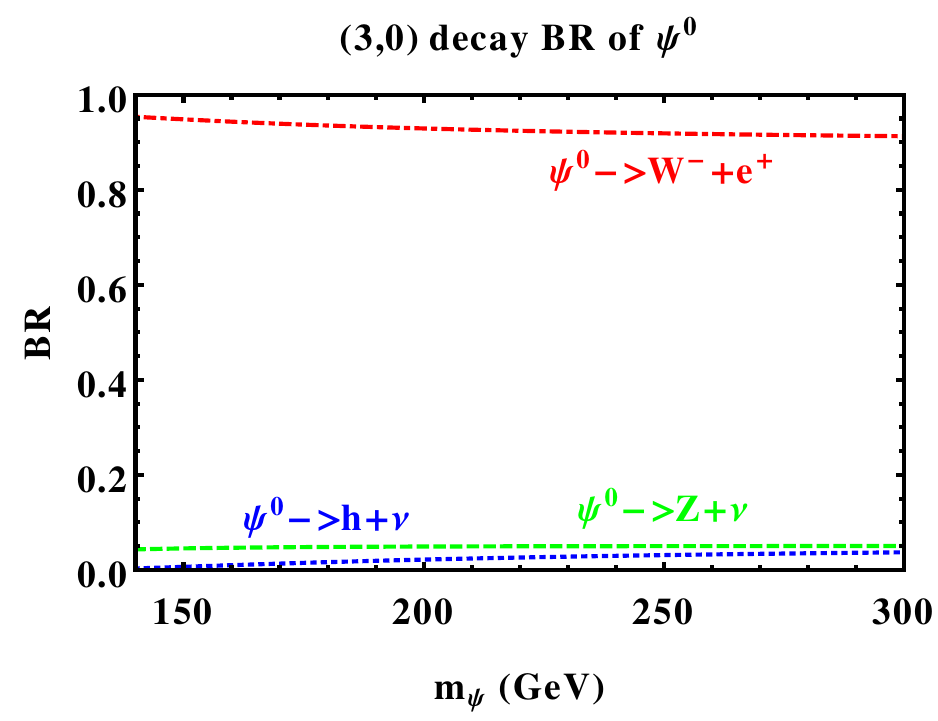}%
\includegraphics[width=0.45 \textwidth]{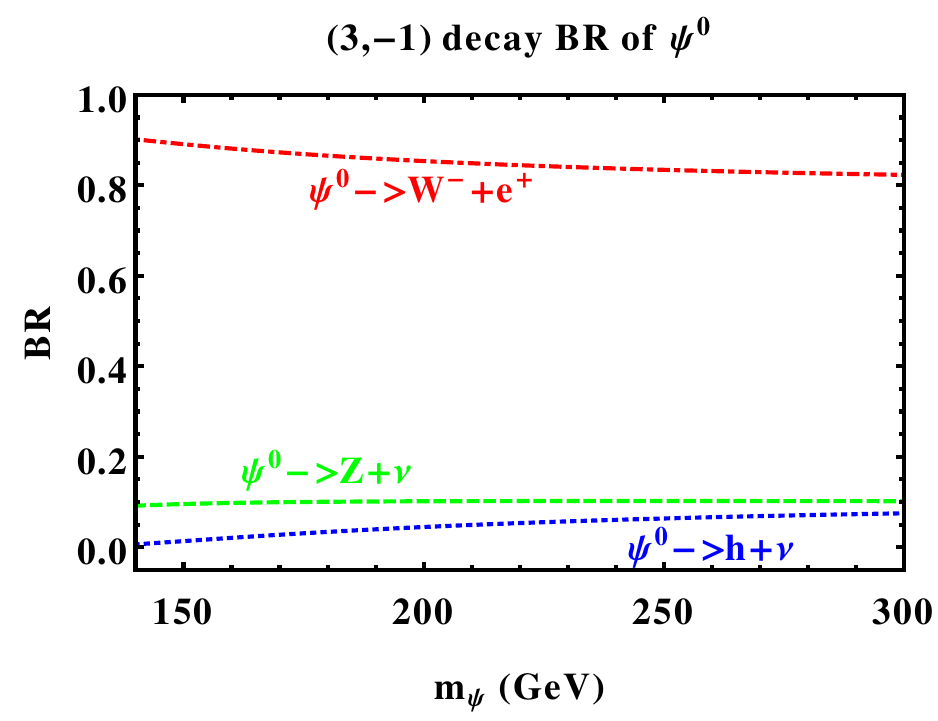}%
\\
\includegraphics[width=0.45 \textwidth]{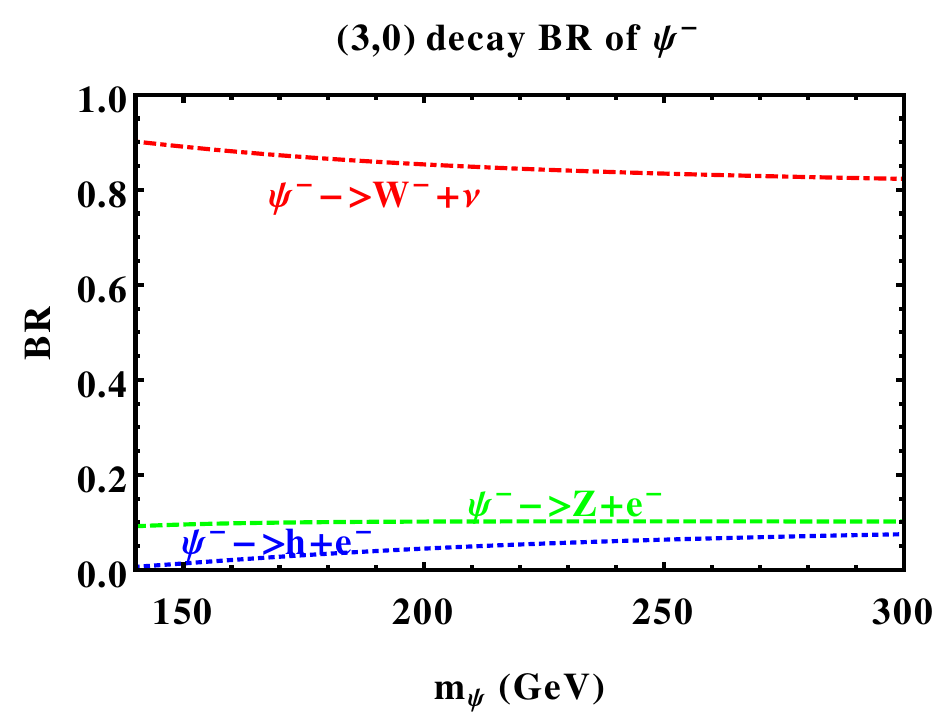}%
\includegraphics[width=0.45 \textwidth]{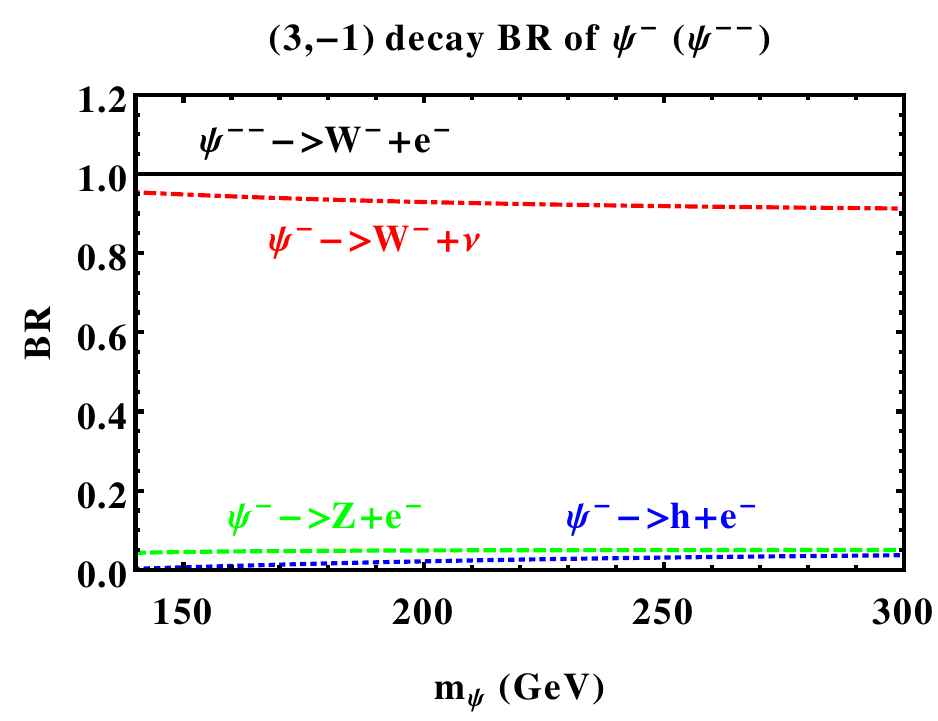}%
\caption{The decay branching ratio of charged and neutral fermions with SM charge $(\mathbf{3}, 0)$ and $(\mathbf{3}, -1)$.}\label{fig:model-tri}
\end{figure}
%

\vspace{5mm}
\hspace{-3mm}{\bf The decay of Z$_2$-even fermions with charge $(\mathbf{3}, -1)$}
\vspace{5mm}

The triplet vector fermion $\psi$ has three components: $\left(\psi ^0,\psi ^-,\psi ^{--}\right)$. We show the branching ratios in the right plot of Fig.~\ref{fig:model-tri}.

\section{The  branching ratios for Z$_2$-odd fermions in 2HDMs}\label{app:oddfermion}
In this appendix, we present and discuss the decay modes of the Z$_2$-odd fermions with electroweak charge $(2,-1/2)$ (Type I 2HDM). First, we note that, since the scalar in this model mixes with the Higgs boson and gets a VEV, the neutral fermion in turn is mixed with DM through the Yukawa interaction $\lambda\meds \bar{\medf}\chi $. The DM doublet fraction should be smaller than $\sim0.2-0.45$ to avoid gamma ray continuum bounds on annihilation to WW. In addition, due to the fermion mixing induced by the scalar messenger's VEV, the neutral fermion component can decay to $\mathrm{DM} + \ZZ$, while the charged fermion can decay to $\mathrm{DM} + \Wpm$. There are also decay channels to DM and an off-shell $\meds$ coming from the same Yukawa interaction $\lambda  \meds  \bar{\medf } \chi$. 

The most direct decay mode is a two-body decay into DM and a gauge/Higgs boson, which is allowed when $\mmedf\gtrsim 200\GeV$. In this mass range, $\medf^\pm$ decays exclusively into $\Wpm+\chi$, while $\medf^0\rightarrow\ZZ+\chi$ when $\mmedf<m_h+m_\chi$; otherwise, the Goldstone equivalence theorem predicts that $\medf^0$ has approximately equal branching fraction into $h+\chi$ and $\ZZ+\chi$.

  For lower masses, $\medf$ has three-body decays, for example into lepton final states through off-shell gauge bosons, as we now discuss. For the neutral fermion, the decays are dominated by Z exchange and are shown in the left pane of Fig.\ref{fig:3bodyneutral}. For the charged fermion, $\Wpm$ mediates similar three-body decays and the relevant branching-ratios are presented in the right pane of Fig.~\ref{fig:3bodyneutral}. The difference between $\tau \overline{\mathit{v}_{\tau }}\chi$ and $s\bar{c}\chi$ and the decays involving 1$^{\rm st}$ generation fermions is due to the contribution from scalar exchange, which is proportional to the different Higgs Yukawa couplings. 


%
\begin{figure}[t]
\centering
\includegraphics[width=0.45 \textwidth]{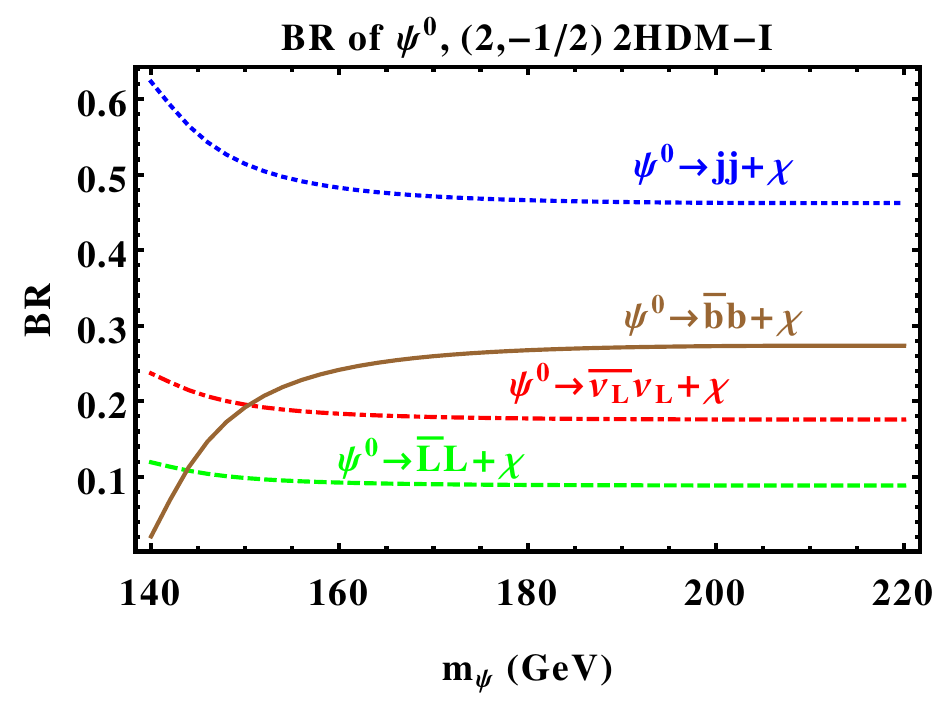}
\includegraphics[width=0.45 \textwidth]{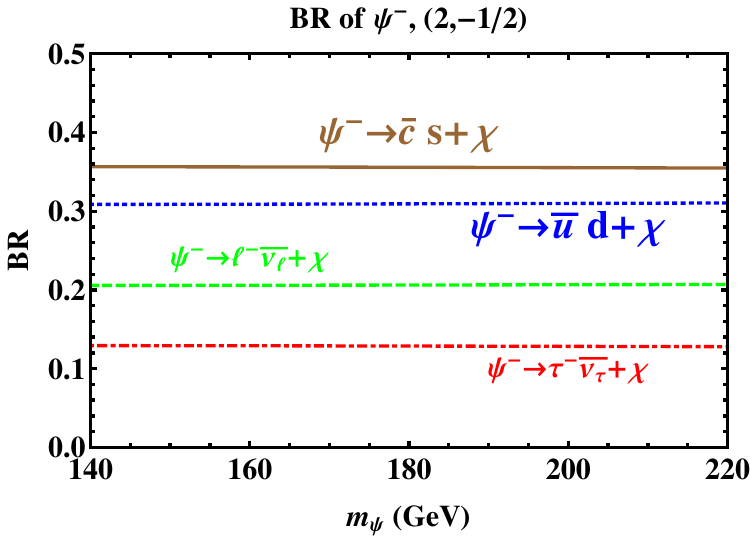}%
\caption{In the left pane (right pane) we depict the three-body decay branching ratio of the neutral fermions $\medf^0$ (charged fermions $\medf^\pm$) with SM charge $(\mathbf{2}, -1/2)$ and odd under Z$_2$. }\label{fig:3bodyneutral}
\end{figure}
%
%

\bibliographystyle{JHEP}
\bibliography{MRayDM-LHC-bib}

\providecommand{\href}[2]{#2}\begingroup\raggedright\begin{thebibliography}{10}

\bibitem{Bagnasco:1993st}
J.~Bagnasco, M.~Dine, and S.~D. Thomas, {\it {Detecting technibaryon dark
  matter}},  {\em Phys. Lett.} {\bf B320} (1994) 99--104,
  [\href{http://xxx.lanl.gov/abs/hep-ph/9310290}{{\tt hep-ph/9310290}}].

\bibitem{Pospelov:2000bq}
M.~Pospelov and T.~ter Veldhuis, {\it {Direct and indirect limits on the
  electro-magnetic form factors of WIMPs}},  {\em Phys. Lett.} {\bf B480}
  (2000) 181--186, [\href{http://xxx.lanl.gov/abs/hep-ph/0003010}{{\tt
  hep-ph/0003010}}].

\bibitem{Sigurdson:2004zp}
K.~Sigurdson, M.~Doran, A.~Kurylov, R.~R. Caldwell, and M.~Kamionkowski, {\it
  {Dark-matter electric and magnetic dipole moments}},  {\em Phys. Rev.} {\bf
  D70} (2004) 083501, [\href{http://xxx.lanl.gov/abs/astro-ph/0406355}{{\tt
  astro-ph/0406355}}].

\bibitem{Gardner:2008yn}
S.~Gardner, {\it {Shedding Light on Dark Matter: A Faraday Rotation Experiment
  to Limit a Dark Magnetic Moment}},  {\em Phys. Rev.} {\bf D79} (2009) 055007,
  [\href{http://xxx.lanl.gov/abs/0811.0967}{{\tt 0811.0967}}].

\bibitem{Masso:2009mu}
E.~Masso, S.~Mohanty, and S.~Rao, {\it {Dipolar Dark Matter}},  {\em Phys.
  Rev.} {\bf D80} (2009) 036009, [\href{http://xxx.lanl.gov/abs/0906.1979}{{\tt
  0906.1979}}].

\bibitem{Cho:2010br}
W.~S. Cho, J.-H. Huh, I.-W. Kim, J.~E. Kim, and B.~Kyae, {\it {Constraining
  WIMP magnetic moment from CDMS II experiment}},  {\em Phys. Lett.} {\bf B687}
  (2010) 6--10, [\href{http://xxx.lanl.gov/abs/1001.0579}{{\tt 1001.0579}}].

\bibitem{An:2010kc}
H.~An, S.-L. Chen, R.~N. Mohapatra, S.~Nussinov, and Y.~Zhang, {\it {Energy
  Dependence of Direct Detection Cross Section for Asymmetric Mirror Dark
  Matter}},  \href{http://xxx.lanl.gov/abs/1004.3296}{{\tt 1004.3296}}.

\bibitem{McDermott:2010pa}
S.~D. McDermott, H.-B. Yu, and K.~M. Zurek, {\it {Turning off the Lights: How
  Dark is Dark Matter?}},  {\em Phys.Rev.} {\bf D83} (2011) 063509,
  [\href{http://xxx.lanl.gov/abs/1011.2907}{{\tt 1011.2907}}].

\bibitem{Chang:2010en}
S.~Chang, N.~Weiner, and I.~Yavin, {\it {Magnetic Inelastic Dark Matter}},
  {\em Phys.Rev.} {\bf D82} (2010) 125011,
  [\href{http://xxx.lanl.gov/abs/1007.4200}{{\tt 1007.4200}}].

\bibitem{Banks:2010eh}
T.~Banks, J.-F. Fortin, and S.~Thomas, {\it {Direct Detection of Dark Matter
  Electromagnetic Dipole Moments}},
  \href{http://xxx.lanl.gov/abs/1007.5515}{{\tt 1007.5515}}.

\bibitem{DelNobile:2012tx}
E.~Del~Nobile, C.~Kouvaris, P.~Panci, F.~Sannino, and J.~Virkajarvi, {\it
  {Light Magnetic Dark Matter in Direct Detection Searches}},
  \href{http://xxx.lanl.gov/abs/1203.6652}{{\tt 1203.6652}}.

\bibitem{Goodman:2010qn}
J.~Goodman, M.~Ibe, A.~Rajaraman, W.~Shepherd, T.~M. Tait, {\em et~al.}, {\it
  {Gamma Ray Line Constraints on Effective Theories of Dark Matter}},  {\em
  Nucl.Phys.} {\bf B844} (2011) 55--68,
  [\href{http://xxx.lanl.gov/abs/1009.0008}{{\tt 1009.0008}}].

\bibitem{Goodman:2010yf}
J.~Goodman, M.~Ibe, A.~Rajaraman, W.~Shepherd, T.~M. Tait, {\em et~al.}, {\it
  {Constraints on Light Majorana dark Matter from Colliders}},  {\em
  Phys.Lett.} {\bf B695} (2011) 185--188,
  [\href{http://xxx.lanl.gov/abs/1005.1286}{{\tt 1005.1286}}].

\bibitem{Goodman:2010ku}
J.~Goodman, M.~Ibe, A.~Rajaraman, W.~Shepherd, T.~M. Tait, {\em et~al.}, {\it
  {Constraints on Dark Matter from Colliders}},  {\em Phys.Rev.} {\bf D82}
  (2010) 116010, [\href{http://xxx.lanl.gov/abs/1008.1783}{{\tt 1008.1783}}].

\bibitem{Weiner:2012cb}
N.~Weiner and I.~Yavin, {\it {How Dark Are Majorana WIMPs? Signals from MiDM
  and Rayleigh Dark Matter}},  {\em Phys.Rev.} {\bf D86} (2012) 075021,
  [\href{http://xxx.lanl.gov/abs/1206.2910}{{\tt 1206.2910}}].

\bibitem{Ackermann:2012qk}
{\bf LAT Collaboration} Collaboration, M.~Ackermann {\em et~al.}, {\it {Fermi
  LAT Search for Dark Matter in Gamma-ray Lines and the Inclusive Photon
  Spectrum}},  {\em Phys.Rev.} {\bf D86} (2012) 022002,
  [\href{http://xxx.lanl.gov/abs/1205.2739}{{\tt 1205.2739}}].

\bibitem{Bringmann:2012vr}
T.~Bringmann, X.~Huang, A.~Ibarra, S.~Vogl, and C.~Weniger, {\it {Fermi LAT
  Search for Internal Bremsstrahlung Signatures from Dark Matter
  Annihilation}},  \href{http://xxx.lanl.gov/abs/1203.1312}{{\tt 1203.1312}}.

\bibitem{Weniger:2012tx}
C.~Weniger, {\it {A Tentative Gamma-Ray Line from Dark Matter Annihilation at
  the Fermi Large Area Telescope}},
  \href{http://xxx.lanl.gov/abs/1204.2797}{{\tt 1204.2797}}.

\bibitem{Tempel:2012ey}
E.~Tempel, A.~Hektor, and M.~Raidal, {\it {Fermi 130 GeV gamma-ray excess and
  dark matter annihilation in sub-haloes and in the Galactic centre}},
  \href{http://xxx.lanl.gov/abs/1205.1045}{{\tt 1205.1045}}.

\bibitem{Su:2012ft}
M.~Su and D.~P. Finkbeiner, {\it {Strong Evidence for Gamma-ray Lines from the
  Inner Galaxy}},  \href{http://xxx.lanl.gov/abs/1206.1616}{{\tt 1206.1616}}.

\bibitem{Hektor:2012jc}
A.~Hektor, M.~Raidal, and E.~Tempel, {\it {Double gamma-ray lines from
  unassociated Fermi-LAT sources revisited}},
  \href{http://xxx.lanl.gov/abs/1208.1996}{{\tt 1208.1996}}.

\bibitem{Hektor:2012kc}
A.~Hektor, M.~Raidal, and E.~Tempel, {\it {An evidence for indirect detection
  of dark matter from galaxy clusters in Fermi-LAT data}},  {\em Astrophys.J.}
  {\bf 762} (2013) L22, [\href{http://xxx.lanl.gov/abs/1207.4466}{{\tt
  1207.4466}}].

\bibitem{Hektor:2012ev}
A.~Hektor, M.~Raidal, and E.~Tempel, {\it {Fermi-LAT gamma-ray signal from
  Earth Limb, systematic detector effects and their implications for the 130
  GeV gamma-ray excess}},  \href{http://xxx.lanl.gov/abs/1209.4548}{{\tt
  1209.4548}}.

\bibitem{Whiteson:2012hr}
D.~Whiteson, {\it {Disentangling Instrumental Features of the 130 GeV Fermi
  Line}},  {\em JCAP} {\bf 1211} (2012) 008,
  [\href{http://xxx.lanl.gov/abs/1208.3677}{{\tt 1208.3677}}].

\bibitem{Finkbeiner:2012ez}
D.~P. Finkbeiner, M.~Su, and C.~Weniger, {\it {Is the 130 GeV Line Real? A
  Search for Systematics in the Fermi-LAT Data}},  {\em JCAP} {\bf 1301} (2013)
  029, [\href{http://xxx.lanl.gov/abs/1209.4562}{{\tt 1209.4562}}].

\bibitem{Rao:2012fh}
K.~Rao and D.~Whiteson, {\it {Where are the Fermi Lines Coming From?}},
  \href{http://xxx.lanl.gov/abs/1210.4934}{{\tt 1210.4934}}.

\bibitem{Buchmuller:2012rc}
W.~Buchmuller and M.~Garny, {\it {Decaying vs Annihilating Dark Matter in Light
  of a Tentative Gamma-Ray Line}},  {\em JCAP} {\bf 1208} (2012) 035,
  [\href{http://xxx.lanl.gov/abs/1206.7056}{{\tt 1206.7056}}].

\bibitem{Cohen:2012me}
T.~Cohen, M.~Lisanti, T.~R. Slatyer, and J.~G. Wacker, {\it {Illuminating the
  130 GeV Gamma Line with Continuum Photons}},  {\em JHEP} {\bf 1210} (2012)
  134, [\href{http://xxx.lanl.gov/abs/1207.0800}{{\tt 1207.0800}}].

\bibitem{Cholis:2012fb}
I.~Cholis, M.~Tavakoli, and P.~Ullio, {\it {Searching for the continuum
  spectrum photons correlated to the 130 GeV gamma-ray line}},  {\em Phys.Rev.}
  {\bf D86} (2012) 083525, [\href{http://xxx.lanl.gov/abs/1207.1468}{{\tt
  1207.1468}}].

\bibitem{Blanchet:2012vq}
S.~Blanchet and J.~Lavalle, {\it {Diffuse gamma-ray constraints on dark matter
  revisited. I: the impact of subhalos}},  {\em JCAP} {\bf 1211} (2012) 021,
  [\href{http://xxx.lanl.gov/abs/1207.2476}{{\tt 1207.2476}}].

\bibitem{Asano:2012zv}
M.~Asano, T.~Bringmann, G.~Sigl, and M.~Vollmann, {\it {The 130 GeV gamma-ray
  line and generic dark matter model building constraints from continuum gamma
  rays, radio and antiproton data}},
  \href{http://xxx.lanl.gov/abs/1211.6739}{{\tt 1211.6739}}.

\bibitem{Weiner:2012gm}
N.~Weiner and I.~Yavin, {\it {UV Completions of Magnetic Inelastic Dark Matter
  and RayDM for the Fermi Line(s)}},  {\em Phys.Rev.} {\bf D87} (2013) 023523,
  [\href{http://xxx.lanl.gov/abs/1209.1093}{{\tt 1209.1093}}].

\bibitem{Dudas:2012pb}
E.~Dudas, Y.~Mambrini, S.~Pokorski, and A.~Romagnoni, {\it {Extra U(1) as
  natural source of a monochromatic gamma ray line}},
  \href{http://xxx.lanl.gov/abs/1205.1520}{{\tt 1205.1520}}.

\bibitem{Cline:2012nw}
J.~M. Cline, {\it {130 GeV dark matter and the Fermi gamma-ray line}},  {\em
  Phys.Rev.} {\bf D86} (2012) 015016,
  [\href{http://xxx.lanl.gov/abs/1205.2688}{{\tt 1205.2688}}].

\bibitem{Choi:2012ap}
K.-Y. Choi and O.~Seto, {\it {A Dirac right-handed sneutrino dark matter and
  its signature in the gamma-ray lines}},  {\em Phys.Rev.} {\bf D86} (2012)
  043515, [\href{http://xxx.lanl.gov/abs/1205.3276}{{\tt 1205.3276}}].

\bibitem{Rajaraman:2012db}
A.~Rajaraman, T.~M. Tait, and D.~Whiteson, {\it {Two Lines or Not Two Lines?
  That is the Question of Gamma Ray Spectra}},
  \href{http://xxx.lanl.gov/abs/1205.4723}{{\tt 1205.4723}}.

\bibitem{Buckley:2012ws}
M.~R. Buckley and D.~Hooper, {\it {Implications of a 130 GeV Gamma-Ray Line for
  Dark Matter}},  {\em Phys.Rev.} {\bf D86} (2012) 043524,
  [\href{http://xxx.lanl.gov/abs/1205.6811}{{\tt 1205.6811}}].

\bibitem{Das:2012ys}
D.~Das, U.~Ellwanger, and P.~Mitropoulos, {\it {A 130 GeV photon line from dark
  matter annihilation in the NMSSM}},  {\em JCAP} {\bf 1208} (2012) 003,
  [\href{http://xxx.lanl.gov/abs/1206.2639}{{\tt 1206.2639}}].

\bibitem{Heo:2012dk}
J.~H. Heo and C.~Kim, {\it {Cosmic ray signatures of Dipole-Interacting
  Fermionic Dark Matter}},  \href{http://xxx.lanl.gov/abs/1207.1341}{{\tt
  1207.1341}}.

\bibitem{Park:2012xq}
J.-C. Park and S.~C. Park, {\it {Radiatively decaying scalar dark matter
  through U(1) mixings and the Fermi 130 GeV gamma-ray line}},
  \href{http://xxx.lanl.gov/abs/1207.4981}{{\tt 1207.4981}}.

\bibitem{Tulin:2012uq}
S.~Tulin, H.-B. Yu, and K.~M. Zurek, {\it {Three Exceptions for Thermal Dark
  Matter with Enhanced Annihilation to Gamma Gamma}},
  \href{http://xxx.lanl.gov/abs/1208.0009}{{\tt 1208.0009}}.

\bibitem{Cline:2012bz}
J.~M. Cline, G.~D. Moore, and A.~R. Frey, {\it {Composite magnetic dark matter
  and the 130 GeV line}},  {\em Phys.Rev.} {\bf D86} (2012) 115013,
  [\href{http://xxx.lanl.gov/abs/1208.2685}{{\tt 1208.2685}}].

\bibitem{Bai:2012qy}
Y.~Bai and J.~Shelton, {\it {Gamma Lines without a Continuum: Thermal Models
  for the Fermi-LAT 130 GeV Gamma Line}},
  \href{http://xxx.lanl.gov/abs/1208.4100}{{\tt 1208.4100}}.

\bibitem{Bringmann:2012mx}
T.~Bringmann and C.~Weniger, {\it {Gamma Ray Signals from Dark Matter:
  Concepts, Status and Prospects}},
  \href{http://xxx.lanl.gov/abs/1208.5481}{{\tt 1208.5481}}.

\bibitem{Bergstrom:2012mx}
L.~Bergstrom, {\it {The 130 GeV Fingerprint of Right-Handed Neutrino Dark
  Matter}},  \href{http://xxx.lanl.gov/abs/1208.6082}{{\tt 1208.6082}}.

\bibitem{Fan:2012gr}
J.~Fan and M.~Reece, {\it {A Simple Recipe for the 111 and 128 GeV Lines}},
  \href{http://xxx.lanl.gov/abs/1209.1097}{{\tt 1209.1097}}.

\bibitem{D'Eramo:2012rr}
F.~D'Eramo, M.~McCullough, and J.~Thaler, {\it {Multiple Gamma Lines from
  Semi-Annihilation}},  \href{http://xxx.lanl.gov/abs/1210.7817}{{\tt
  1210.7817}}.

\bibitem{Rajaraman:2012fu}
A.~Rajaraman, T.~M. Tait, and A.~M. Wijangco, {\it {Effective Theories of
  Gamma-ray Lines from Dark Matter Annihilation}},
  \href{http://xxx.lanl.gov/abs/1211.7061}{{\tt 1211.7061}}.

\bibitem{Fan:2013qn}
J.~Fan and M.~Reece, {\it {Probing Charged Matter Through Higgs Diphoton Decay,
  Gamma Ray Lines, and EDMs}},  \href{http://xxx.lanl.gov/abs/1301.2597}{{\tt
  1301.2597}}.

\bibitem{Lee:2012ph}
H.~M. Lee, M.~Park, and V.~Sanz, {\it {Interplay between Fermi gamma-ray lines
  and collider searches}},  \href{http://xxx.lanl.gov/abs/1212.5647}{{\tt
  1212.5647}}.

\bibitem{Kopp:2013mi}
J.~Kopp, E.~T. Neil, R.~Primulando, and J.~Zupan, {\it {From gamma ray line
  signals of dark matter to the LHC}},
  \href{http://xxx.lanl.gov/abs/1301.1683}{{\tt 1301.1683}}.

\bibitem{Fox:2011pm}
P.~J. Fox, R.~Harnik, J.~Kopp, and Y.~Tsai, {\it {Missing Energy Signatures of
  Dark Matter at the LHC}},  {\em Phys.Rev.} {\bf D85} (2012) 056011,
  [\href{http://xxx.lanl.gov/abs/1109.4398}{{\tt 1109.4398}}].

\bibitem{Abreu:1999qr}
{\bf DELPHI} Collaboration, P.~Abreu {\em et~al.}, {\it {Search for charginos
  nearly mass - degenerate with the lightest neutralino}},  {\em Eur.Phys.J.}
  {\bf C11} (1999) 1--17, [\href{http://xxx.lanl.gov/abs/hep-ex/9903071}{{\tt
  hep-ex/9903071}}].

\bibitem{Feng:1999fu}
J.~L. Feng, T.~Moroi, L.~Randall, M.~Strassler, and S.-f. Su, {\it {Discovering
  supersymmetry at the Tevatron in wino LSP scenarios}},  {\em Phys.Rev.Lett.}
  {\bf 83} (1999) 1731--1734,
  [\href{http://xxx.lanl.gov/abs/hep-ph/9904250}{{\tt hep-ph/9904250}}].

\bibitem{Gunion:1999jr}
J.~F. Gunion and S.~Mrenna, {\it {A Study of SUSY signatures at the Tevatron in
  models with near mass degeneracy of the lightest chargino and neutralino}},
  {\em Phys.Rev.} {\bf D62} (2000) 015002,
  [\href{http://xxx.lanl.gov/abs/hep-ph/9906270}{{\tt hep-ph/9906270}}].

\bibitem{Ibe:2006de}
M.~Ibe, T.~Moroi, and T.~Yanagida, {\it {Possible Signals of Wino LSP at the
  Large Hadron Collider}},  {\em Phys.Lett.} {\bf B644} (2007) 355--360,
  [\href{http://xxx.lanl.gov/abs/hep-ph/0610277}{{\tt hep-ph/0610277}}].

\bibitem{Asai:2008sk}
S.~Asai, T.~Moroi, and T.~Yanagida, {\it {Test of Anomaly Mediation at the
  LHC}},  {\em Phys.Lett.} {\bf B664} (2008) 185--189,
  [\href{http://xxx.lanl.gov/abs/0802.3725}{{\tt 0802.3725}}].

\bibitem{FileviezPerez:2008bj}
P.~Fileviez~Perez, H.~H. Patel, M.~Ramsey-Musolf, and K.~Wang, {\it {Triplet
  Scalars and Dark Matter at the LHC}},  {\em Phys.Rev.} {\bf D79} (2009)
  055024, [\href{http://xxx.lanl.gov/abs/0811.3957}{{\tt 0811.3957}}].

\bibitem{Buckley:2009kv}
M.~R. Buckley, L.~Randall, and B.~Shuve, {\it {LHC Searches for Non-Chiral
  Weakly Charged Multiplets}},  {\em JHEP} {\bf 1105} (2011) 097,
  [\href{http://xxx.lanl.gov/abs/0909.4549}{{\tt 0909.4549}}].

\bibitem{Kanemura:2011kx}
S.~Kanemura, K.~Tsumura, and H.~Yokoya, {\it {Multi-tau-lepton signatures at
  the LHC in the two Higgs doublet model}},  {\em Phys.Rev.} {\bf D85} (2012)
  095001, [\href{http://xxx.lanl.gov/abs/1111.6089}{{\tt 1111.6089}}].

\bibitem{Randall:1998uk}
L.~Randall and R.~Sundrum, {\it {Out of this world supersymmetry breaking}},
  {\em Nucl.Phys.} {\bf B557} (1999) 79--118,
  [\href{http://xxx.lanl.gov/abs/hep-th/9810155}{{\tt hep-th/9810155}}].

\bibitem{Giudice:1998xp}
G.~F. Giudice, M.~A. Luty, H.~Murayama, and R.~Rattazzi, {\it {Gaugino mass
  without singlets}},  {\em JHEP} {\bf 9812} (1998) 027,
  [\href{http://xxx.lanl.gov/abs/hep-ph/9810442}{{\tt hep-ph/9810442}}].

\bibitem{Alwall:2011uj}
J.~Alwall, M.~Herquet, F.~Maltoni, O.~Mattelaer, and T.~Stelzer, {\it {MadGraph
  5 : Going Beyond}},  {\em JHEP} {\bf 1106} (2011) 128,
  [\href{http://xxx.lanl.gov/abs/1106.0522}{{\tt 1106.0522}}].

\bibitem{Sjostrand:2006za}
T.~Sjostrand, S.~Mrenna, and P.~Z. Skands, {\it {PYTHIA 6.4 Physics and
  Manual}},  {\em JHEP} {\bf 0605} (2006) 026,
  [\href{http://xxx.lanl.gov/abs/hep-ph/0603175}{{\tt hep-ph/0603175}}].

\bibitem{pgs4}
J.~Conway {\em et~al.} {\em
  http://www.physics.ucdavis.edu/$\sim$conway/research/software/pgs/pgs4-general.htm}
  (2012).

\bibitem{Eriksson:2009ws}
D.~Eriksson, J.~Rathsman, and O.~Stal, {\it {2HDMC: Two-Higgs-Doublet Model
  Calculator Physics and Manual}},  {\em Comput.Phys.Commun.} {\bf 181} (2010)
  189--205, [\href{http://xxx.lanl.gov/abs/0902.0851}{{\tt 0902.0851}}].

\bibitem{ATLAS:2012ht}
{\bf ATLAS} Collaboration, G.~Aad {\em et~al.}, {\it {Search for Supersymmetry
  in Events with Large Missing Transverse Momentum, Jets, and at Least One Tau
  Lepton in 7 TeV Proton-Proton Collision Data with the ATLAS Detector}},  {\em
  Eur.Phys.J.} {\bf C72} (2012) 2215,
  [\href{http://xxx.lanl.gov/abs/1210.1314}{{\tt 1210.1314}}].

\bibitem{Kolb:1990vq}
E.~W. Kolb and M.~S. Turner, {\it {The Early universe}},  {\em Front.Phys.}
  {\bf 69} (1990) 1--547.

\bibitem{Aprile:2012nq}
{\bf XENON100} Collaboration, E.~Aprile {\em et~al.}, {\it {Dark Matter Results
  from 225 Live Days of XENON100 Data}},  {\em Phys.Rev.Lett.} {\bf 109} (2012)
  181301, [\href{http://xxx.lanl.gov/abs/1207.5988}{{\tt 1207.5988}}].

\bibitem{Han:1997wn}
T.~Han and R.~Hempfling, {\it {Messenger sneutrinos as cold dark matter}},
  {\em Phys.Lett.} {\bf B415} (1997) 161--169,
  [\href{http://xxx.lanl.gov/abs/hep-ph/9708264}{{\tt hep-ph/9708264}}].

\bibitem{Hall:1997ah}
L.~J. Hall, T.~Moroi, and H.~Murayama, {\it {Sneutrino cold dark matter with
  lepton number violation}},  {\em Phys.Lett.} {\bf B424} (1998) 305--312,
  [\href{http://xxx.lanl.gov/abs/hep-ph/9712515}{{\tt hep-ph/9712515}}].

\bibitem{TuckerSmith:2001hy}
D.~Tucker-Smith and N.~Weiner, {\it {Inelastic dark matter}},  {\em Phys. Rev.}
  {\bf D64} (2001) 043502, [\href{http://xxx.lanl.gov/abs/hep-ph/0101138}{{\tt
  hep-ph/0101138}}].

\bibitem{ATLAS:2012jp}
{\bf ATLAS} Collaboration, G.~Aad {\em et~al.}, {\it {Search for direct
  chargino production in anomaly-mediated supersymmetry breaking models based
  on a disappearing-track signature in $pp$ collisions at $\sqrt{s}=7$ TeV with
  the ATLAS detector}},  {\em JHEP} {\bf 1301} (2013) 131,
  [\href{http://xxx.lanl.gov/abs/1210.2852}{{\tt 1210.2852}}].

\bibitem{Fox:2012ee}
P.~J. Fox, R.~Harnik, R.~Primulando, and C.-T. Yu, {\it {Taking a Razor to Dark
  Matter Parameter Space at the LHC}},  {\em Phys.Rev.} {\bf D86} (2012)
  015010, [\href{http://xxx.lanl.gov/abs/1203.1662}{{\tt 1203.1662}}].

\bibitem{Ali:1999we}
A.~Ali and D.~London, {\it {Profiles of the unitarity triangle and CP violating
  phases in the standard model and supersymmetric theories}},  {\em
  Eur.Phys.J.} {\bf C9} (1999) 687--703,
  [\href{http://xxx.lanl.gov/abs/hep-ph/9903535}{{\tt hep-ph/9903535}}].

\bibitem{Buras:2000dm}
A.~Buras, P.~Gambino, M.~Gorbahn, S.~Jager, and L.~Silvestrini, {\it {Universal
  unitarity triangle and physics beyond the standard model}},  {\em Phys.Lett.}
  {\bf B500} (2001) 161--167,
  [\href{http://xxx.lanl.gov/abs/hep-ph/0007085}{{\tt hep-ph/0007085}}].

\bibitem{D'Ambrosio:2002ex}
G.~D'Ambrosio, G.~Giudice, G.~Isidori, and A.~Strumia, {\it {Minimal flavor
  violation: An Effective field theory approach}},  {\em Nucl.Phys.} {\bf B645}
  (2002) 155--187, [\href{http://xxx.lanl.gov/abs/hep-ph/0207036}{{\tt
  hep-ph/0207036}}].

\bibitem{ATLAS-CONF-2012-154}
{\it Search for direct production of charginos and neutralinos in events with
  three leptons and missing transverse momentum in 13.0 fb-1 of pp collisions
  at sqrt(s)=8 tev with the atlas detector},  Tech. Rep. ATLAS-CONF-2012-154,
  CERN, Geneva, Nov, 2012.

\bibitem{CMS-PAS-SUS-12-022}
{\it Search for direct ewk production of susy particles in multilepton modes
  with 8tev data},  Tech. Rep. CMS-PAS-SUS-12-022, 2012.

\bibitem{Andreev:2004qq}
Y.~Andreev, S.~Bityukov, and N.~Krasnikov, {\it {Sleptons at post-WMAP
  benchmark points at LHC(CMS)}},  {\em Phys.Atom.Nucl.} {\bf 68} (2005)
  340--347, [\href{http://xxx.lanl.gov/abs/hep-ph/0402229}{{\tt
  hep-ph/0402229}}].

\bibitem{ATLAS:2012mn}
{\bf ATLAS} Collaboration, G.~Aad {\em et~al.}, {\it {Search for anomalous
  production of prompt like-sign lepton pairs at $\sqrt{s}=7$ TeV with the
  ATLAS detector}},  {\em JHEP} {\bf 1212} (2012) 007,
  [\href{http://xxx.lanl.gov/abs/1210.4538}{{\tt 1210.4538}}].

\bibitem{CMS-PAS-SUS-12-026}
{\it A search for anomalous production of events with three or more leptons
  using $9.2\,\text{fb}^-1$},  Tech. Rep. CMS-PAS-SUS-12-026, 2012.

\bibitem{Blank:1997qa}
T.~Blank and W.~Hollik, {\it {Precision observables in SU(2) x U(1) models with
  an additional Higgs triplet}},  {\em Nucl.Phys.} {\bf B514} (1998) 113--134,
  [\href{http://xxx.lanl.gov/abs/hep-ph/9703392}{{\tt hep-ph/9703392}}].

\bibitem{Chen:2006pb}
M.-C. Chen, S.~Dawson, and T.~Krupovnickas, {\it {Higgs triplets and limits
  from precision measurements}},  {\em Phys.Rev.} {\bf D74} (2006) 035001,
  [\href{http://xxx.lanl.gov/abs/hep-ph/0604102}{{\tt hep-ph/0604102}}].

\bibitem{Glashow:1976nt}
S.~L. Glashow and S.~Weinberg, {\it {Natural Conservation Laws for Neutral
  Currents}},  {\em Phys.Rev.} {\bf D15} (1977) 1958.

\bibitem{Mahmoudi:2009zx}
F.~Mahmoudi and O.~Stal, {\it {Flavor constraints on the two-Higgs-doublet
  model with general Yukawa couplings}},  {\em Phys.Rev.} {\bf D81} (2010)
  035016, [\href{http://xxx.lanl.gov/abs/0907.1791}{{\tt 0907.1791}}].

\bibitem{Craig:2012pu}
N.~Craig, J.~A. Evans, R.~Gray, C.~Kilic, M.~Park, {\em et~al.}, {\it
  {Multi-Lepton Signals of Multiple Higgs Bosons}},  {\em JHEP} {\bf 1302}
  (2013) 033, [\href{http://xxx.lanl.gov/abs/1210.0559}{{\tt 1210.0559}}].

\bibitem{Holzner:2001tv}
A.~G. Holzner, {\it {Searches for charged Higgs bosons at LEP}},
  \href{http://xxx.lanl.gov/abs/hep-ex/0105045}{{\tt hep-ex/0105045}}.

\bibitem{LEPHiggsWorking:2001ab}
{\bf ALEPH collaboration, DELPHI collaboration, L3 collaboration, OPAL
  Collaboration, and the LEP Higgs Working Group} Collaboration, LEP, {\it
  {Searches for the neutral Higgs bosons of the MSSM: Preliminary combined
  results using LEP data collected at energies up to 209-GeV}},
  \href{http://xxx.lanl.gov/abs/hep-ex/0107030}{{\tt hep-ex/0107030}}.

\bibitem{ATLAS-CONF-2013-012}
{\it Measurements of the properties of the higgs-like boson in the two photon
  decay channel with the atlas detector using 25 $\mathrm{fb}^{-1}$ of
  proton-proton collision data},  Tech. Rep. ATLAS-CONF-2013-012, CERN, Geneva,
  Mar, 2013.

\bibitem{Kusenko:2009up}
A.~Kusenko, {\it {Sterile neutrinos: The Dark side of the light fermions}},
  {\em Phys.Rept.} {\bf 481} (2009) 1--28,
  [\href{http://xxx.lanl.gov/abs/0906.2968}{{\tt 0906.2968}}].

\bibitem{Fox:2011fx}
P.~J. Fox, R.~Harnik, J.~Kopp, and Y.~Tsai, {\it {LEP Shines Light on Dark
  Matter}},  {\em Phys.Rev.} {\bf D84} (2011) 014028,
  [\href{http://xxx.lanl.gov/abs/1103.0240}{{\tt 1103.0240}}].

\bibitem{Ade:2013zuv}
{\bf Planck Collaboration} Collaboration, P.~Ade {\em et~al.}, {\it {Planck
  2013 results. XVI. Cosmological parameters}},
  \href{http://xxx.lanl.gov/abs/1303.5076}{{\tt 1303.5076}}.

\bibitem{Curtin:2012nn}
D.~Curtin, P.~Jaiswal, and P.~Meade, {\it {Charginos Hiding In Plain Sight}},
  \href{http://xxx.lanl.gov/abs/1206.6888}{{\tt 1206.6888}}.

\end{thebibliography}\endgroup

\end{document}